\documentclass[sigconf,nonacm,letterpaper,10pt]{acmart}
\pdfoutput=1
\usepackage[english]{babel}
\usepackage{blindtext}
\usepackage{booktabs}
\usepackage{threeparttable}
\definecolor{strawberry}{rgb}{1.0, 0.26, 0.64}
\definecolor{ruby}{rgb}{0.88, 0.07, 0.37}
\definecolor{princetonorange}{rgb}{1.0, 0.56, 0.0}
\usepackage{hyperref}
\usepackage{enumitem}
\usepackage{epsfig,endnotes}
\usepackage[export]{adjustbox}
\usepackage{tabularx}
\usepackage{appendix}
\usepackage{url}
\usepackage{lineno}
\usepackage{pifont}
\usepackage[linesnumbered,ruled,noline,noend]{algorithm2e}
\usepackage{amsmath,amsfonts,amsthm}
\usepackage{cleveref}
\usepackage{mathtools}
\DeclarePairedDelimiter\ceil{\lceil}{\rceil}

\usepackage{rotating}
\usepackage{multirow}
\usepackage{hhline}
\usepackage{color}
\usepackage{xcolor}
\usepackage{epstopdf}
\usepackage{subfig}
\usepackage{colortbl}
\long\def\comment#1{}

\definecolor{lightgrey}{RGB}{144,144,144}

\newcommand{\TODO}[1]{{\color{red}[#1]}}
\newcommand{\mmm}[1]{{\color{blue}[#1]}}

\newcommand{\sys}{MOCC\xspace}

\newcommand{\mocc}{MOCC\xspace}

\newcommand{\paraspace}{\vspace{0.05in}}
\newcommand{\parab}[1]{\paraspace\noindent{\bf #1} }

\newenvironment{icompact}{
	\begin{list}{$\bullet$}{
			\parsep 1pt plus 1pt
			\partopsep 1pt plus 1pt
			\topsep 1pt plus 2pt minus 1pt
			\itemsep 1.5pt plus 1pt
			\parskip 0pt plus 2pt
			\leftmargin 0.15in}
	}
	{\normalsize\end{list}}

\def\ie{i.e.\xspace}

\makeatletter
\let\c@subfigure\relax
\let\c@subtable\relax
\let\@listsubcaptions\relax
\let\@dottedxxxline\relax
\let\l@subfigure\relax
\let\c@lofdepth\relax
\let\l@subtable\relax
\let\c@lotdepth\relax

\let\subfloat@label\relax
\let\sf@@sub@label\relax

\makeatother
\usepackage[tight]{subfigure}

\usepackage{breakurl}           % break too-long urls in refs
\usepackage{url}                % allow \url in bibtex for clickable links
\usepackage{xcolor}             % color definitions, to be use for...
\usepackage[]{hyperref}         % ...clickable refs within pdf...

\usepackage{tabularx}
\usepackage{booktabs}

\begin{document}\sloppy

\title{Multi-Objective Congestion Control}
%\author{Yiqing Ma$^1$~~~~Han Tian$^1$~~~~Xudong Liao$^1$~~~~Junxue Zhang$^1$~~~~Weiyan Wang$^1$~~~~~~~Kai Chen$^1$~~~~~Xin Jin$^2$}
\affiliation{
Yiqing Ma$^1$~~~Han Tian$^1$~~~Xudong Liao$^1$~~~Junxue Zhang$^1$~~~Weiyan Wang$^1$~~~Kai Chen$^1$~~~Xin Jin$^2$\\
 $^1$iSING Lab, Hong Kong University of Science and Technology\\
 $^2$Peking University \\
}

\maketitle
\thispagestyle{empty}
\section*{Abstract}
%\begin{abstract}
Decades of research on Internet congestion control (CC) has produced a plethora
of algorithms that optimize for \emph{different} performance objectives.
Applications face the challenge of choosing the most suitable algorithm based on
their needs, and it takes tremendous efforts and expertise to customize CC
algorithms when new demands emerge. In this paper, we explore a basic question:
can we design a \emph{single} CC algorithm to satisfy \emph{different}
objectives?

We propose \sys, the \emph{first} multi-objective congestion control algorithm
that attempts to address this challenge. The core of \sys is a novel
\emph{multi-objective reinforcement learning} framework for CC that can automatically
learn the correlations between different application requirements and the
corresponding optimal control policies. Under this framework, \sys further
applies \emph{transfer learning} to transfer the knowledge from past
experience to new applications, quickly adapting itself to a new objective even if it is \emph{unforeseen}. We provide both user-space and kernel-space implementation of \sys. Real-world experiments and extensive simulations show that \sys well supports multi-objective, competing or outperforming the
best existing CC algorithms on individual objectives, and quickly adapting to new
applications (e.g., $14.2\times$ faster than prior work) without compromising old ones.

\vspace{-2mm}
\section{Introduction}\label{sec:intro}%\vspace{-1mm}

Congestion control (CC) is a fundamental, enduring topic in networking
research. Decades of study on this topic have produced a plethora of CC
algorithms~\cite{vegas, orca, cubic, newreno, compound, bbr, copa, fast, remy,
allegro, vivace, aurora, chen2013towards, zeng2019congestion}. These algorithms are motivated by new applications
that impose different demands on network performance, as well as new
technologies that change the underlying Internet infrastructure. The confluence
of these two factors requires a CC algorithm to be deliberately designed to
optimize for a particular performance objective.

Consequently, applications face the challenge of choosing the most suitable CC
algorithm based on their needs. This choice is definitely not easy given the
wide range of options, and the subtle differences between them that oftentimes
require deep understanding of TCP minutiae. At the same time, whenever new
applications with different demands emerge, it takes tremendous efforts and
expertise to customize CC algorithms for their new requirements.

In this paper, we explore a basic question: can we design a \emph{single} CC
algorithm to satisfy \emph{different} objectives? Traditional CC
algorithms~\cite{vegas, cubic, newreno, compound, bbr, copa, fast} are
\emph{hand-crafted}. They rely on certain assumptions about the network, and
hardwire packet-level events to pre-defined control rules based on human
experience. Recent learning-based CC algorithms~\cite{aurora,orca} relieve the burden by applying deep reinforcement learning to automatically \emph{learn} an optimal control policy for a given objective.
Yet, satisfying different objectives requires us to maintain one copy for each
algorithm (either traditional or learning-based), and pay excessive time to design or train an algorithm each time when a new application with a
different objective arrives ($\S$\ref{sec:motivation}).

We propose \sys, the \emph{first} multi-objective CC algorithm
that attempts to address this question ($\S$\ref{sec:MORL}). The core of \sys is a novel
\emph{multi-objective reinforcement learning} framework for CC to automatically
learn the correlations between different application requirements and their
corresponding optimal control policies. \sys explicitly incorporates the
performance objective into both the state input and the dynamic reward function,
and leverages a new policy neural network with \emph{a preference sub-network}
to correlate different objectives with optimal control policies ($\S$\ref{subsec:architecture}).
This allows \sys to effectively establish {\em a single correlation model} to
support different performance objectives. Under this framework, \sys further
applies \emph{transfer learning} to quickly transfer the knowledge learned from past
experience to new applications, and optimizes the CC algorithm for a given
objective, even if it is \emph{unforeseen}.

\sys achieves its goal by a combination of offline training ($\S$\ref{subsec:offline}) and online
adaptation ($\S$\ref{subsec:online}). In offline training, \sys is trained over a set of
well-distributed landmark objectives to learn the base correlations between application
requirements and optimal policies. Then, whenever a new application arrives, \sys can {\em immediately} provide a moderate policy using the offline trained model by
correlating the application's objective with the landmark objectives, even if
it is unforeseen. Meanwhile, \sys activates online adaptation to
transfer the knowledge from the base correlation model to the new application.
With transfer learning, \sys can quickly converge to the optimal policy
within just a few training iterations, orders of magnitude faster than
training from scratch. In addition, to avoid forgetting the learned policies,
we customize the loss function of \sys online adaptation for both new
arrival and old (sampled) applications. This enables \sys to learn and apply
optimal policies for new applications without compromising old ones. 

We fully implement \sys ($\S$\ref{sec:implement}) with a user-space implementation based on UDT~\cite{udt} and a kernel-space implementation based on CCP~\cite{ccp}. We
leverage OpenAI Gym~\cite{openai} and Aurora~\cite{aurora} to implement the
training and adaptation components, and use parallel training to reduce training
time. For better portability, we encapsulate all \sys's functions into one
library that is plug-and-play and readily deployable with any networking data
paths that include, but not limited to, our user-space and kernel-space
implementations.

We evaluate \sys with extensive simulations and real-world Internet experiments ($\S$\ref{sec:eva}). We show that \sys well supports multiple objectives, competing or outperforming the best existing CC algorithms (including both traditional ones and recent learning-based ones) on individual objectives ($\S$\ref{subsec:AA}), and can quickly adapt to new application objective in 288 seconds, $14.2\times$ faster than prior solution ($\S$\ref{subsec:QA}). We further demonstrate the benefits of \mocc with three real Internet applications in $\S$\ref{subsec:realapp}, and inspect the fairness and friendliness of \mocc in $\S$\ref{subsec:FF}. Finally, we deep-dive into various design choices of \sys and its overhead in $\S$\ref{subsec:DD}.

\vspace{-2mm}
%\input{sections/2background.tex}\vspace{-2mm}
%\vspace{-0.05in}
\section{Background and Motivation}\label{sec:motivation}

\subsection{Diverse Application Requirements}\label{subsec:appreq}
%\vspace{-0.1in}
Internet applications have diverse performance
requirements for the network, typically characterized by metrics such as
throughput, latency, jitter, and packet loss
rate~\cite{qos,ar,vr,vsnetwork,mmnetwork,kenyon1992audiovisual,silveira1999multimedia,szuprowicz1995multimedia}. 
Throughput is the main metric for many applications, in which minimum bandwidth
is required to provide good user experience, e.g., HDTV requires ($>$34Mbps) to play
high-definition video without rebuffering~\cite{qos}. On the other hand,
real-time interactive applications usually require low latency, e.g., autonomous
driving requires low latency ($<$15ms) to react to  immediate environment
signals~\cite{qos1}. For some real-time applications, temporal packet loss is
also important, e.g., online video/audio conferencing can only tolerate
($<$$0.1\%/1\%$) packet loss rate~\cite{fluckiger1995understanding}. Emerging
Internet applications such as augmented/virtual reality may have tight requirements
on several metrics simultaneously~\cite{2017vr}. 

To summarize, these application demands pose different requirements on Internet CC algorithms. Ideally, the CC algorithm should be {\em multi-objective} to support diverse application requirements simultaneously. However, as we will show subsequently ($\S$\ref{subsec:problem}), none of existing CC solutions can do this.

%\vspace{-0.1in}
\subsection{In Pursuit of Multi-Objective CC}\label{subsec:problem}

\begin{table}[t]
\centering
%\vspace{-0.1in}
%\resizebox{0.5\textwidth}{12mm}{
\begin{tabular}{|c|c|}
\hline
\rowcolor{black}
{\color{white}{Algorithm}} & {\color{white}{Objective}} \\
\hline
PCC Allegro~\cite{allegro}
    & \begin{math}T - \delta RTT\end{math} \\
\hline
PCC Vivace~\cite{vivace}
    & \begin{math}T^{t} - b \times \frac{d(RTT)}{dt} -c \times L \end{math} \\
\hline
Aurora~\cite{aurora}
    & \begin{math} \alpha T - \beta RTT- \gamma L\end{math} \\
\hline
Orca~\cite{orca}
    & \begin{math} \frac{T-\varepsilon L}{RTT} / (\frac{T_{max}}{RTT_{min}}) \end{math}\\
\hline
\end{tabular}%\vspace{-0.1in}
\caption{Performance objectives in learning-based CC. $T$ is throughput, $RTT$ is latency, and $L$ is loss rate.}\label{table:reward}
%\vspace{-0.45in}
\end{table}

\begin{figure*}[t]
%\vspace{-0.2in}
\centering
\subfigure[Support for throughput-intensive application] {
    \includegraphics[width=0.32\textwidth]{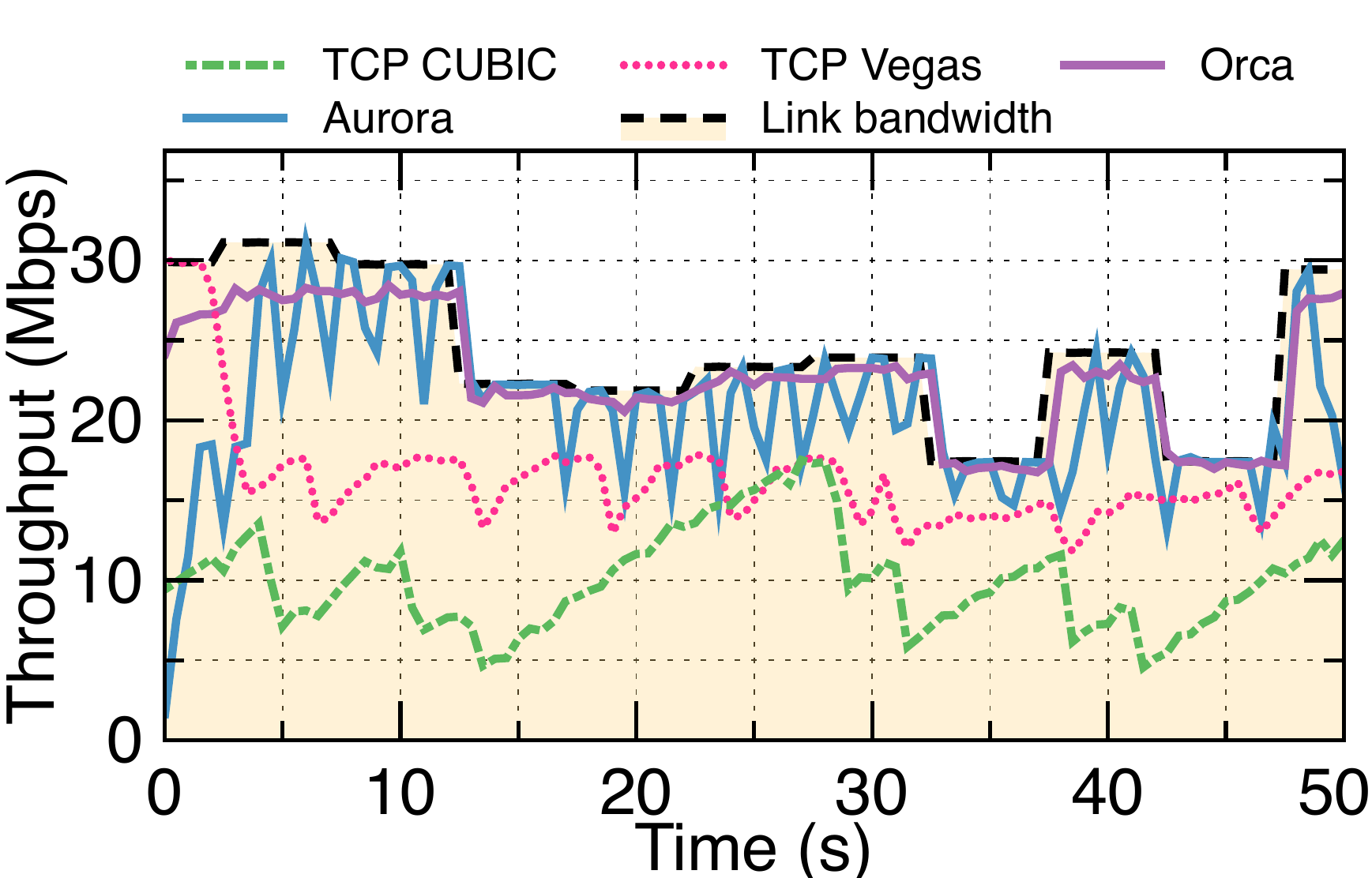}
  }
\subfigure[Support for different objectives]{
    \includegraphics[width=0.32\textwidth]{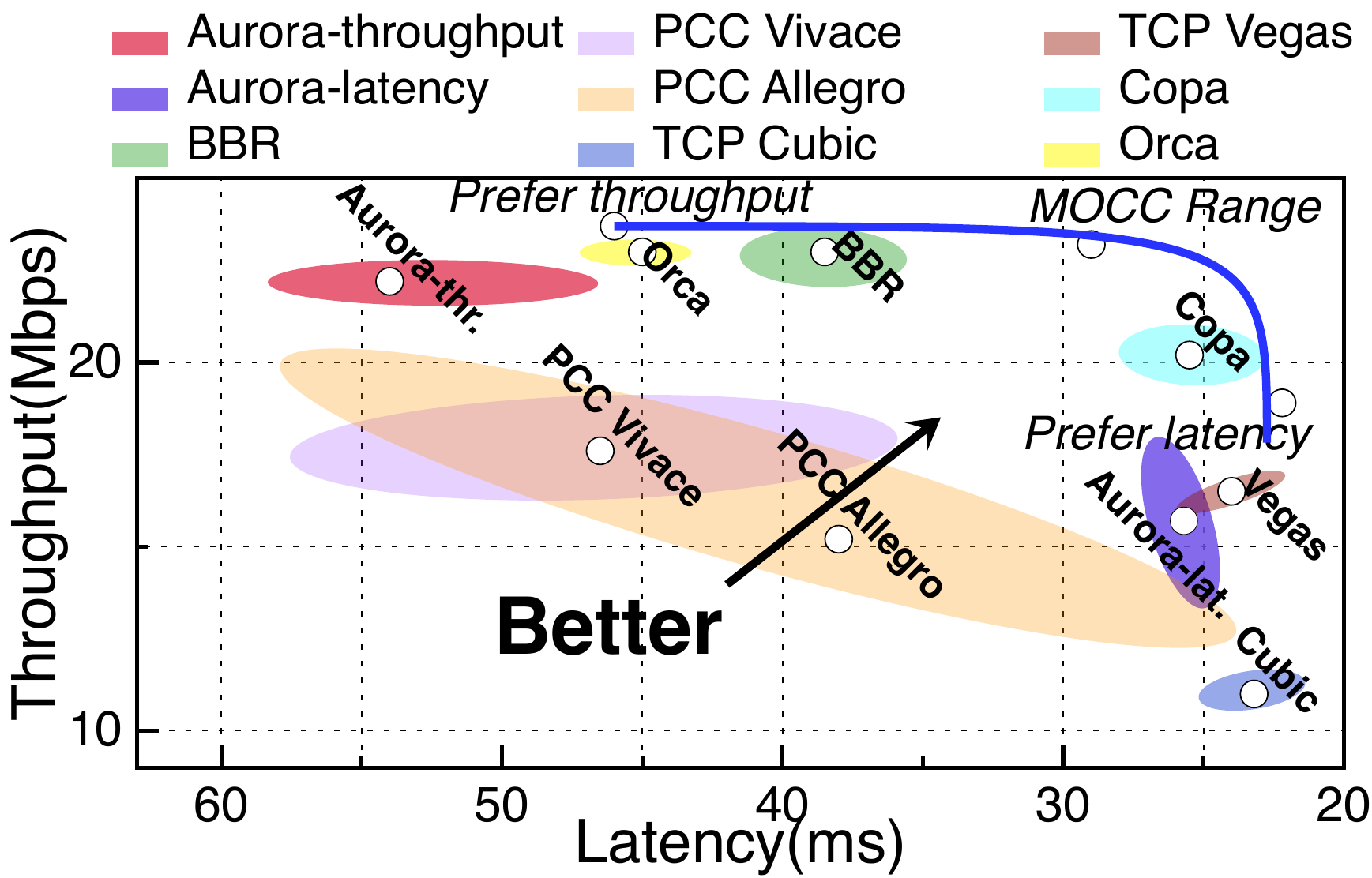}
  }
\subfigure[Time cost for re-training a new objective]{
    \includegraphics[width=0.32\textwidth]{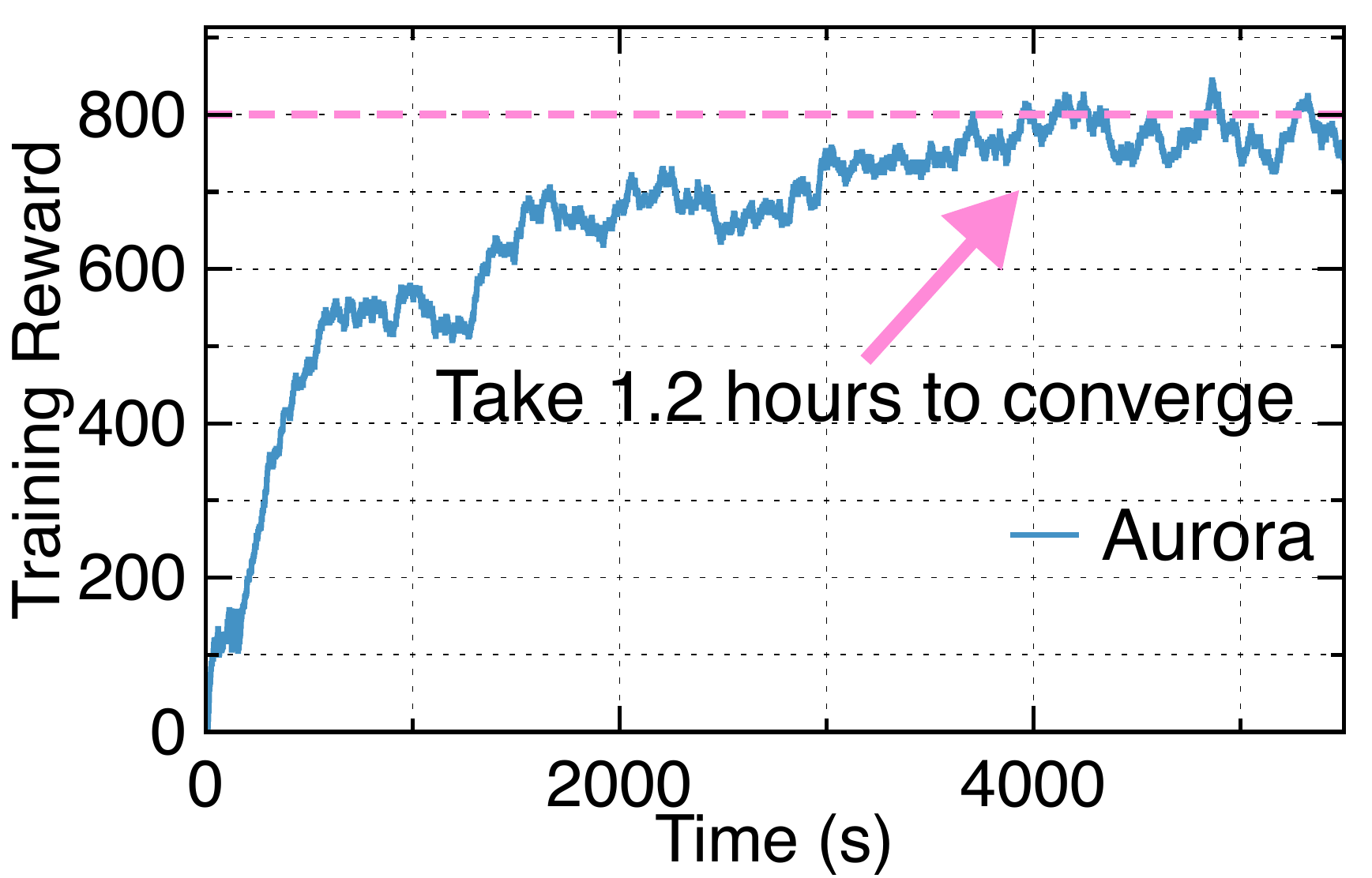}
  }
%\vspace{-0.2in}
\caption{(a) Learning-based CC algorithms can explicitly optimize for a particular
objective, and perform better than traditional CC algorithms. (b) Existing
learning-based CC cannot support multiple objectives. (c) Existing
learning-based CC such as Aurora takes a long time to re-train the
model when the objective changes.}
\label{fig:moti}
%\vspace{-0.2in}
\end{figure*}

We broadly classify existing Internet CC algorithms into two main categories: hand-crafted~\cite{vegas, cubic, newreno, compound, bbr, copa,fast} and learning-based~\cite{orca,aurora,vivace,allegro}. Traditional CC algorithms hardwire packet-level events to pre-defined control rules based on human experience. The performance objective is implicitly encoded in the mapping, and in many cases, it is hard to infer what is exactly being optimized.

\parab{Existing learning-based CC can optimize for a given objective.}
Recent learning-based CC algorithms can address the above problem of hand-crafted heuristics by explicitly encoding the performance objectives in the reward/utility function and maximizing it through machine learning from network environments. Table~\ref{table:reward} lists several reward/utility functions used by state-of-the-art learning-based CC algorithms.The reward function is typically expressed as a combination of metrics such as throughput, latency and loss rate. The coefficient parameters ($\alpha,\beta,\gamma,\delta,\varepsilon, b, c $) can express the relative importance of these metrics based on application requirements explicitly. Thus, learning-based CC is able to perform well for a particular objective.

We use a simple simulation to showcase this. The setup follows that in Orca~\cite{orca}. Specifically, we simulate a network in which the one-way delay is 20 ms, the bottleneck link bandwidth varies between 20--30Mbps, and the loss ratio is 0.02\%.
We compare two traditional CC algorithms (TCP CUBIC and Vegas) and two learning-based CC algorithms (Aurora and Orca). As shown in Figure~\ref{fig:moti}(a), CUBIC and Vegas under-utilize the bandwidth and do not perform well when the link bandwidth changes. In contrast, Aurora and Orca are trained by assigning high weight to
throughput in the reward function. As a result, they achieve higher throughput
than CUBIC and Vegas, and the throughput benefits are consistent under
changing network conditions.

\parab{But, these learning-based CC cannot support multiple objectives.} As listed
in Table~\ref{table:reward}, the current learning-based algorithms set relative
importance of throughput, latency and loss rate to realize different performance
objectives. These coefficient parameters are fixed during training. Thus the
trained agent can only optimize for a particular objective at a time. To show this, we reuse the
above simulated network setting to evaluate different CC algorithms (Aurora, PCC-Allegro, PCC-Vivace, BBR, Cubic, Vegas, Copa). For Aurora, we apply two
models, one trained for throughput (Aurora-throughput) and the other trained for
latency (Aurora-latency). We present throughput-delay plot for each CC in Figure~\ref{fig:moti}(b). We take each individual 60-second run as one point, and then compute the $1-\sigma$ elliptic contour of the maximum-likelihood 2D
Gaussian distribution that explains the points. As shown in the figure, from
right to left and bottom to top, these algorithms trace out a path from most
latency-optimized to most throughput-optimized. Aurora-throughput provides
higher throughput, while Aurora-latency provides lower latency. But each of
them can only optimize for one particular objective. In comparison, we propose
\sys, a multi-objective CC that can support different application
requirements. The ideal performance of \sys is shown in the blue line. By dynamic adjusting the relative importance of its reward function, \sys is expected to accommodate different objectives.

For learning-based approach, hypothetically, one can train a custom model for each
performance objective. However, this is undesirable: given the diverse
application requirements and with new applications emerging every year, it is
hard to exhaust all performance objectives. And even if possible, one may need
to install a copy of algorithm in each device and train the
algorithm at real time when the performance objective changes. However, existing
learn-based CC algorithms are not \emph{quick-adaptive}. As an example,
Figure~\ref{fig:moti}(c) shows that re-training the model of Aurora~\cite{aurora} for a new
objective takes more than one hour to converge. Besides these drawbacks, from a
scientific point of view, we would like to explore {\em whether it is possible to
design a single CC algorithm to satisfy multiple objectives}.

\begin{figure*}[ht]
%\vspace{-0.2in}
\centering
\subfigure[Single-objective CC (Aurora~\cite{aurora})] {
	\includegraphics[width=0.42\textwidth]{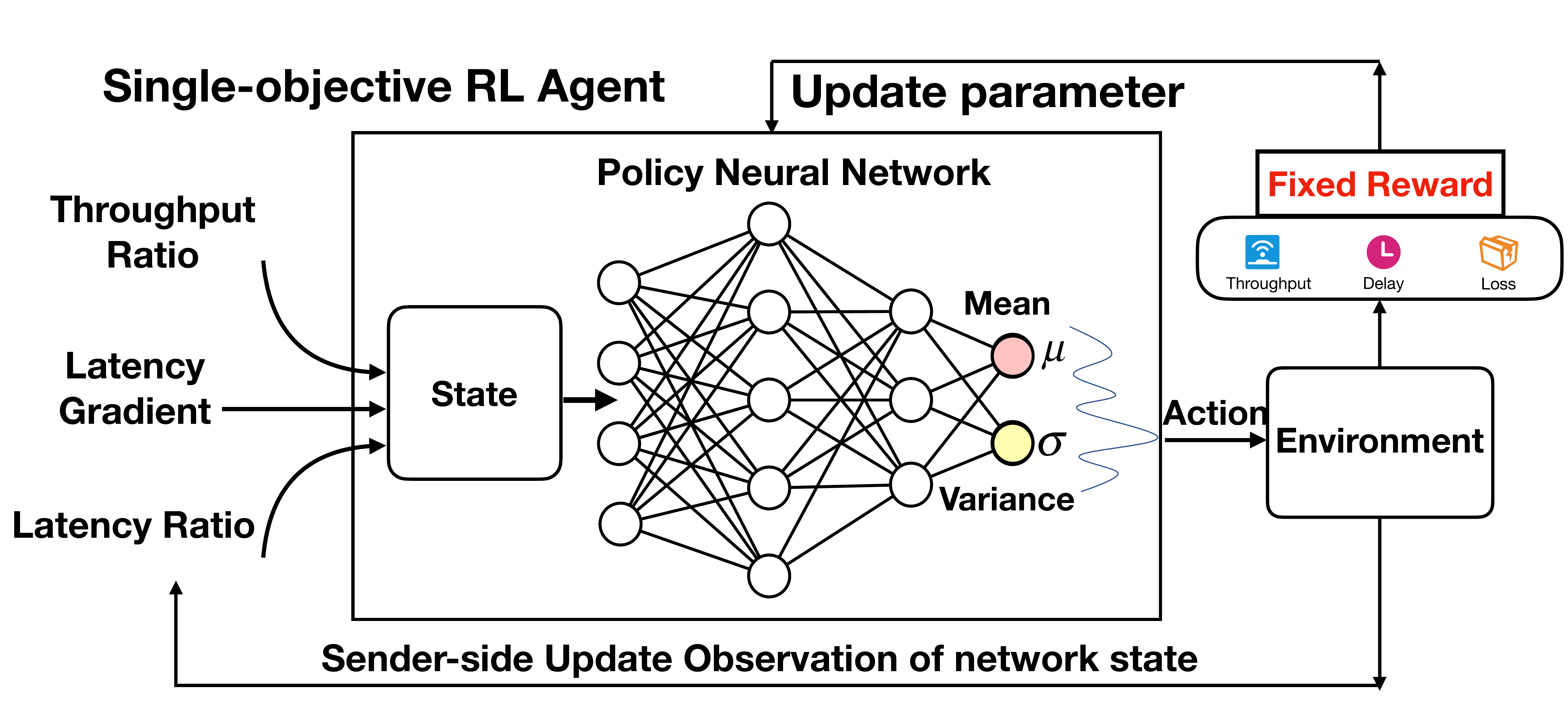}
}
\subfigure[Multi-objective CC (MOCC)]{
	\includegraphics[width=0.5\textwidth]{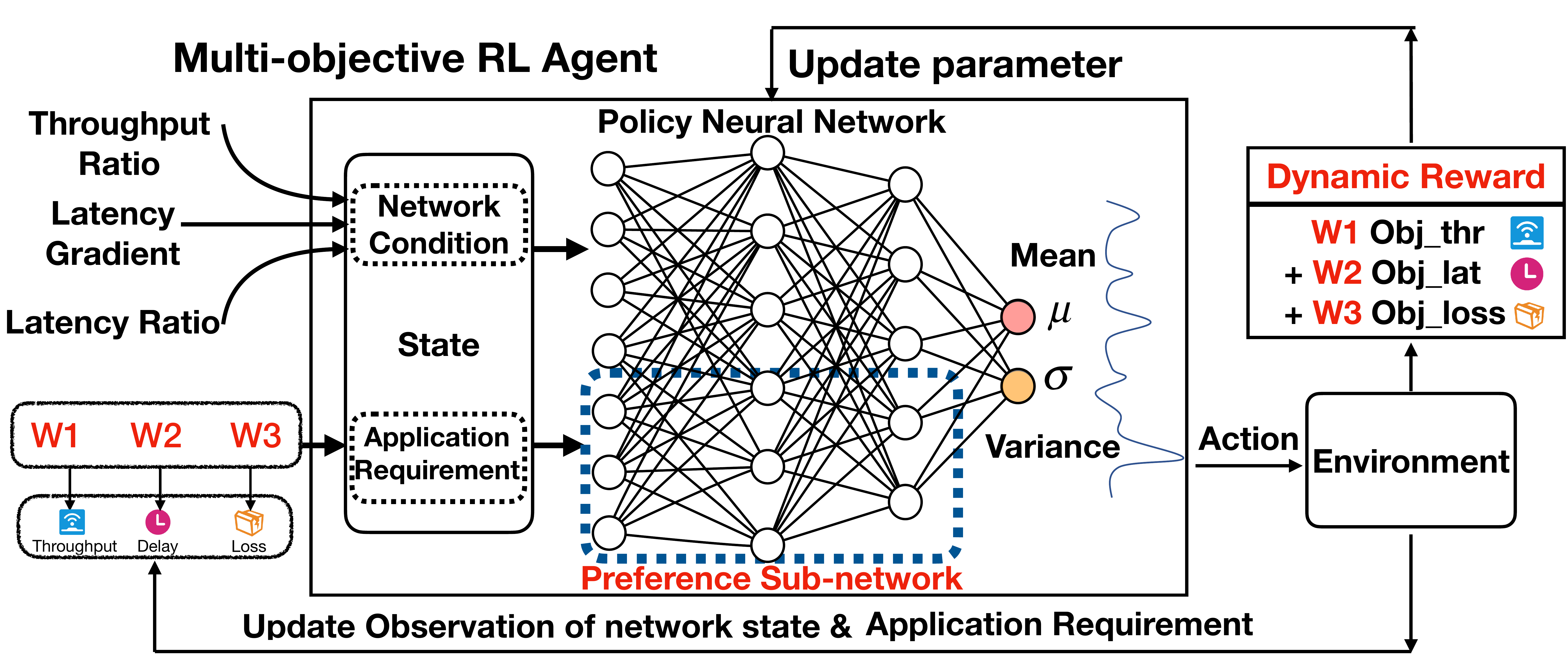}
}
%\vspace{-0.2in}
\caption{From single-objective CC to multi-objective CC: Incorporating preference sub-network into MOCC with application requirements explicitly used in both state input and dynamic reward function enables \sys to learn (and memorize) the correlations between application requirements and the corresponding optimal rate control policies, thus realizing multi-objective, i.e., one single \sys model can support multiple applications.}
\label{fig:somo}
%\vspace{-0.2in}
\end{figure*}

\parab{QUIC:} QUIC~\cite{quic} is a user-space transport protocol on top of UDP
to improve the transport performance of Internet applications and to enable
application-specific customizations. QUIC itself, however, is not tied to a
particular CC algorithm. It only provides the \emph{mechanism} to implement
application-specific CC algorithms, and an application still needs to specify
which CC it uses, which can be either a traditional or learning-based algorithm,
to achieve its performance objective. As such, our work is orthogonal to QUIC,
and more importantly, we show that we only need a single CC to satisfy different
objectives.

%\vspace{-0.2in}
\subsection{Design Goals}\label{subsec:goal}
%\vspace{-0.1in}

We seek {\em one single CC algorithm} satisfying all the following three goals simultaneously.
\begin{icompact}
%\vspace{-2mm}

\item \textbf{Multi-objective}: The algorithm can support different
applications with different performance objectives, and provide optimal control policies for individual applications.

\item \textbf{Quick-adaptive}: The algorithm can quickly adapt to new
applications with unseen requirements, without compromising performance of old
applications.

\item \textbf{Consistent high-performance}: The algorithm should maintain high-performance in various network conditions without any pre-assumption.
\end{icompact}

We are inspired by recent trend (notably, Aurora~\cite{aurora} and Orca~\cite{orca})
to adopt RL for CC, which can readily achieve consistent high-performance.
However, the challenge is how to simultaneously support multiple application objectives
and quickly accommodate new arrival ones.

\vspace{-2mm}
%\vspace{-0.12in}
\section{Multi-Objective Learning for CC}\label{sec:MORL}\vspace{-1mm}
%\vspace{-0.12in}

We formulate CC as a sequential decision-making problem under the  RL framework. Consider a general RL setting where an agent interacts with an environment. At each time step $t$, the agent observes some state $s_t$, and chooses an action $a_t$. After applying the action, the state of the environment transits from $s_{t}$ to $s_{t+1}$ and the agent receives a reward $r_t$ in ~$\S$\ref{subsec:appreq}. The state transitions and rewards are stochastic and Markovian~\cite{mdp}. The goal of RL learning is to maximize the expected cumulative discounted reward $E[\Sigma_{t=0}^{\infty}\gamma^{t}r_{t}]$, where $\gamma \in (0, 1]$ is a factor discounting future rewards.

Figure~\ref{fig:somo}(a) describes the standard way how to apply RL for CC, which reflects the state-of-the-art work Aurora/Orca~\cite{aurora,orca}. Basically, in each time interval, the agent (i.e., sender) observes a set of network metrics such as throughput, latency, and packet-level events, etc., and feeds these values into the neural network, which outputs the action, i.e., the sending rate for the next interval. In the meanwhile, the resulting network performance (e.g., throughput and latency) is measured and passed back to the agent as a reward, which will be used to train and improve the neural network model.

While the above standard RL shows early promise, it has a key shortcoming: the algorithm can only optimize for a \emph{single objective} at a time. The crux is that the model (Figure~\ref{fig:somo}a) has no way to recognize and differentiate among multiple different applications. As a result, supporting multiple applications requires multiple different models, and furthermore, adapting to a new application entails retraining the model from scratch which takes time, making it neither multi-objective nor quick-adaptive. 

We seek one algorithm to simultaneously support multiple application objectives while quickly adapting to new arrival ones. To this end, we extend the existing single-objective RL approach and establish a multi-objective RL\footnote{Multi-objective RL (MORL) is a fast-developing topic in machine learning community, Appendix~\ref{ap:morl} provides some more background on it.} framework for CC (\sys) that meets all our design goals stated in $\S$\ref{subsec:goal}. By contrasting Figure~\ref{fig:somo}(a), our \sys framework in Figure~\ref{fig:somo}(b) illustrates how it works out. From model structure perspective, we make two important changes: (1) we expand the policy neural network by incorporating a {\em preference sub-network} that explicitly takes application requirements, each denoted by a weight vector of performance metrics, as state input, making our model aware of different objectives\footnote{Similarly, researchers have adopted feature vectors to represent multi-user requirements in Video Streaming Dash Approach and achieved significantly better personalized QoE~\cite{qoe1,qoe2}.} in addition to network conditions; and (2) we dynamically parameterize the reward function with the weight vector of application currently under training, which enables our model to learn the optimal policy for the corresponding objective. As a result, \sys automatically learns the correlations between application requirements and the corresponding optimal rate control policies, thus achieving multi-objective (more details in $\S$\ref{subsec:architecture}).

We train our \sys model through offline pre-training ($\S$\ref{subsec:offline}) and online adaptation ($\S$\ref{subsec:online}). In particular, we leverage transfer learning techniques~\cite{transferrl,neighbour,schaul2015universal,dynamic,meta} to speedup offline pre-training as well as adapting to new applications in an online manner. In the offline phase, we pre-train our model with a well-distributed set of landmark weight vectors to learn the correlations between application requirements and optimal policies. This brings two important benefits to the online phase. First, for a new application, \sys can {\em immediately} provide a reasonable policy even it is unforeseen, maintaining performance during the transition. Second, transferring from such base correlation model, \sys is able to quickly converge to the optimal policy for the new application with just a few RL iterations, much faster than learning from scratch (e.g., $14.2\times$ in our evaluation $\S$\ref{sec:eva}). Furthermore, to avoid forgetting the already learned policies for old applications, we modify the loss function of the online learning by optimizing for both the new arrival and sampled history applications, so that our \sys can recall the learned policies for previous applications. 

To summarize, by deliberately architecting and training the model as above, our \sys framework is able to learn, remember, and apply optimal rate control policies for multiple applications simultaneously while adapting to new ones on-the-fly. In $\S$\ref{sec:design} below, we will go deeper to design details.

\vspace{-2mm}
\section{Design}\label{sec:design}
%\vspace{-0.15in}

%This section describes the design of \mocc. 
We start by introducing the model architecture that enables \sys to achieve the multi-objective property (\S\ref{subsec:architecture}). Then, we describe our offline training (\S\ref{subsec:offline}) and online adaptation (\S\ref{subsec:online}) that can quickly adapt \sys to new applications. 

%we explain the implementation of \mocc and its integration with real-world transports (\S\ref{subsec:implement}).

%\vspace{-3mm}
%\vspace{-0.12in}
\subsection{Model Architecture}\label{subsec:architecture}
%\vspace{-0.12in}

%To apply RL for our multi-objective CC problem, we firstly formalize the congestion control task as a multi-objective RL~\cite{morl} problem.  Here we give the detailed definitions of state, action, reward and application objectives in \mocc and the strucuture and functionality of our neural network.

%The key of \mocc is a multi-objective RL~\cite{morl}. To help readers better understand \mocc, we first give the definitions of state, action and application objectives and then introduce the structure and function of the neural networks used in our multi-objective RL.

To enable multi-objective, \sys makes two main changes upon the standard RL-based CC: 1) incorporating a preference sub-network into the policy network, and 2) including application requirements in both state input and dynamic reward function. In this way, \sys can establish the correlations between various application requirements and the corresponding optimal rate control policies.   

%\parab{State}
%The state contains statistical information of the network condition collected from received packet acknowledgements, which can help the sender makes decision of sending rate. We divide time into constant consecutive intervals so the sender can interact with the network in a turn-based fashion. \yq{what is turn-based} To fully cover the neccesary features to detect the trends of network dynamics, \mocc defines the state of the network at time interval $t$ as a fixed $k$ length history of the statistics vectors $S_{t} = <s_{t-k},s_{t-k+1}, ...,s_{t}>$, where $s_{t}=(\sigma_{t},\tau_{t},\upsilon_{t})$. $\sigma_{t}$ is the network throughput measurements for this time interval, which equals the packets sent divide by the packets acknowledged by the receiver. $\tau_{t}$ is the network delay measurements, which equals to the mean latency of this interval divide by the minimum observed latency in history. $\upsilon_{t}$ is the latency gradient measurements for this interval. This is the derivative of latency with respect to time.

\parab{States:} State inputs to \sys include both application requirements and network conditions. To express application requirements, we use weight vector $\vec{w}=$$<$$w_{thr}, w_{lat}, w_{loss}$$>$ which contains the relative weights of three main performance metrics\footnote{Note that our \sys framework can generalize to any other objectives.} in CC algorithm: throughput, latency, and packet loss rate. The range of each weight $w_i \in (0,1)$ and $\Sigma_i{w_i}=1$. For example, $<$$0.8,0.1,0.1$$>$ means that the application desires high throughput, and $<$$0.4,0.5,0.1$$>$ indicates the application is latency-sensitive but still needs certain throughput. 

For network conditions, similar to prior work~\cite{aurora,vivace,remy}, we use statistics vector $\vec{g_t}=$$<$$l_{t},p_{t},q_{t}$$>$ to express the network status at time interval $t$. Specifically, $l_{t}$ is sending ratio, defined as packets sent by sender over packets acknowledged by receiver; $p_{t}$ is latency ratio, the ratio of mean latency of the current time interval $t$ to the minimum observed mean latency in the history; and $q_{t}$ is latency gradient, the derivative of latency with respect to time. Furthermore, to capture the trends and changes of network dynamics, we use a fixed-length history of network statistics instead of the most recent one (i.e., $\vec{g}_{(t,\eta )}=$$<$$\vec{g}_{t-\eta },\vec{g}_{t-\eta +1}, ...,\vec{g}_{t}$$>$ with length $\eta >0$) as network state input. This improves \mocc by reacting to network dynamics more appropriately~\cite{aurora}.

\parab{Actions:} Upon observing state $s_{t}$=$(\vec{w}, \vec{g}_{(t,\eta )})$, the RL agent chooses an action $a_t$. Then the \sys sender takes the output $a_t$ to change its sending rate from $x_t$ to $x_{t+1}$ for the next time interval $t+1$ as follows:
\begin{equation}
\vspace{-0.1in}
\label{eq:send}
x_{t}=\left\{
\begin{aligned}
x_{t-1}*(1+\alpha a_{t}) & \quad  & a_{t} > 0 \\
x_{t-1}/(1-\alpha a_{t}) & \quad  & a_{t} < 0
\end{aligned}
\right.
%\vspace{-1em}
\end{equation}
Here $\alpha$ is a scaling factor used to dampen oscillations. Instead of discrete sending rate adjustment, we choose a continuous sending rate adjustment to improve model robustness and achieve faster convergence.

%After experimenting with several options, we choose to express actions as the proportional change to the sending rate.

\parab{Rewards:} The \sys reward function $r_t$ is dynamically parameterized with the weight vector $\vec{w}$ of application under training, so that the RL agent can capture the requirement of the application. Specifically,
%To define the reward function, we apply the weighted-sum approach~\cite{weightsum} shown in Equation~\ref{eq:reward}. The calculation of each objective is shown in Equation~\ref{eq:objective} to reflect the requirements of application in \S\ref{subsec:appilication}.
\begin{equation}\label{eq:reward}
Reward:~r_t = w_{thr}* O_{thr} + w_{lat}* O_{lat} + w_{loss}* O_{loss}
\end{equation}
%\begin{equation}\label{eq:objective}
%\left\{
%\begin{aligned}
%O_{thr} &= \frac{\text{Measured Throughput}}{\text{Available Link Bandwidth}}  \\
%O_{lat} &= \frac{\text{Link Latency}}{\text{Measured Latency}}  \\
%O_{loss} &= 1- \frac{\text{Lost Packets}}{\text{Total Packets}}
%\end{aligned}
%\right.
%\end{equation}
in which $O_{thr}$=$\frac{\text{Measured Throughput}}{\text{Link Capacity}}$, $O_{lat}$=$\frac{\text{Base Link Latency}}{\text{Measured Latency}}$, and $O_{loss}$=$1-\frac{\text{Lost Packets}}{\text{Total Packets}}$ are three performance measures on throughput, latency and packet loss rate. They are configured to positively relate to the final reward, and normalized to $[0,1]$ to ensure fairness among each other. We use measured maximum throughput and minmum delay to estimate the Link Capacity and Base Link Latency in the online phase.

\begin{figure}[t]
%\vspace{-0.6in}
	\centering
	\includegraphics[width=0.45\textwidth]{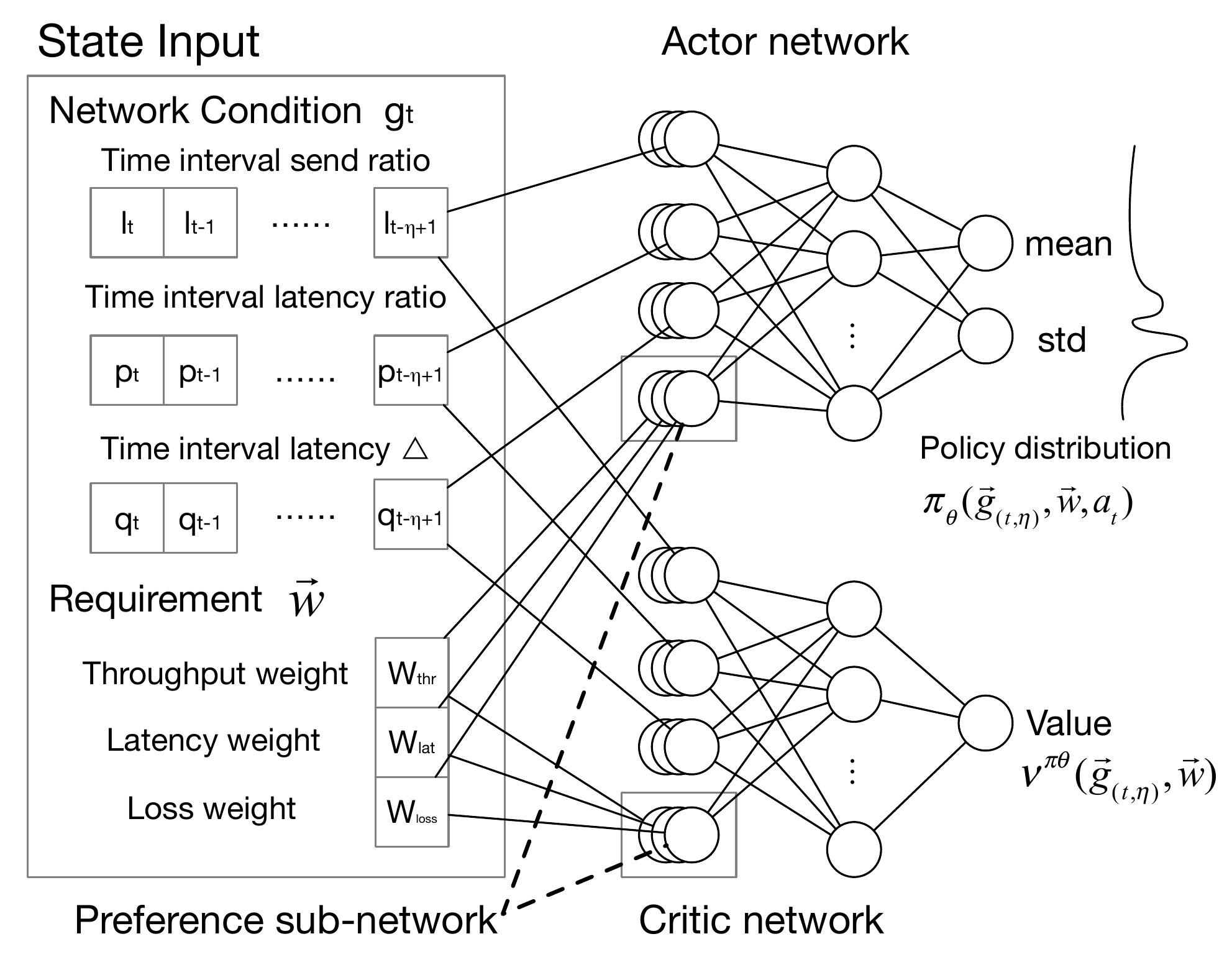}
%	\vspace{-0.2in}
	\caption{The Actor-Critic model that \mocc uses to generate CC rate control policies.}
	\label{fig:ppo}
%	\vspace{-0.3in}
\end{figure}

\parab{Model structure:}\mocc adopts the actor-critic method~\cite{sutton1998introduction}, a basic approach to train policy network in RL. The actor-critic method uses two neural networks: the actor network and the critic network (Figure~\ref{fig:ppo}). The actor network is used to represent the policy $\pi_{\theta}$ that maps application requirements and network conditions to action distribution $\pi_{\theta}$: $\pi_{\theta}(\vec{g}_{(t,\eta)},\vec{w}, a_{t})$$\rightarrow$$[0,1]$, where $\theta$ represents the adjustable model parameters. The critic network is to evaluate the results of actor network during training by the output value $V^{\pi_{\theta}}(\vec{g}_{(t,\eta)},\vec{w})$. After training, the actor network is used as the policy network of \mocc.

%The critic network evaluate the value of states following actor network policy and appliaction objective $v^{\pi_{\theta}}(S_t,W)$. The function of critic network is to provide evaluation information for the actor network during training. After training, the actor network is used as our congestion control policy network. We depict the model structure in Figure \ref{fig:ppo}. 

To support multiple objectives, \mocc extends both the actor network and critic network with a preference sub-network (PN). PN takes the application weight vector $\vec{w}$ as input, and performs feature transformation to concatenate with the network state $\vec{g}_{t,\eta}$ to feed both networks. Then, the actor network outputs a distribution of the action space for choosing the proper action. 

By incorporating the PN, both the decisions made by the actor network and the evaluation given by the critic network are not only based on the network conditions, but also taking the application requirements into consideration. In other words, our \sys model adopts neural network structure that can recognize different application requirements/preferences, and correlate them with the corresponding optimal policies. As a result, \mocc can learn and apply optimal rate control polices for multiple applications simultaneously, enabling multi-objective.

%the actor network can make decisions and the critic network evaluate not only on the current state, but also the given objective criteria. With the objective as feature, \mocc can learn how to differentiate the optimal policies for different applications even at the same state. The training process of two networks is introduced in detail in $\S$\ref{subsec:offline} and $\S$\ref{subsec:online}.

%\vspace{-3mm}
%\vspace{-0.1in}
\subsection{Offline Training}\label{subsec:offline}
%\vspace{-0.12in}

%We now describe our training algorithm. We introduce the environment of training, a network simulator, the training methodology, and PPO (Proximal Policy Optimization), one of the state-of-art policy optimization algorithms to train our policy network.

Our goal of offline pre-training is to learn the correlations between application requirements and optimal rate control polices, in order to make \sys quickly adapt to new applications with high accuracy during deployment. In this section, we introduce our two-phase training strategy as well as the policy optimization algorithm.

\parab{Two-phase training:} To train a multi-objective RL, one straightforward way is to decompose it into multiple single-objective RLs~\cite{liu2014multiobjective}. If we can enumerate all possible objectives and train each of them iteratively, the multi-objective RL can achieve the optimal Convex Converge Set. However, in our case of \sys, there are infinite possible objectives, i.e., any weight vector that satisfies $w_{thr} + w_{lat} + w_{loss}$=$1$, $w_{i}\in (0,1)$. The problem becomes intractable.

To efficiently train \sys, instead of exploring the whole objective space, we train on a subset of landmark objectives, say $\omega$, that can produce a satisfying model\footnote{In $\S$\ref{subsec:DD}, we deep-dive $\omega$ and find that $\omega=36$ achieves a good performance.}. However, even a moderate $\omega$ with tens of objectives will take several days to train. To speedup, we introduce a two-phase training: bootstrapping and fast traversing. In bootstrapping phase, we build a base model by selecting just a small number of objectives to train. In our implementation, we chose 3 to bootstrap with, and our base model can take hours to converge. 

\begin{figure}[t]
	\centering
	%\vspace{-0.5in}
	\includegraphics[width=0.35\textwidth]{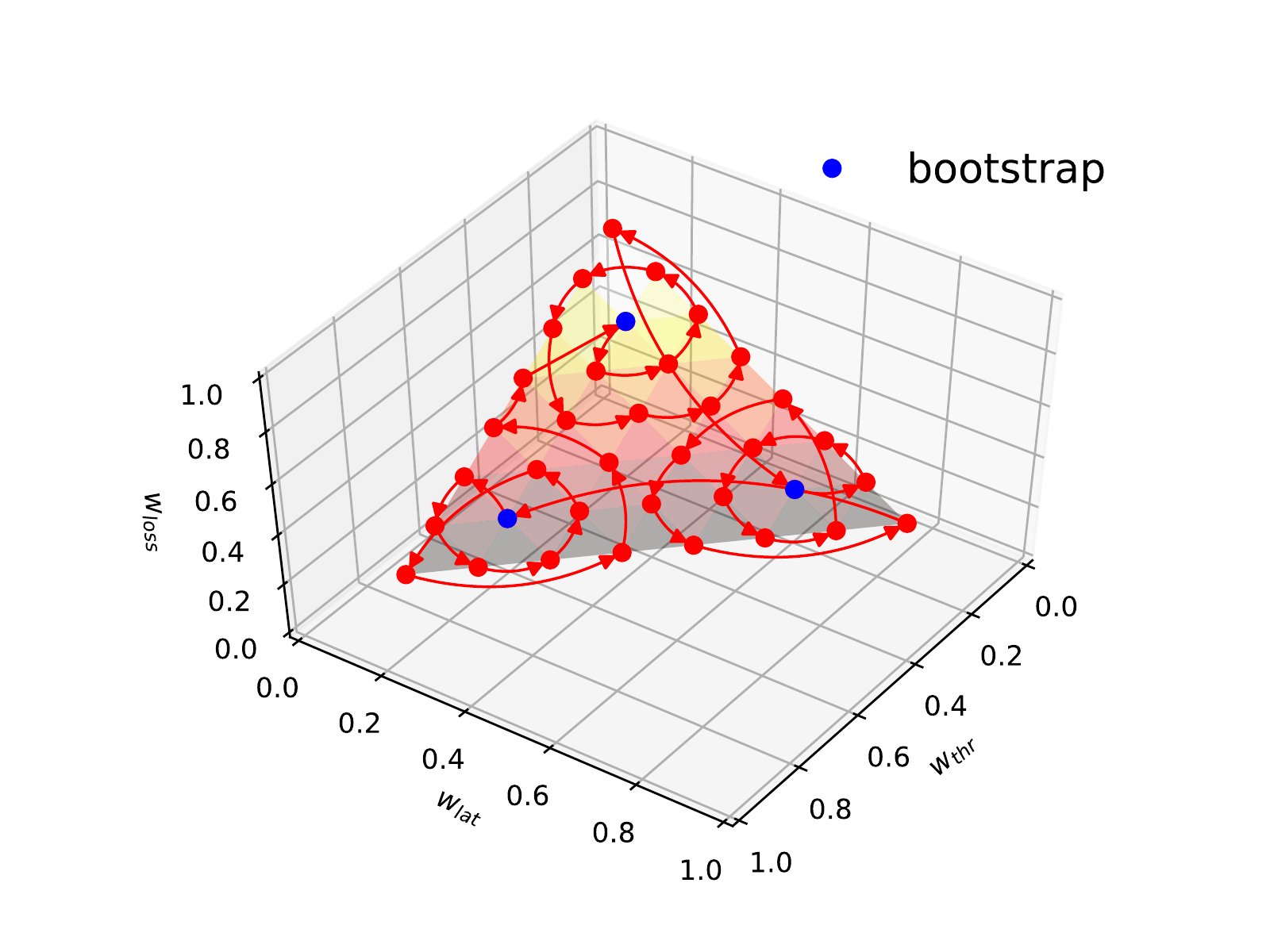}
%	\vspace{-0.25in}
	\caption{The objective training trajectory for fast traversing, generated by neighbourhood-based objective sorting algorithm (details in Appendix~\ref{ap:neighbor}).}
	\label{fig:nei}
	%\vspace{-0.3in}
\end{figure}

Then, in the fast traversing phase, building on the base model, we accelerate the training of the remaining $\omega$-$3$ objectives by adopting a neighborhood-based transfer learning strategy~\cite{neighbour}. This method is based on the observation that when two RLs have close objectives (i.e., similar weight vectors), their optimal solutions are close. Thus, when training a RL, we can speedup by leveraging the solutions of its neighboring RLs. 
To do this, we purposely arrange the $\omega$ objectives in a neighborhood-based way as shown in Figure~\ref{fig:nei}. We train from one objective to its neighbor iteratively and traverse all the objectives in a cyclic way. Note that each time we do not train an objective until convergence but only for a few steps in order to achieve balanced improvement on all objectives. The whole training completes when the model converges on all objectives. We explain why such two-phase training achieves near-optimal solution in Appendix~\ref{ap:twophase}. Deep-dive in $\S$\ref{subsec:DD}  shows it effectively speedups the training by $18\times$.

\parab{Policy optimization algorithm:} Among a variety of different algorithms for training RL~\cite{ppo,dqn,a3c}, we adopt Proximal Policy Optimization (PPO) \cite{ppo} as the policy optimization algorithm to train \mocc. It is a policy gradient method updating the model with estimated gradient to maximize the expected total reward. We chose PPO because: 1) it is the state-of-the-art approach and easy to tune; and 2) it performs particularly well on continuous control problem~\cite{ppo}, which makes it suitable for deciding the sending rates.

Instead of directly optimizing the expected total reward, PPO optimizes on its lower bound, a surrogate objective function (The lower bound proof is given in \cite{ppo} and \cite{schulman2015trust}):
\begin{equation}\label{eq:originalppo}
\small
L^{C L I P}(\theta,\vec{w})_t=\hat{\mathbb{E}}_{t}\left[\min \left(r_{t}(\theta), \operatorname{clip}\left(r_{t}(\theta), 1-\epsilon, 1+\epsilon\right)\right) \hat{A}_{t}\right]
\end{equation}
where $r_t(\theta)=\frac{\pi_{\theta}(a_t|\vec{v}_{(t,\eta )}, \vec{w})}{\pi_{\theta_{old}}(a_t|\vec{v}_{(t,\eta )}, \vec{w})}$ denotes the probability ratio of action $a_t$ compared to the current policy. The term ${c l i p}(r_t(\theta),1-\epsilon,1+\epsilon)$ clips the probability ratio to the range $[1-\epsilon,1+\epsilon]$.
$\hat{A}_t$ represents the advantage of a specific action over the current policy. It is defined as the difference between empirical total reward applying the action $a_t$ and expected total reward applying policy $\pi_{\theta}$:
\begin{equation}\label{eq:advantage}
\vspace{-0.1in}
\hat{A}(\vec{g}_{(t,\eta )}, \vec{w},a_t)=\sum_{t}\gamma^tr_t-V^{\pi_{\theta}}(\vec{g}_{(t,\eta )}, \vec{w})
\end{equation}
where $V^{\pi_{\theta}}(\vec{g}_{(t,\eta )}, \vec{w})$ is estimated by the critic network. 
%where $v^{\pi_{\theta}}(s_t,\vec{w})$ represents the expected cumulative discounted reward executing policy network $\pi_{\theta}$  from state $s_t$ in terms of objective $\vec{w}$. 

To encourage exploration of policy network, as suggested in past works~\cite{a3c}, we add an entropy regularization term to the objective function $L^{C L I P}$:
\begin{equation}\label{eq:ppo}
L^{C L I P+E}_t(\theta,\vec{w})=L^{C L I P}(\theta,\vec{w})+\beta H(\pi_{\theta}(\cdot|\vec{g}_{(t,\eta )}, \vec{w})),
\end{equation}
where $H(\cdot)$ is the entropy function of the probability distribution over actions at each time step. Thus action distribution with higher entropy is preferred, exploring a more diverse set of possible actions.

During offline training, for each step, \mocc's RL agent performs the policy of actor network to generate a network trace for a short period of time. With the trace and collected empirical rewards, we update the critic network following the standard \emph{Temporal Difference} method \cite{rlintro}. Then, the critic network provides $V^{\pi_{\theta}}(\vec{g}_{(t,\eta )}, \vec{w})$ for computing the advantage function according to Equation \ref{eq:advantage}. Finally, the actor network is updated with gradients computed to maximize Equation \ref{eq:ppo}. Because the surrogate objective is the lower bound of the expected total reward, optimization on it guarantees the improvement of the policy network on gained reward. As a result, model parameters are updated such that the new policy assigns higher probability to state-action pairs resulting in positive reward advantages, moving towards the optimal policy.

%\parab{Parallel training}
%\TODO{Move to implementation?}
%To further speed up the offline training, \mocc adopt parallel training introduced in \cite{ray}. We configures multiple environments with different network conditions and application requirements. Accordingly, we spawns the same number of agents parallelly interacting with them. The trajectories are pulled to the central agent to train a single model, and the updated model is then pushed back to agents. The process is asynchronous and the training speed is enhanced with faster production of training trajectories.

%\vspace{-3mm}
%\vspace{-0.12in}
\subsection{Online Adaptation}\label{subsec:online}
%\vspace{-0.12in}

%We adopt an online learning algorithm to continuously adapt to new applications' objectives and achieve optimal performance under the personalized application context of the sender. As showed in $S$\ref{subsec:offline}, initialized with the offline trained policy network, \mocc can provide moderate policy for new application requirements. Transferring from such a moderate policy, a few more training steps are sufficient to converge to the new objective.

Our offline pre-trained model from $\S$\ref{subsec:offline} effectively correlates the application requirements with the optimal policies. This brings two key benefits to \sys's online adaptation. First, for a new application, \sys can generate a moderate policy of rate control even the requirement is unforeseen, providing reasonable performance for the new application at the beginning. Second, starting from such moderate model, with transfer learning, \sys is able to quickly converge to the optimal model for the new application with just a few RL iterations, which is much faster than learning from scratch (e.g., we see over 10 times faster in $\S$\ref{subsec:QA}). These two benefits enable \sys to adapt to any new applications on-the-fly. 

%As discussed in \S\ref{subsec:offline}, after offline training, \mocc can provide optimal solution for a large set of objectives. Furthermore, to grant \mocc the ability to adapt to any new applications with any objectives, \mocc uses an online learning algorithm to continuously learn the optimal solution of different objectives.

%Faced with a new application, \mocc can already provide moderate solution for the application based on its well trained neural network. \mocc uses the online algorithm to transfer from the moderate solution to optimal solutions, and a few more training steps are sufficient to converge.

However, there is one issue: we do not want to compromise the performance of old applications while adapting to new ones. Unlike offline training where all objectives are artificially generated and uniformly distributed, the objective distribution in real environment may have bias: some applications are very frequent, some are rare. Under such a bias, the traditional RL algorithm will overfit to those new frequent applications but gradually forget those old rare ones, which is undesirable.

%However, unlike training environment, where all objectives are uniformly distributed and generated, the distribution of applications in real environment has dramatic bias (some applications are very frequent, some are very rare). Under such a bias, the traditional RL algorithm will overfit to those frequent applications and gradually forget those rare applications, which is not desired in \mocc.

To avoid this problem, \sys uses a requirement replay learning algorithm~\cite{dynamic}. During the online learning, \mocc stores encountered applications (weight vectors) in a long period of time. For each online training step, the model is trained on both the current objective and an old objective drawn uniformly at random from the pool of the stored applications. We define the online learning objective to be:
\begin{equation}
\vspace{-0.1in}
L_{online}(\theta)= \frac{1}{2}* [ L^{C L I P+E}(\theta,\vec{w}_i)+L^{C L I P+E}(\theta,\vec{w}_j)\label{eq:online}
%\vspace{-0.1in}
\end{equation}
where $\vec{w}_i$ refers to the current application requirement,  $\vec{w}_j$ refers to a sampled old application requirement,
and $L^{C L I P+E}$ is the PPO surrogate objective function defined in Equation \ref{eq:ppo}. In this way, \mocc not only learns new applications, but also recalls old applications and reinforces previously learned policies. Thus, \mocc can preserve the learned policies of old applications while adapting to new applications. Our evaluation in $\S$\ref{subsec:QA} confirms this property.

\vspace{-2mm}
\section{Implementation}\label{sec:implement}
%\vspace{-0.15in}
%We explain the implementation of \mocc and its integration with real-world transports. 
Our implementation of \mocc mainly consists of two components: 1) offline training, and 2) online deployment.

%three major parts: 1) offline training simulator; 2) \mocc online adaption library; 3) application integration layer. and more details will be covered in \S\ref{subsec:implement}.

\parab{Offline training:} Directly training \mocc in real environment is slow considering the actual time cost in real control loops of CC with complex network dynamics~\cite{chen2018auto,aurora,pensieve}. To enable efficient training, we train \sys in a networking simulator that faithfully mimics Internet links with various characteristics. Our simulator is based on OpenAI Gym~\cite{openai} and Aurora~\cite{aurora}, and further incorporates new design elements in $\S$\ref{subsec:architecture}--$\S$\ref{subsec:offline} that are essential to \sys, such as the encapsulation of application requirements as state input and dynamic reward functions.

\mocc policy network uses a fully-connected MLP(Multi-layer perceptron) with two hidden layers of 64 and 32 units, respectively, and \texttt{tanh} activation function to output the mean and standard deviations of the Gaussian distribution of action. The critic network uses the same neural network structure to estimate the scalar value function. We control the entropy factor $\beta$ to decay from 1 to 0.1 over 1000 iterations, and set clipping threshold $\epsilon=0.2$.
For the learning rate, we adopt Adam~\cite{adam}, a famous adaptive learning rate optimization algorithm, switch consistently outperforms standard SGD method. Important training parameter settings are listed in Table~\ref{tab:para}. We implement our model architecture with TensorFlow 1.14.0. For PPO implementation, we use an open-source implementation of several reinforcement learning baselines\footnote{https://github.com/hill-a/stable-baselines}.

To further accelerate \mocc's exploration towards optimal solutions for massive objectives, in addition to the two-phase training introduced in $\S$\ref{subsec:offline}, we also adopt parallel training. We implemented this architecture using Ray~\cite{ray} and RLlib~\cite{rllib} to build the multiple parallel environments. For compatibility, we leverage Ray API to declare the neural network during both training and testing.

\begin{table}[t]
\small
%\vspace{-0.2in}
  \begin{center}
   \begin{tabularx}{0.8\columnwidth}{c|c}
    \hline
    Parameter & Value\\
    \hline
    \hline
       Discount factor ($\gamma$) & 0.99 \\
       Learning rate ($\epsilon$) & 0.001 \\
       Action scale factor ($\alpha$) & 0.025 \\
       History length ($\eta$)& 10 \\
       Landmark objectives \# ($\omega$) & 36 \\
    \hline
   \end{tabularx}
   % \vspace{-0.1in}
   \caption{Parameter settings}
    \label{tab:para}
  %  \vspace{-0.3in}
  \end{center}
%\vspace{-0.2in}
\end{table}

\parab{Online Deployment:} After the \mocc model was offline trained in the simulator, it needs to be online deployed with the real Internet applications. For better portability, we encapsulate all \mocc's functions into one library. Our library provides three main functions:
\begin{icompact}
	\item \texttt{Register($w$)}. Before using \mocc, we should register with it by providing the requirement/preference (weight vector $w$) of the application.
	\item \texttt{ReportStatus($s_t$)}. At each time interval, we should report the latest networking status ($s_t$) to \mocc.
	\item \texttt{GetSendingRate( )}. When sending packets, we use this function to obtain the sending rate calculated by \mocc.
\end{icompact}
With clean encapsulation, \mocc becomes an easy-to-use module and can be deployed with any networking datapaths.

In our implementation, we integrate \mocc with UDT~\cite{udt} and CCP~\cite{ccp} to build user-space \mocc and kernel-space \mocc. UDT is a widely used user-space implementation~\cite{allegro,vivace,aurora,orca,pantheon}. The \texttt{shim-helper} in UDT will interact with \mocc library and obtain the sending rate. CCP is a more general solution and enables congestion control outside datapath such as Linux kernel networking stack. We integrate \mocc with CCP for more general-purpose applications. In $\S$\ref{subsec:realapp}, we use \sys to support 3 real Internet applications: video streaming, real-time communications and bulk data transfer, and we will introduce more implementation details there. Furthermore, we note that \mocc with CCP achieves much lower CPU overhead than that with UDT ($\S$\ref{subsec:DD}).

\vspace{-2mm}
%!TEX root = main.tex
%\vspace{-0.05in}
\section{Evaluation}\label{sec:eva}
%\vspace{-0.05in}

\begin{figure*}
\centering
%\vspace{-0.2in}
\begin{minipage}{0.11\textwidth}{\includegraphics[height=4.9cm]{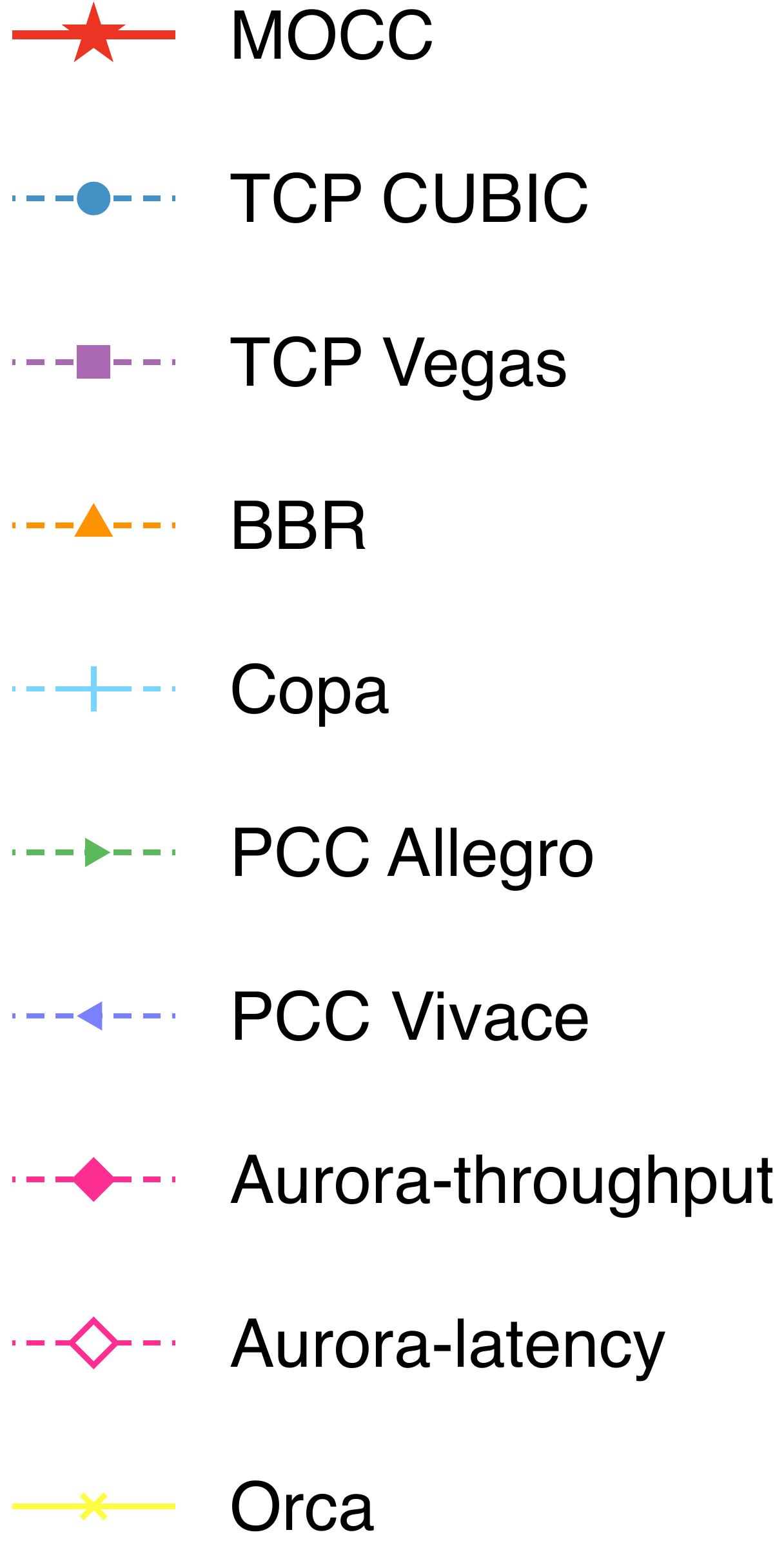}
%\vspace{-0.1in}
}\end{minipage}
\hfill
\begin{minipage}{0.86\textwidth}{
\subfigure[Varying bandwidth.]{\includegraphics[height=2.5cm] {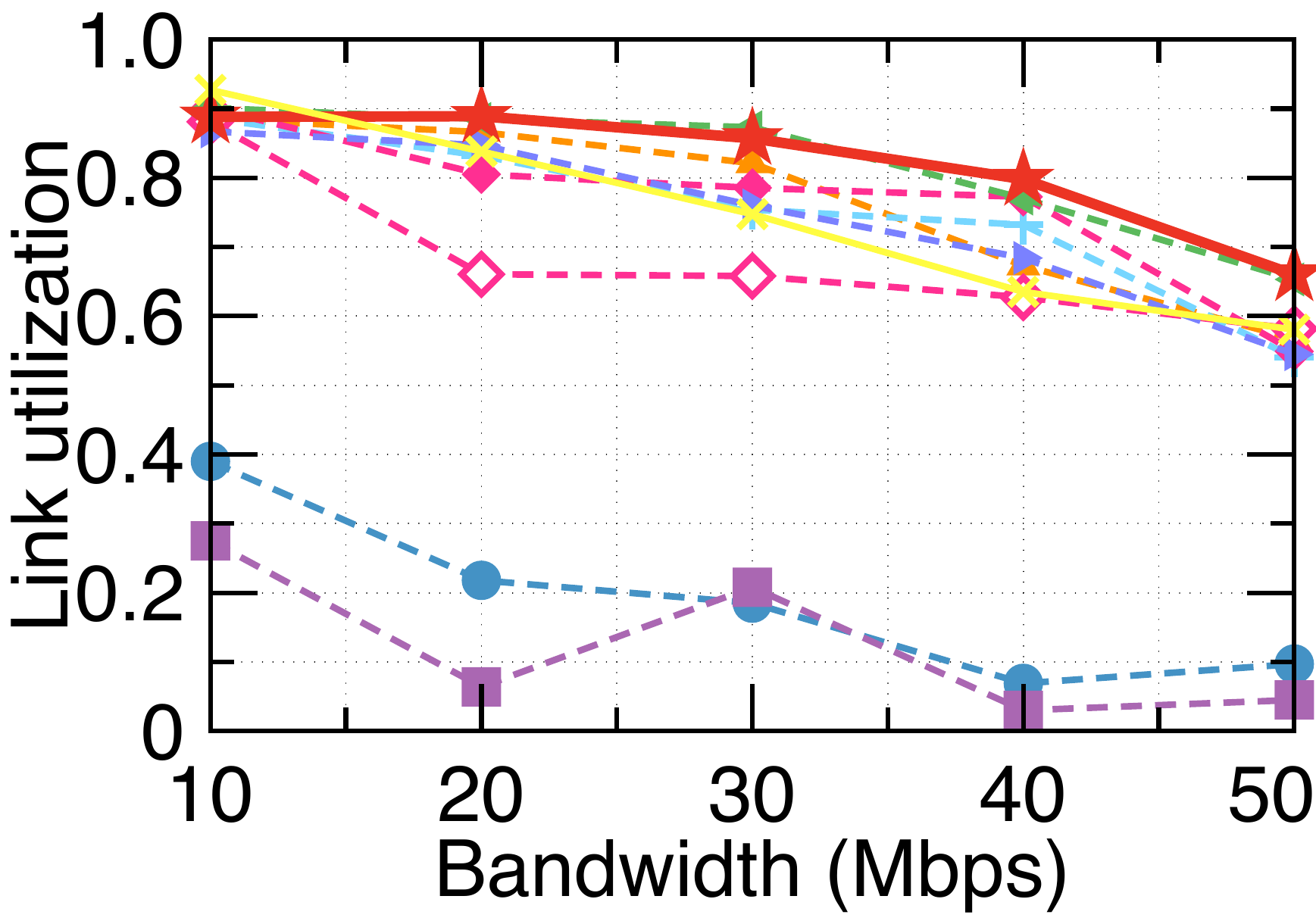}\label{fig:multi11}}
\subfigure[Varying latency.] {\includegraphics[height=2.5cm]{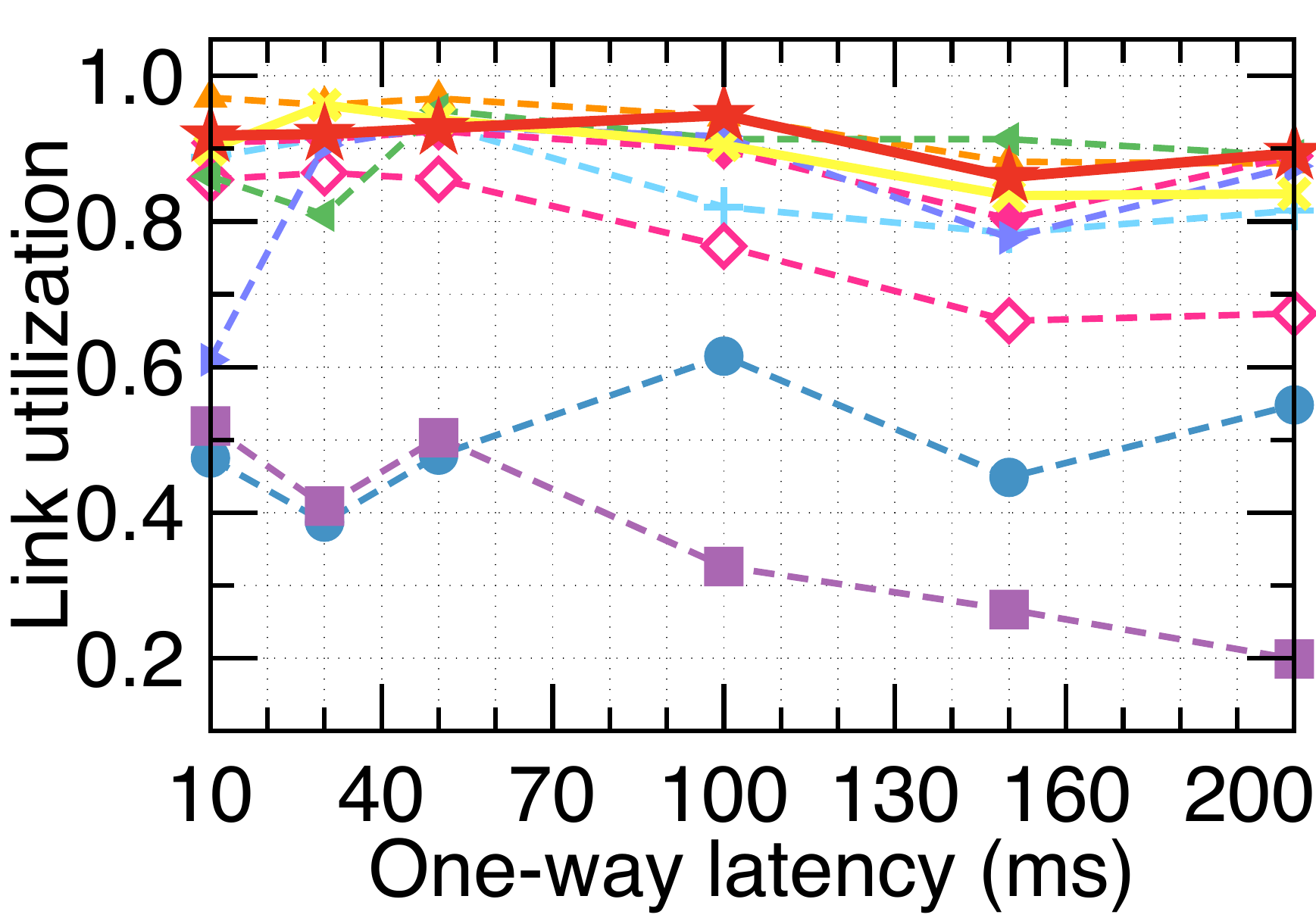}\label{fig:multi12}}
\subfigure[Varying random loss.] {\includegraphics[height=2.5cm]{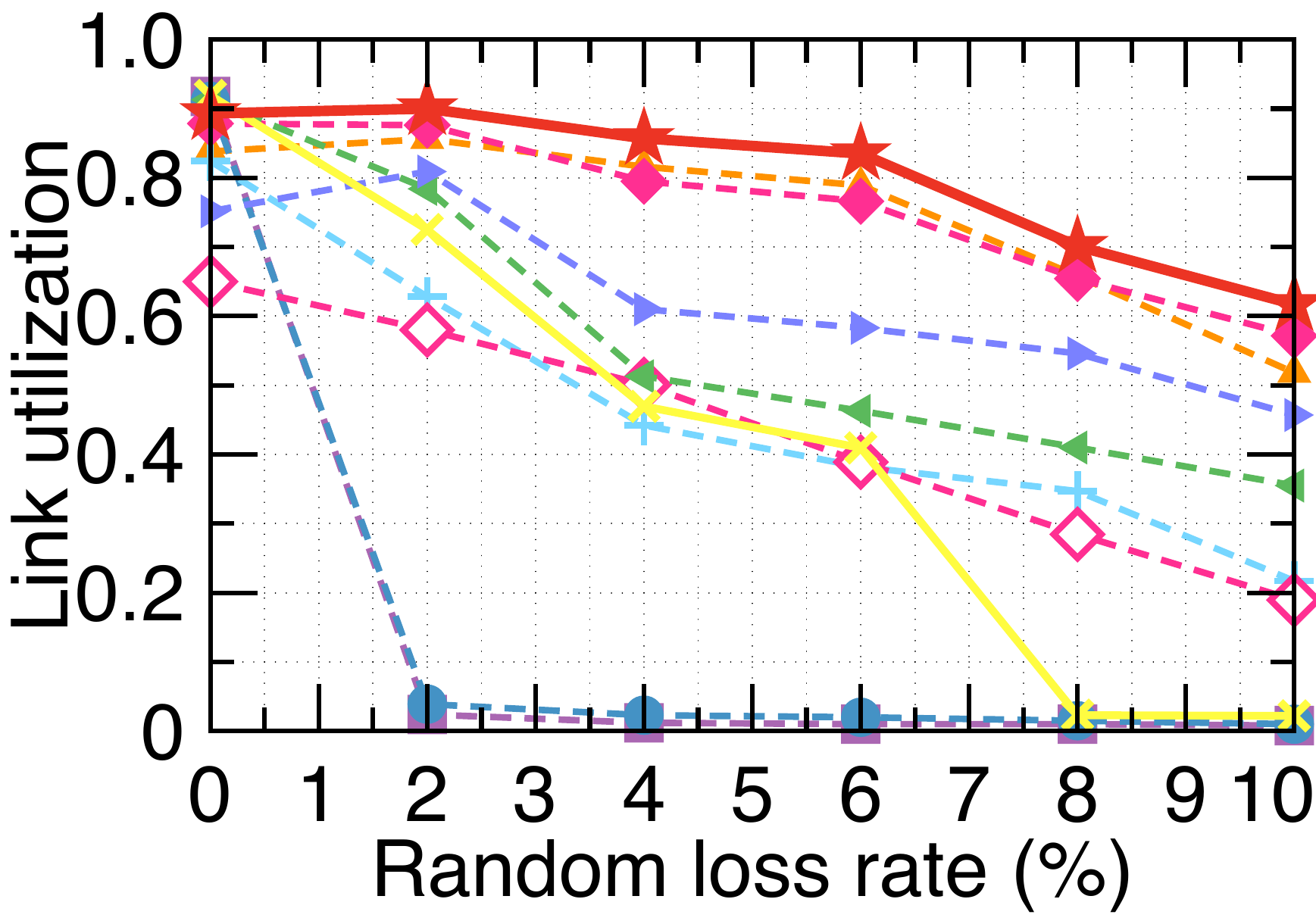}\label{fig:multi13}}
\subfigure[Varying buffer size.] {\includegraphics[height=2.5cm]{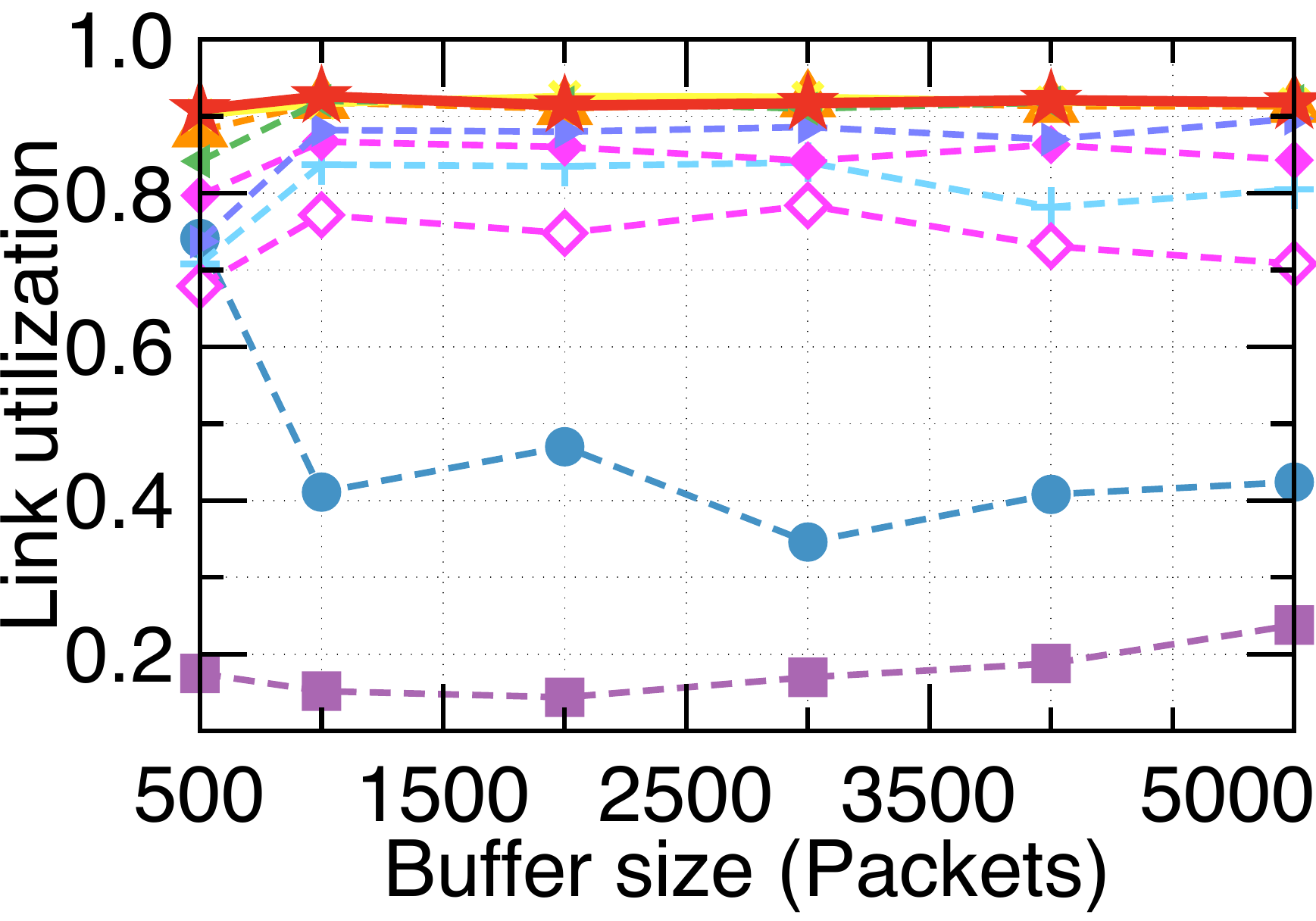}\label{fig:multi14}}
%\vspace{-0.1in}

\subfigure[Varying bandwidth.]{\includegraphics[height=2.5cm]{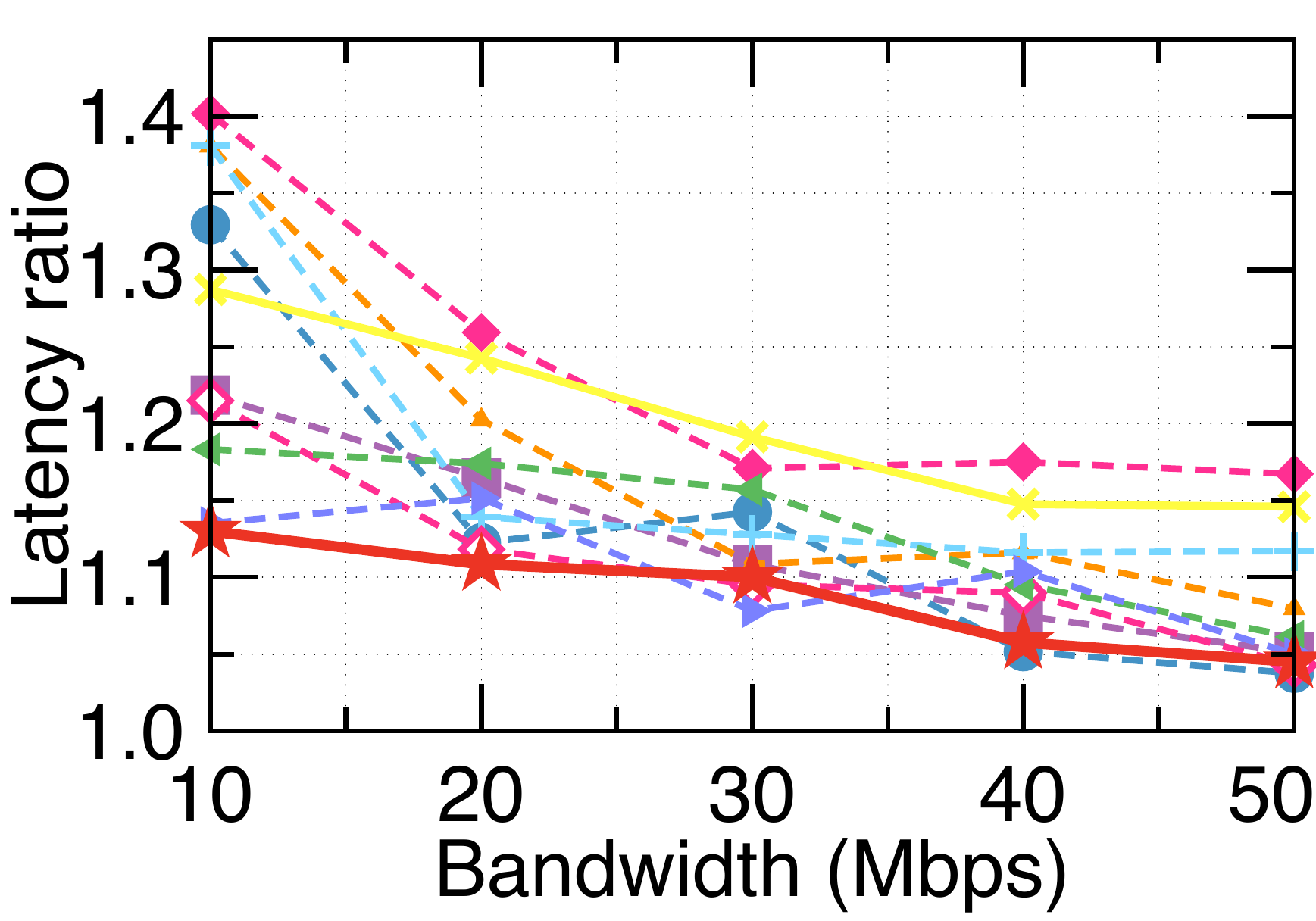}\label{fig:multi21}}
\subfigure[Varying latency.] {\includegraphics[height=2.5cm]{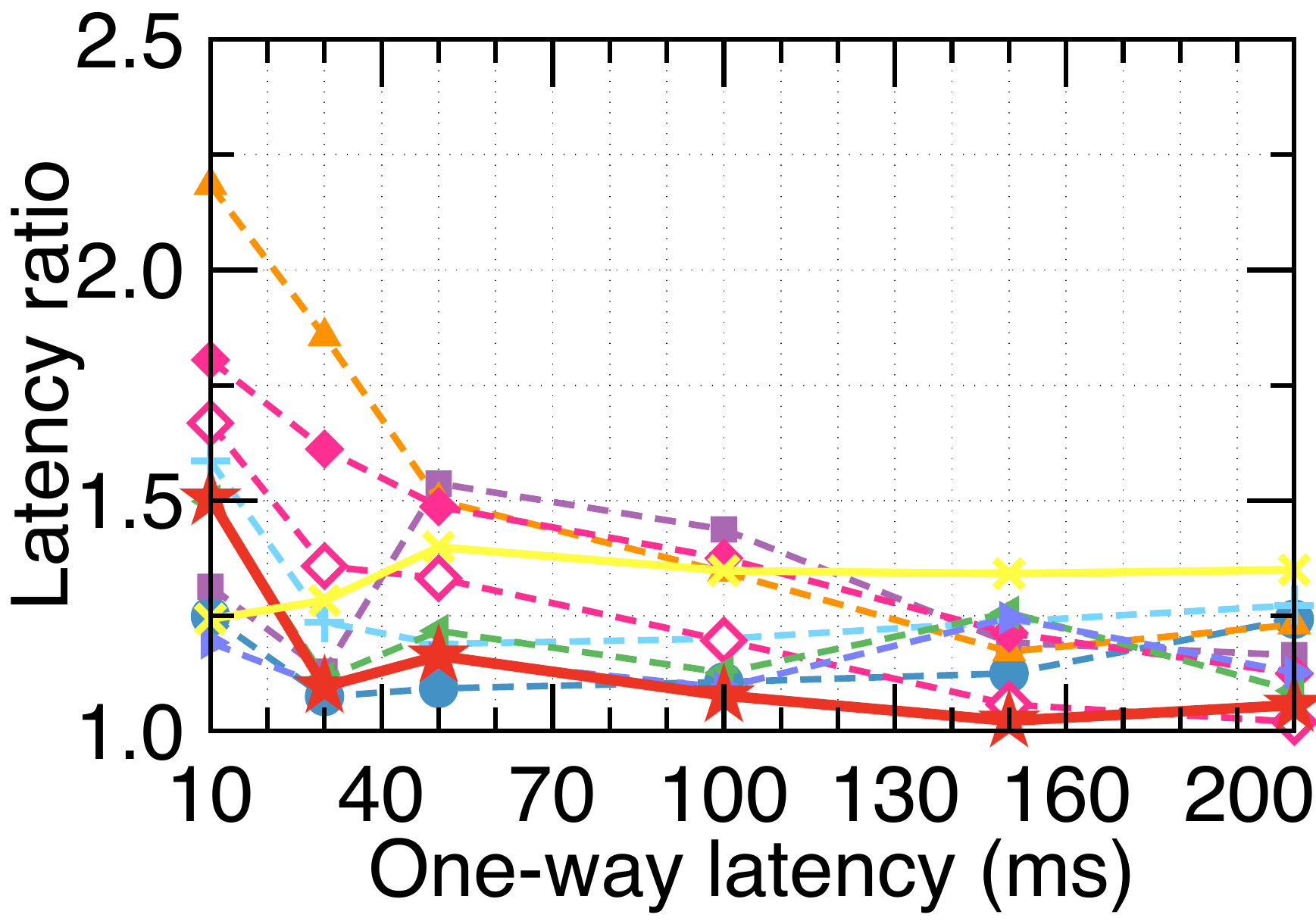}\label{fig:multi22}}
\subfigure[Varying random loss.]{\includegraphics[height=2.5cm]{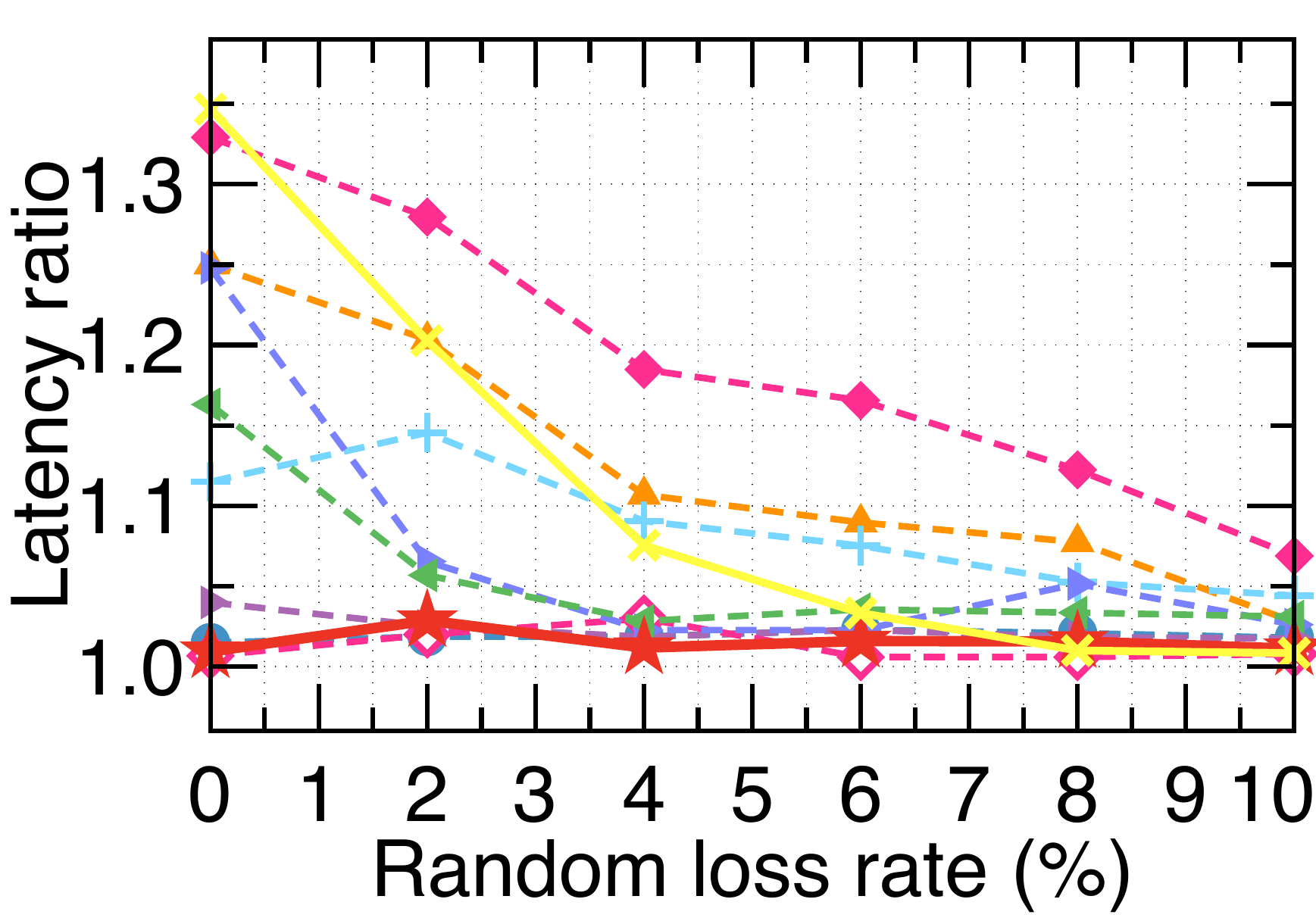}\label{fig:multi23}}
\subfigure[Varying buffer size.]{\includegraphics[height=2.5cm]{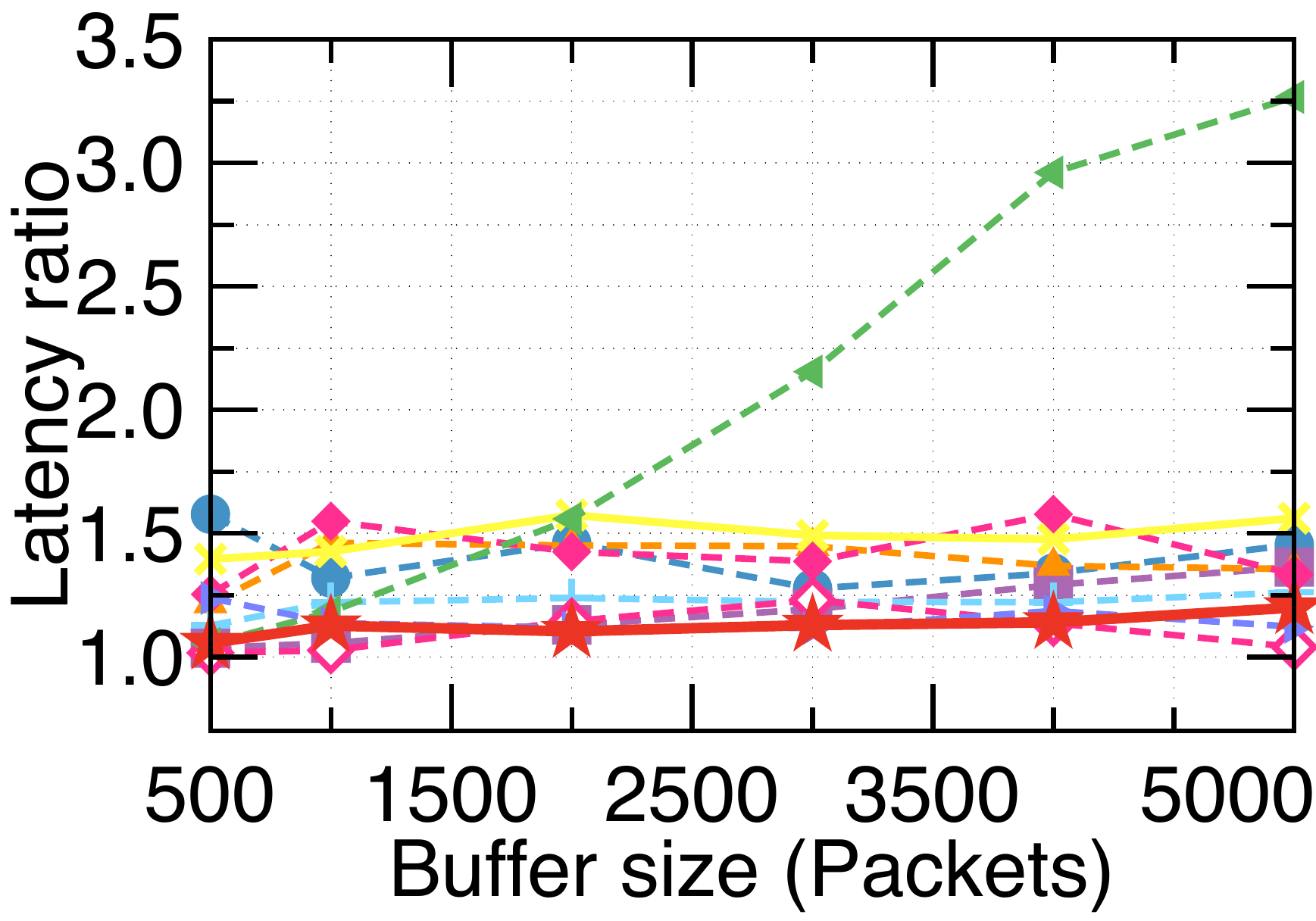}\label{fig:multi24}}
%\vspace{-0.1in}
}
\end{minipage}
%\vspace{-0.1in}
\caption{The multi-objective performance of \sys in terms of throughput (a-d) and latency (e-h), under various network conditions. Note that the network conditions under evaluation are far beyond the environment where \sys was trained, demonstrating both its high performance and robustness when adopted in practice.}\label{fig:multi1}
%\vspace{-0.2in}
\end{figure*}

We evaluate \mocc with extensive simulations as well as real Internet experiments. Our key results are as follows:
\begin{icompact}
%\item \textbf{Multi-objective:} We compare \mocc against a variety of heuristic and learning-based CC algorithms in terms of meeting different application requirements ($\S$\ref{subsec:AA}). Our results show that \mocc well enables different applications and is able to rival or outperform the best existing schemes under different scenarios (Figure~\ref{fig:multi1} and~\ref{fig:cdf}).

\item \textbf{Multi-objective ($\S$\ref{subsec:AA}):} Compared with a series of heuristic/learning CC algorithms, \mocc demonstrates its multi-objective performance by competing or outperforming the best existing schemes in supporting 2 common objectives: high throughput and low latency applications (Figure~\ref{fig:multi1}), as well as a generalized 100-objective setting (Figure~\ref{fig:cdf}).

%\item \textbf{Quick-adaptation:} We compare \mocc with the state-of-the-art RL CC algorithm Aurora~\cite{aurora} in terms of adapting to new applications ($\S$\ref{subsec:QA}). We find that, for a new application, \mocc can start from a good policy and quickly converge to the optimal policy after 4.8 minutes, $14.2\times$ faster than Aurora (Figure~\ref{fig:quick}a). We also observe that \mocc does not compromise old applications when adapting to the new one, while Aurora does, significantly (Figure~\ref{fig:quick}b).

\item \textbf{Quick-adaptation ($\S$\ref{subsec:QA}):} Compared with the state-of-the-art RL CC algorithm Aurora~\cite{aurora}, \sys can adapt to a new application in 4.8 minutes, $14.2\times$ faster than Aurora (Figure~\ref{fig:quick}a). Furthermore, \mocc does not compromise old applications while adapting to the new one, whereas Aurora does, significantly (Figure~\ref{fig:quick}b).

%\item \textbf{Real Internet applications:} We showcase the performance of \mocc with 3 real Internet applications: video streaming, real-time communications (RTC), and bulk data transfer ($\S$\ref{subsec:realapp}). Our results demonstrate that \mocc is able to provide high bitrate/throughput for video streaming (Figure~\ref{fig:vs}) and bulk data transfer (Figure~\ref{fig:bdt}), while delivering the lowest inter-packet latency for RTC (Figure~\ref{fig:rtc}).

\item \textbf{Real Internet applications ($\S$\ref{subsec:realapp}):} Among all the algorithms compared, \mocc is the only one that can simultaneously provide high bitrate/throughput for video streaming (Figure~\ref{fig:vs}) and bulk data transfer (Figure~\ref{fig:bdt}), while delivering the lowest inter-packet latency for real-time communications (Figure~\ref{fig:rtc}).

%\item \textbf{\mocc Deep-dive:} Finally, we deep-dive into \mocc ($\S$\ref{subsec:DD}) from a series of other aspects such as the choices of hyperparameters (Figure~\ref{fig:finegrain}), learning algorithm (Figure~\ref{fig:model}), training speedup techniques (Figure~\ref{fig:speedup}), etc., to validate its design decisions. Finally, we inspect \mocc's fairness (Figure~\ref{fig:fair}) and TCP-friendliness (Figure~\ref{fig:friend}).
\item \textbf{Fairness and Friendliness ($\S$\ref{subsec:FF}):} \mocc with the same weight achieves fair share (Figure~\ref{fig:thrdynamics}, ~\ref{fig:jainfair2}), and \mocc variants with different weights grab different bandwidth according to $weight_{thr}$ (Figure~\ref{fig:twoweight}). \sys is friendly among its own variants (Figure~\ref{fig:moccfriend}) and achieves comparable TCP-friendliness as other CC schemes (Figure~\ref{fig:friendrtt}).

\item \textbf{Deep-dive ($\S$\ref{subsec:DD})} into \mocc from different aspects such as hyperparameter setting (Figure~\ref{fig:finegrain}), CPU overhead (Figure~\ref{fig:overhead}), learning algorithm selection (Figure~\ref{fig:magic2}) and training speedup (Figure~\ref{fig:speedup}) has validated its design efficiency.
\end{icompact}

%\begin{figure*}
%\vspace{-0.2in}
% \includegraphics[width=\textwidth] {image/label.pdf}
% \center
%    \label{fig:label}
%\end{figure*}

%[width=0.2\textwidth]
%\begin{figure*}
%\centering
%\begin{minipage}{0.09\textwidth}
%{\includegraphics[height=5cm]{image/label_left.pdf}}
%\end{minipage}
%\hfill
%\begin{minipage}{0.85\textwidth}{
%  \includegraphics[height=2.5cm] {image/bw-utilization.pdf}\label{fig:multi11}
%  \includegraphics[height=2.5cm]{image/delay-utilization.pdf}\label{fig:multi12}
%  \includegraphics[height=2.5cm]{image/loss-utilization.pdf}\label{fig:multi13}
%  \includegraphics[height=2.5cm] {image/buffer-utilization.pdf}\label{fig:multi14}%

%  \includegraphics[height=2.5cm]{image/bw-latency.pdf}\label{fig:multi21}
%  \includegraphics[height=2.5cm]{image/delay-latency.pdf}\label{fig:multi22}
%  \includegraphics[height=2.5cm]{image/loss-latency.pdf}\label{fig:multi23}
%  \includegraphics[height=2.5cm]{image/buffer-latency.pdf}\label{fig:multi24}
%}
%\end{minipage}
%\caption{The multi-objective performance of \sys in terms of throughput (a-d) and latency (e-h), under varied network conditions. Note that the network conditions under evaluation are far beyond the environment where \sys was trained, demonstrating both its high performance and robustness when adopted in practice.}\label{fig:multi1}
%\end{figure*}

%\iffalse
%\begin{figure*}
%  \centering
%  \vspace{-6pt}
%  \includegraphics[width=\textwidth]{image/loss-good.pdf}
%  \label{fig:lossgood}
%  \caption{packet loss ratio in different network condtions}
%\end{figure*}
%\fi

\parab{Settings:} Following $\S$\ref{sec:implement}, we use CCP to deploy our \mocc in both real Internet and Pantheon~\cite{pantheon} emulated environment. We train \mocc with varied bandwidth, latency, queue size and loss rate to cover a wide range of network conditions following the settings of prior works~\cite{aurora,pantheon}. The key parameters used in both training and evaluation are shown in Table~\ref{table:range}. In particular, when evaluating \mocc, we use a much wider parameter range beyond training to show the robustness of \sys.

\begin{table}[t]
\small
\begin{tabular}{c|c|c|c|c}
  \hline 
   & Bandwidth & Latency & Queue size & Loss rate\\
  \hline 
  \hline
  Training & 1-5 Mbps & 10-50ms & 0-3000 pkts & 0-3\% \\
  \hline 
  Testing & 10-50 Mbps & 10-200ms & 500-5000 pkts & 0-10\% \\
  \hline 
\end{tabular}
\caption{Training/testing parameters}\label{table:range}
%\vspace{-0.4in}
\end{table}

%To inspect the robustness of our congestion control algorithm to real-world the variety of network conditions, the testing suite have a range 10x larger than the training environment as showed in the table. 

\parab{Schemes compared:} We compare \mocc with various CC algorithms, including both handcrafted and learning-based:
\begin{enumerate}
\item Aurora~\cite{aurora}: RL based CC algorithm, single-objective RL, Aurora-throughput and Aurora-latency basically use two separate models.
\item Orca~\cite{orca}: RL based CC algorithm, single-objective RL combined with the classic CC (CUBIC) to achieve low overhead and high performance.
\item PCC Allegro~\cite{allegro}: learning-based, performs \emph{micro experiments} to continuously explore and learn the target sending rate.
\item PCC Vivace~\cite{vivace}: learning-based, extends upon Allegro to achieve better performance.
\item BBR~\cite{bbr}: model-based heuristic, builds an explicit model based on available bandwidth and RTT, and uses the model to control congestion window.
\item Copa~\cite{copa}, delay-based heuristic, computes the target sending rate by estimating minimum delay.
\item TCP CUBIC~\cite{cubic}, loss-based heuristic, when packets are dropped, CUBIC modulates its congestion window based on a CUBIC function.
\item TCP Vegas~\cite{vegas}, delay-based heuristic, uses RTT as congestion signal and controls congestion window to maintain desired RTT.
%\item Comound,\TODO{not compared but introduced}, loss and delay-based handcrafted heuristic, Compound reacts to delay signals and packet loss events and adopts a scalable increasing rule on cwnd in response to changes inn the RTTs.
%\item Remy, \TODO{not compared but introduced}, learning-based, RemyCC formalizes the multiuser CC problem as the POMDP and learns the optimum offline policy under a wide range of network environments.
%\item PCC Allegro and Vivace~\cite{allegro,vivace}, learning-based. PCC continuously performs \emph{micro-experiments} to learn the target sending rate.
% adaptively adjusts its sending rate based on continuously carried out "micro-experiments", and online choose the action which could maximize the PCC utility funcntion.
%Aurora proposed to use RL to design a TCP CC scheme, adjusting the sending rate according to the policy which could optimize the reward function. Here we trained two version of Aurora. We adjusted the reward function to make one model much prefer to capture throughput, another model much prefer to avoid increasing delay.
\end{enumerate}

\vspace{-0.1in}
\subsection{Multi-objective Performance}\label{subsec:AA}

\begin{figure*}[t]
%\vspace{-0.2in}
\centering
\begin{minipage}[t]{0.29\textwidth}
\vspace{0pt}
\includegraphics[width=1.1\textwidth]{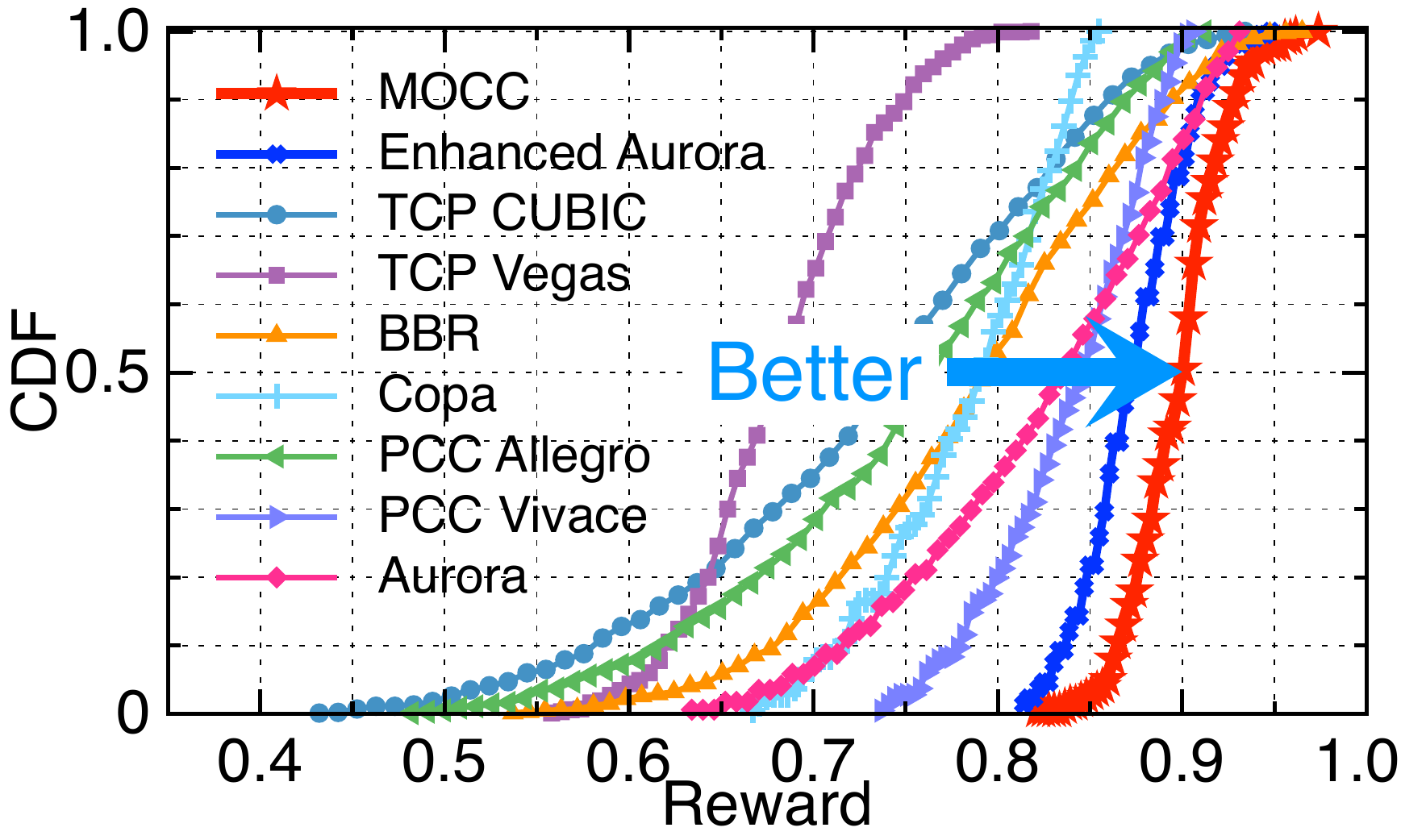}
%\vspace{-0.2in}
\caption{Quantitative CDF of 100-objective rewards for all CC algorithms compared.}
\label{fig:cdf}
\end{minipage}
  \hfill
\begin{minipage}[t]{0.70\textwidth}
%\vspace{-6pt}
\centering
\subfigure[Quickly adapt to new application] {\includegraphics[width=0.45\textwidth]{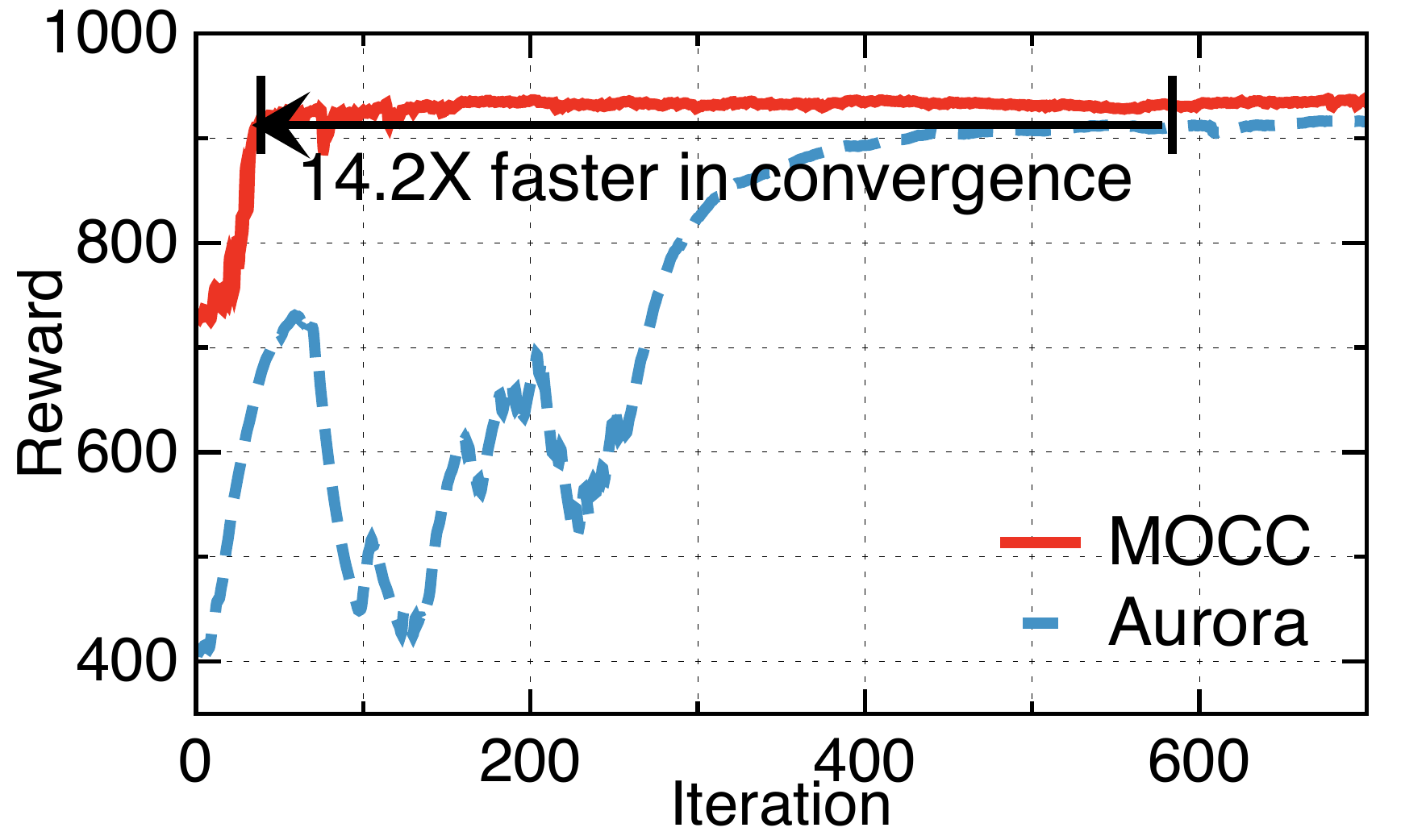}}
\subfigure[Not compromising old applications]{\includegraphics[width=0.45\textwidth]{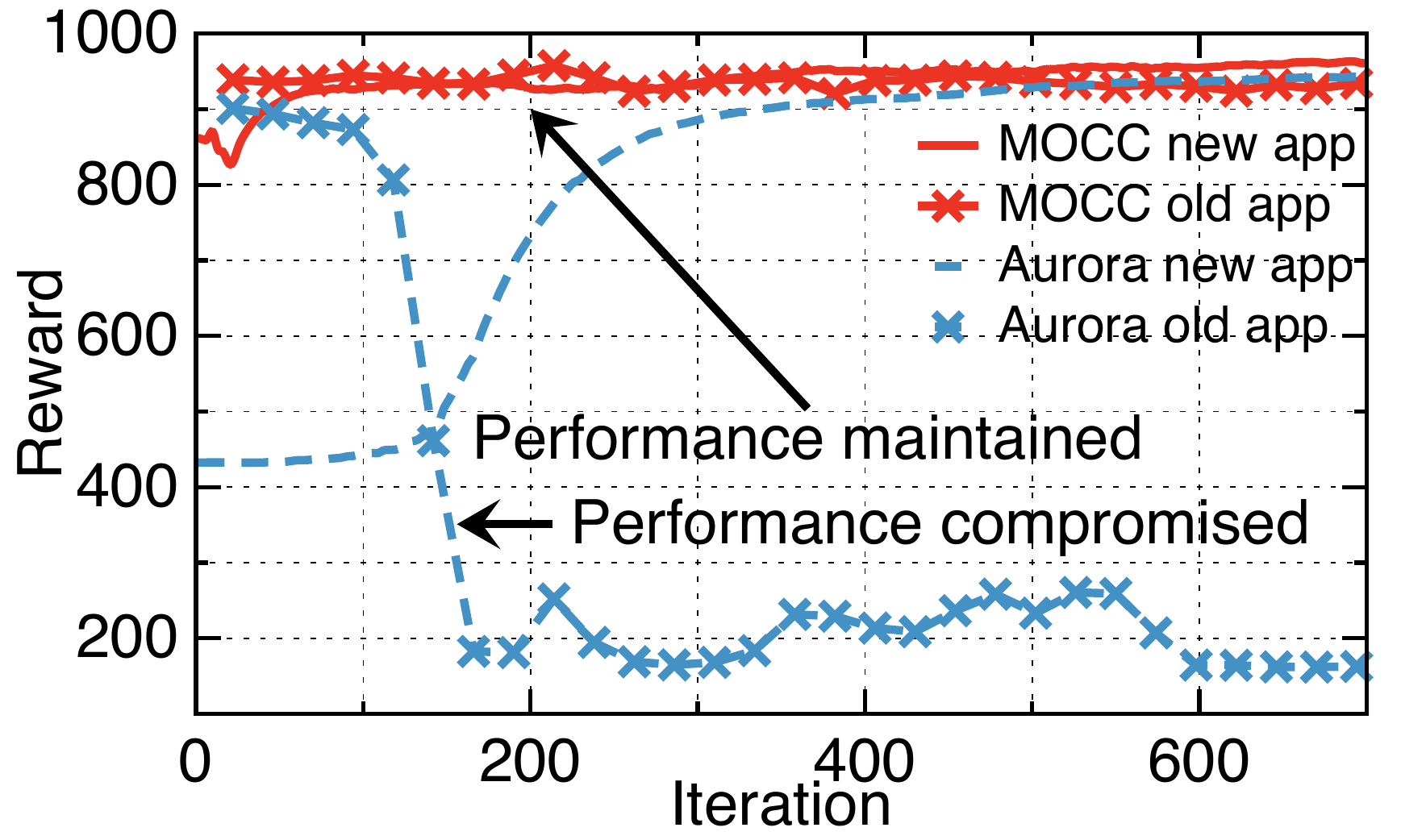}}
\vspace{-0.15in}
\caption{The quick-adaptive property of \sys. Note that we set the maximum iteration step to be 1000, so the maximum reward gain is 1000 on y-axis.}
\label{fig:quick}
\end{minipage}
%\vspace{-5mm}
\end{figure*}

To evaluate \sys's multi-objective performance in supporting different application requirements, we compare it against the above CC algorithms across different network conditions. 

\parab{2-objective:} We first consider a simple case with two common objectives: high throughput and low latency. Even so, as high throughput and low latency typically conflict with each other, it is not easy to achieve both at the same time in prior solutions ($\S$\ref{subsec:problem}). However, we show \mocc can achieve both objectives simultaneously, with weight vectors $\vec{w_1}=$$<$$0.8,0.1,0.1$$>$ and $\vec{w_2}=$$<$$0.1,0.8,0.1$$>$ respectively\footnote{Note that we only used these two particular weight vectors as example, and any vectors with similar weight settings would work.}.

The detailed results are shown in Figure~\ref{fig:multi1}: (a) to (d) show the bottleneck link utilization with varied bandwidth, one-way RTT, loss rate and buffer size when the application prefers high throughput; (e) to (h) show the latency ratio~\cite{selfinflict} when the application demands low latency.

In general, \mocc can compete or outperform the best existing CC algorithms and show consistent high performance. First, for hand-crafted CC schemes, \mocc can at least rival them in one objective and outperform them in the other, or even both. For example, \mocc achieves comparable throughput as BBR, while delivering up to $18.8\%$ (1.12 to 1.38 in Figure~\ref{fig:multi21}) lower latency. Furthermore, \mocc outperforms CUBIC in both throughput (at least $1.5\times$ from 0.95 to 0.62 in Figure~\ref{fig:multi12}) and latency ($15\%$ lower from 1.13 to 1.33 in Figure~\ref{fig:multi21}). The reason is that handcrafted CC algorithms generally adopt hardwired policies based on pre-assumptions of network conditions and human experiences, thus hard to achieve optimal application-specific performance, not to mention multi-objective. In contrast, MOCC explicitly considers application requirements in both state input and reward function, effectively solving the problem with multi- objective RL.

Second, we find that learning-based CC algorithms (non-RL), such as PCC Vivace and PCC Allegro, which essentially use online greedy optimization method, could lead to local optimization. \mocc uses RL to avoid this problem, and thus outperforms them with up to $1.43\times$ better throughput (0.83 to 0.58 in Figure~\ref{fig:multi13}) and $63.2\%$ latency reduction (1.20 to 3.27 in Figure~\ref{fig:multi24}) respectively.

Third, compared to RL-based Aurora/Orca, we see that \mocc outperforms Aurora-throughput in terms of latency ($21.6\%$ lower from 1.16 to 1.48 in Figure~\ref{fig:multi22}) while exceeding Aurora-latency in terms of throughput ($1.3\times$ from 0.89 to 0.66 in Figure~\ref{fig:multi11}). Default Orca~\cite{orca} shows similar trend except in the random loss case, due partially to the effect of CUBIC (its heuristic part). These results are expected because they use single-objective RL, which cannot simultaneously optimize for both throughput and latency, thus leading to degraded performance. On the contrary, MOCC uses multi-objective RL to simultaneously support both objectives with one model.

The readers may wonder: {\em can we pre-train a few variants of Aurora/Orca (with different weights) to achieve multi-objective?}

\parab{100-objective:} To answer the above question, we consider a more generalized, uniformly-distributed 100-objective setting. To make our results visually clear, we unify the performance metrics using {\em reward} calculated by Equation~\ref{eq:reward}. In this experiment, \sys only used offline trained model without online adaptation although with which we believe is even better. We enhanced Aurora with 10 pre-trained models that best suit these 100 objectives (Orca shares similar property in terms of single-objective RL). 

We run \mocc, enhanced-Aurora, and other CC algorithms under 10 different network conditions with 100 objectives, resulting in 1000 different scenarios. Figure~\ref{fig:cdf} presents CDFs of rewards of these 1000 cases for all the algorithms compared. It is evident that \mocc outperforms all other CC schemes (including enhanced-Aurora) in satisfying various objectives across different network conditions. We find that enhanced-Aurora with 10 pre-trained models is secondary to \sys, but vanilla Aurora with single model cannot perform well. The handcrafted heuristics, as expected, cannot well meet the multi-objective requirements because they are designed with no explicit application requirements in mind and their control policies are hardwired with pre-assumptions. 

\subsection{Quick Adaptation}\label{subsec:QA}

\begin{figure*}[t]
%\vspace{-0.2in}
\centering
\begin{minipage}[t]{0.32\textwidth}
\vspace{0pt}
 \includegraphics[width=\textwidth]{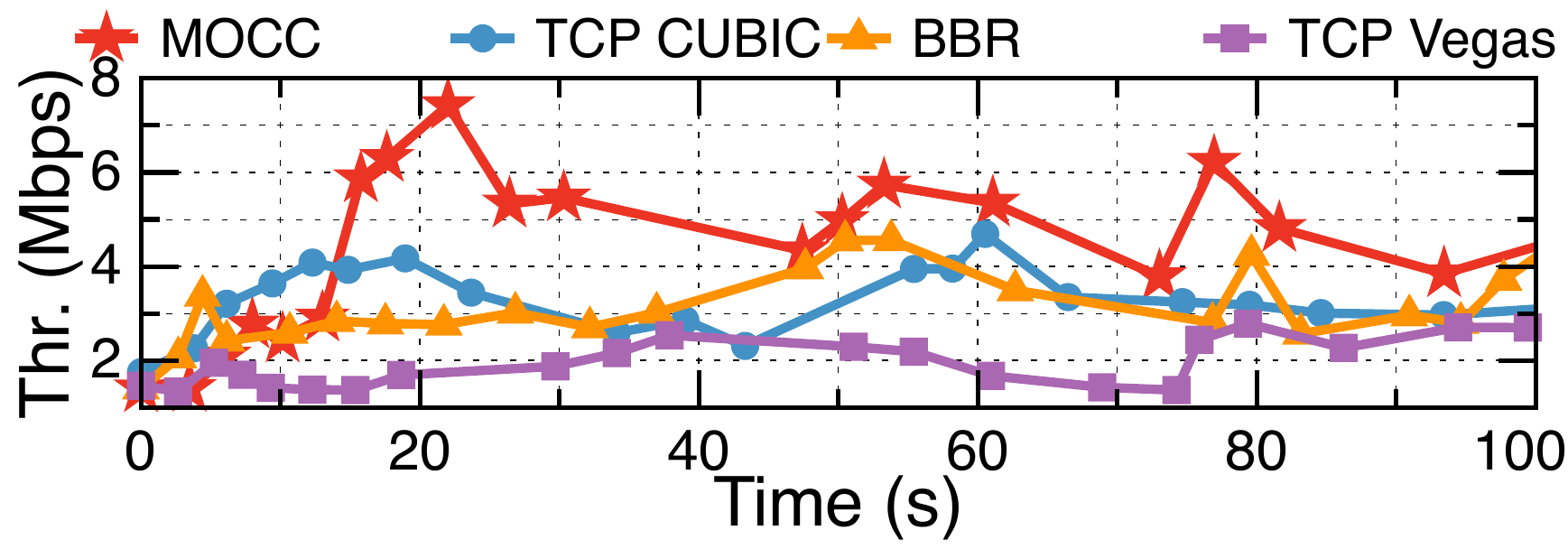}
\includegraphics[width=\textwidth]{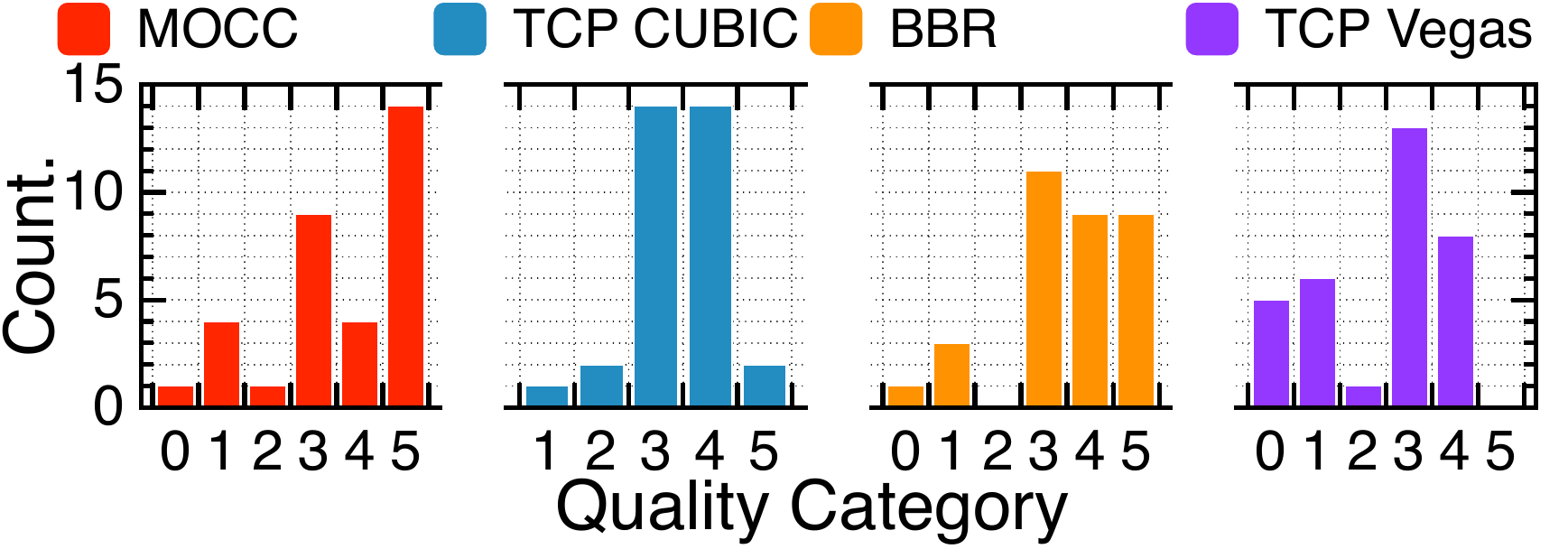}
\vspace{-0.3in}
\caption{Video streaming}\label{fig:vs}
\end{minipage}%
\hfill
\begin{minipage}[t]{0.32\textwidth}
\vspace{0pt}
\includegraphics[width=\textwidth]{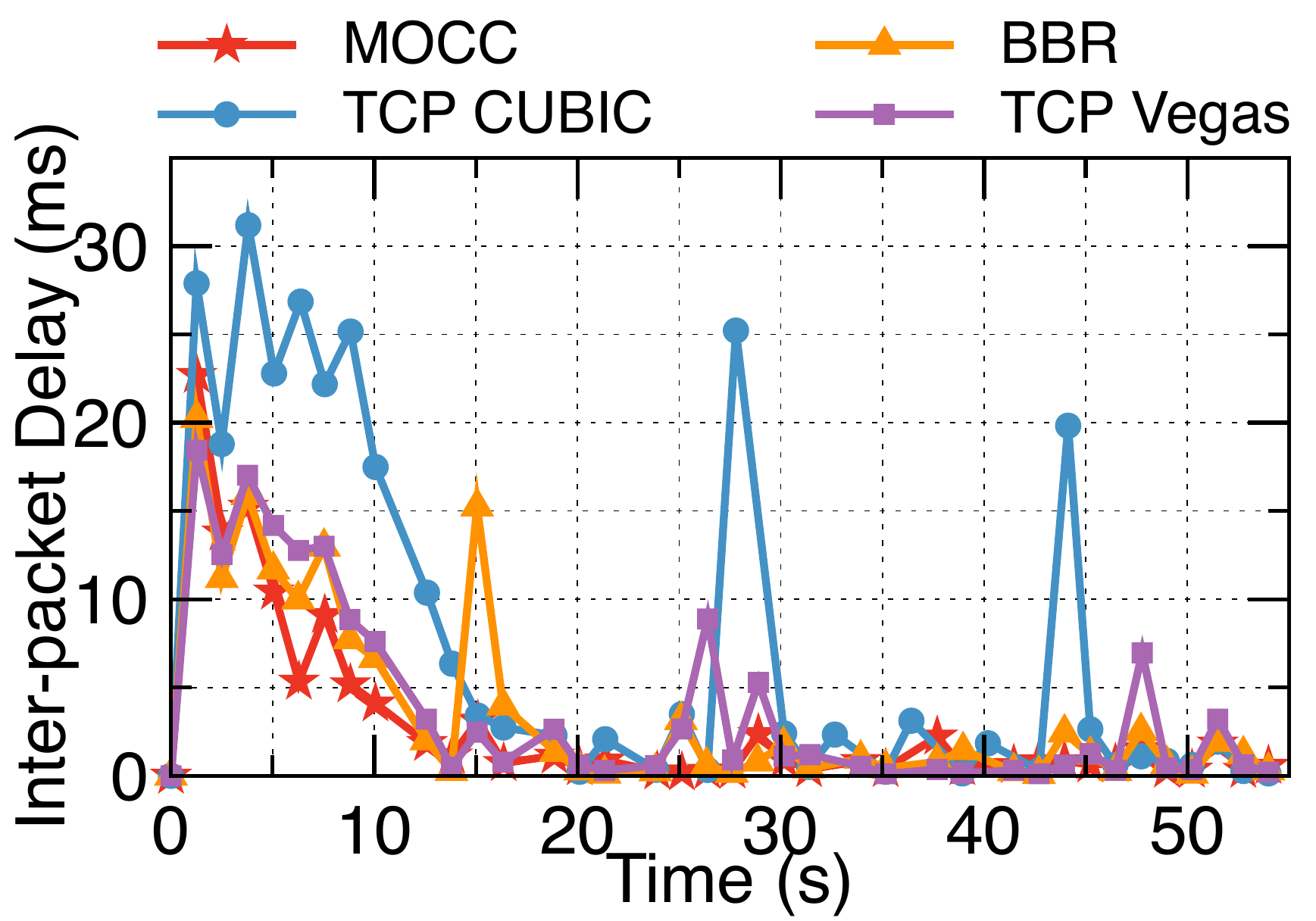}
\vspace{-0.3in}
\caption{Real time communications}\label{fig:rtc}
\end{minipage}
\hfill
\begin{minipage}[t]{0.32\textwidth}
\vspace{0pt}
\includegraphics[width=\textwidth]{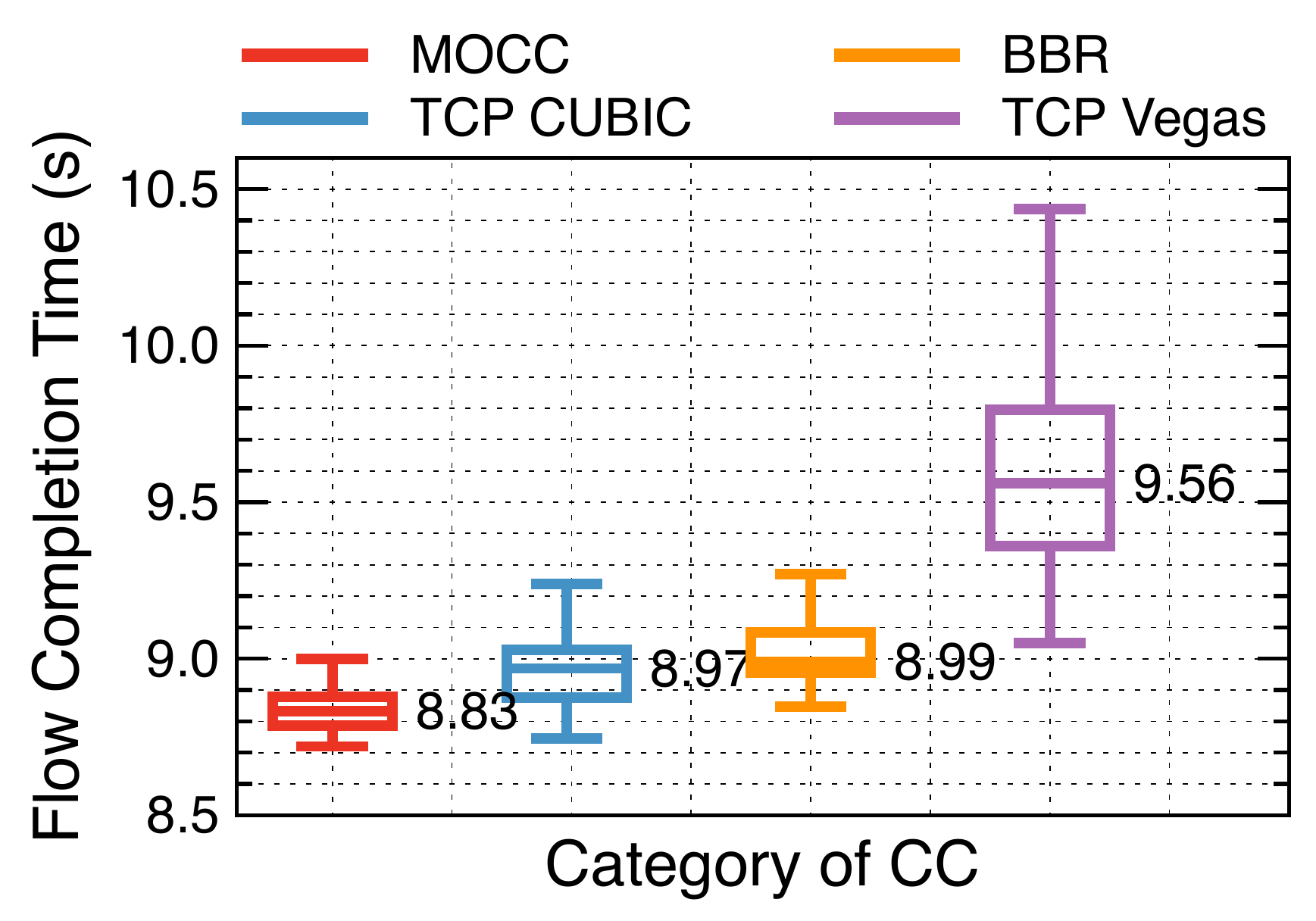}
\vspace{-0.3in}
\caption{Bulk data transfer}\label{fig:bdt}
\end{minipage}
%\vspace{-0.2in}
\end{figure*}

To show how quickly \sys adapts to new applications, we compare it against Aurora. We define the convergence point as 99\% of the maximum reward gain.

Figure~\ref{fig:quick}(a) plots the trend of both algorithms in adapting to new applications. The x-axis denotes the number of iterations and y-axis the gained reward. First, we observe that \mocc achieves $1.8\times$ higher initial performance for a new application over Aurora. This suggests that by learning correlations between application requirements and optimal policies, \mocc can provide moderately good polices for applications with unseen requirements. Meanwhile, with transfer learning, such base correlation knowledge brings a $14.2\times$ faster convergence speed (639 down to 45 iterations). This result confirms that \mocc can quickly adapt to new applications, whereas the single-objective Aurora, without the correlation knowledge, re-trains the model from scratch which takes long time.

Figure~\ref{fig:quick}(b) checks whether \mocc will degrade the performance of old applications while adapting to new ones. To do so, we snapshot the models of \mocc and Aurora every 8 iterations, and apply them to the old application to compute the rewards. The curves are illustrative. We observe that \mocc well preserves the performance of the old application with reward loss $<$5\%. This is because \sys has the correlation model and applies the requirement replay algorithm ($\S$\ref{subsec:online}) to recall the old application during online training. In contrast, Aurora, as a single-objective CC model, gradually forget the old application and degrades the performance greatly (916.1 to 156.1) while serving the new application.

\subsection{Real Internet Applications}\label{subsec:realapp}

\begin{figure*}[t]
%\vspace{-0.2in}
\begin{minipage}{0.66\textwidth}
\centering
  \subfigure{\includegraphics[width=0.235\textwidth] {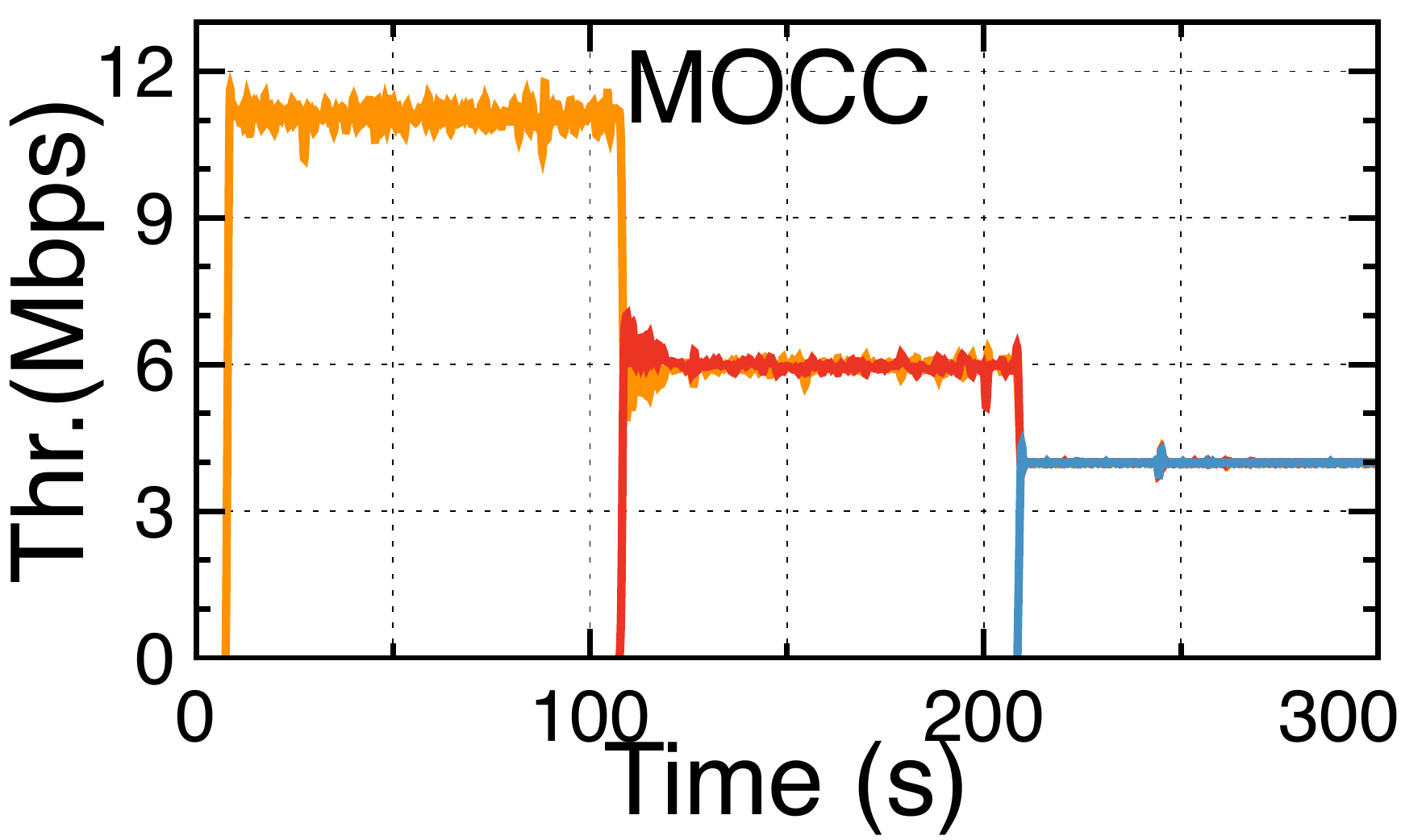}}
  \subfigure{\includegraphics[width=0.235\textwidth] {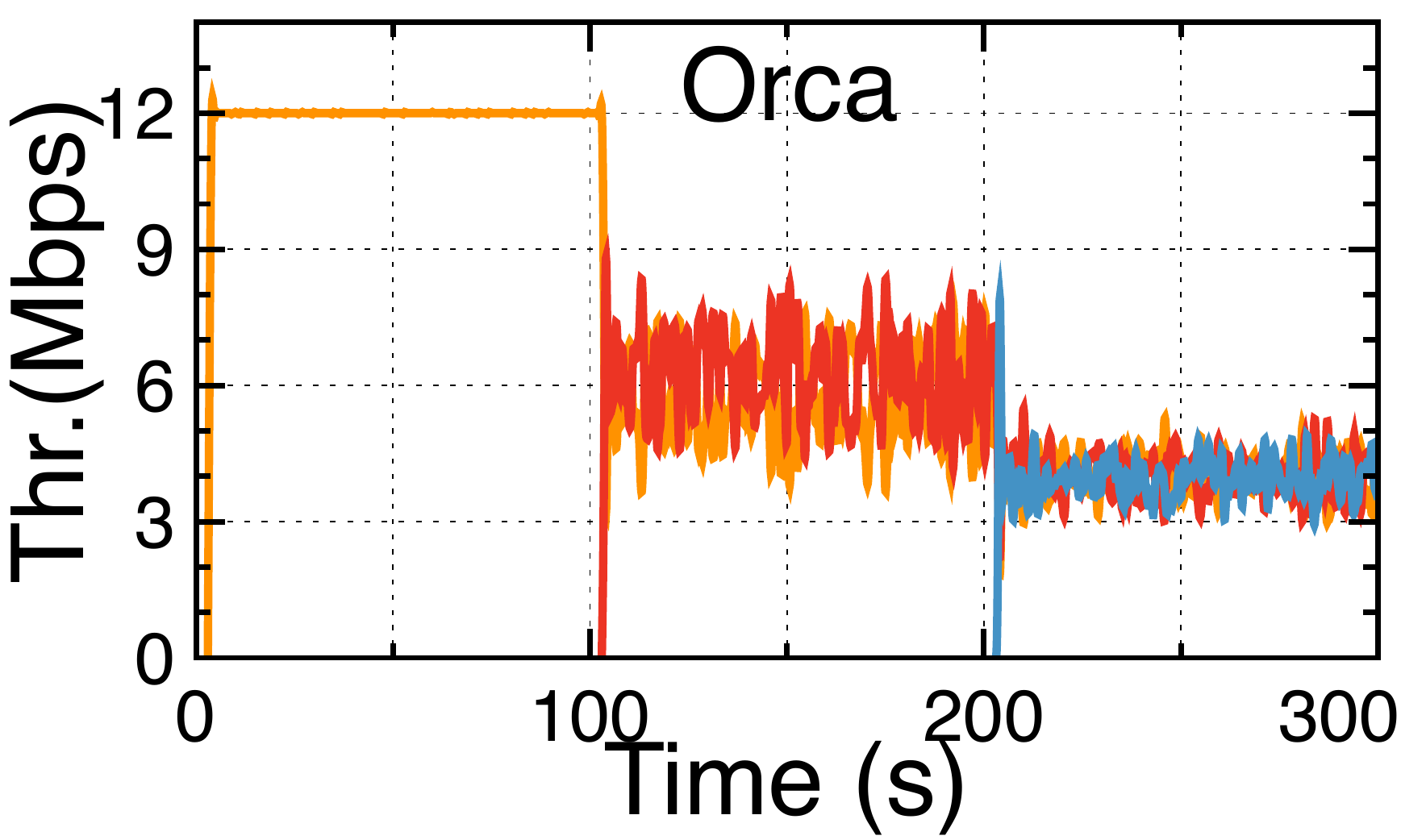}}
  \subfigure{\includegraphics[width=0.235\textwidth] {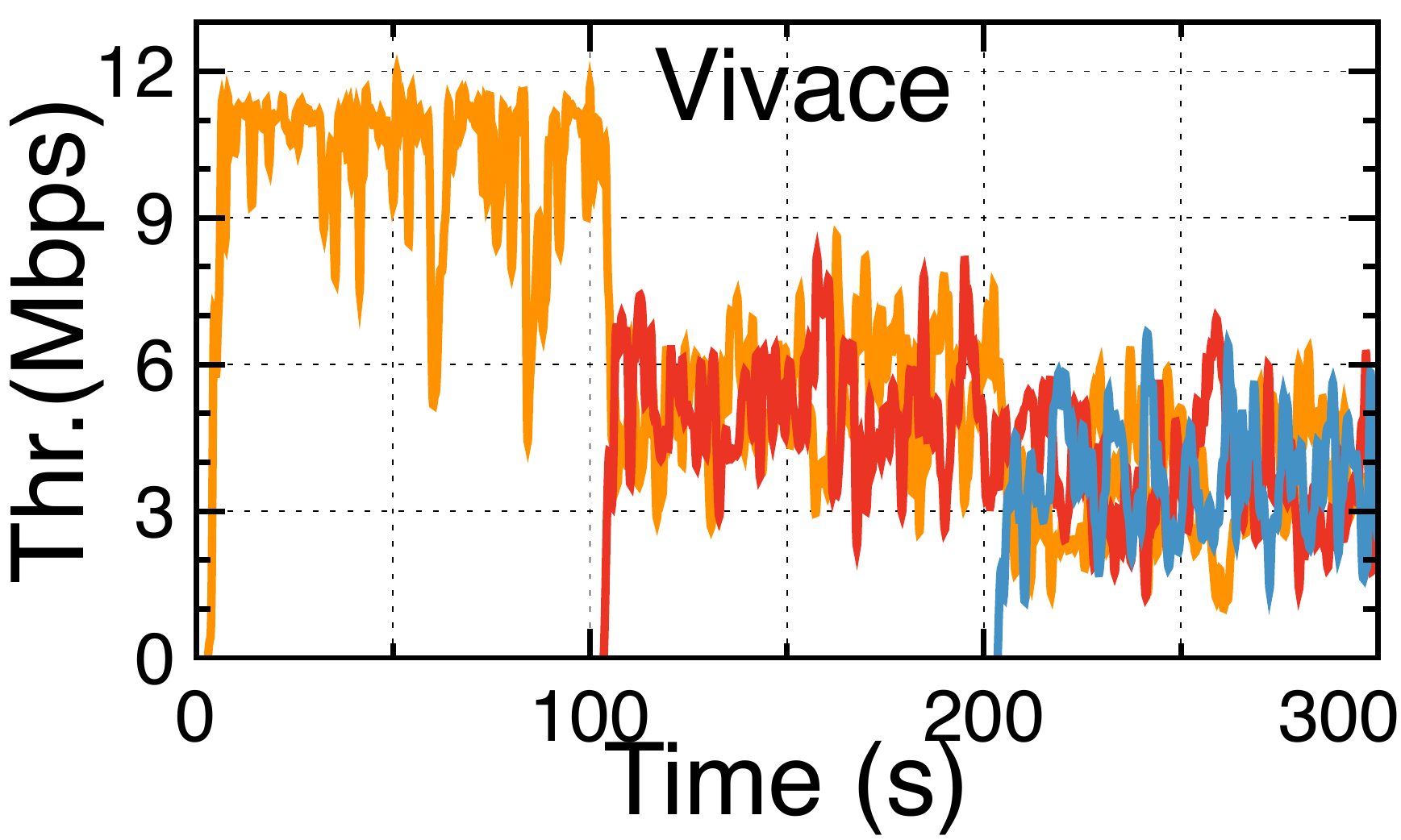}}
  \subfigure{\includegraphics[width=0.235\textwidth] {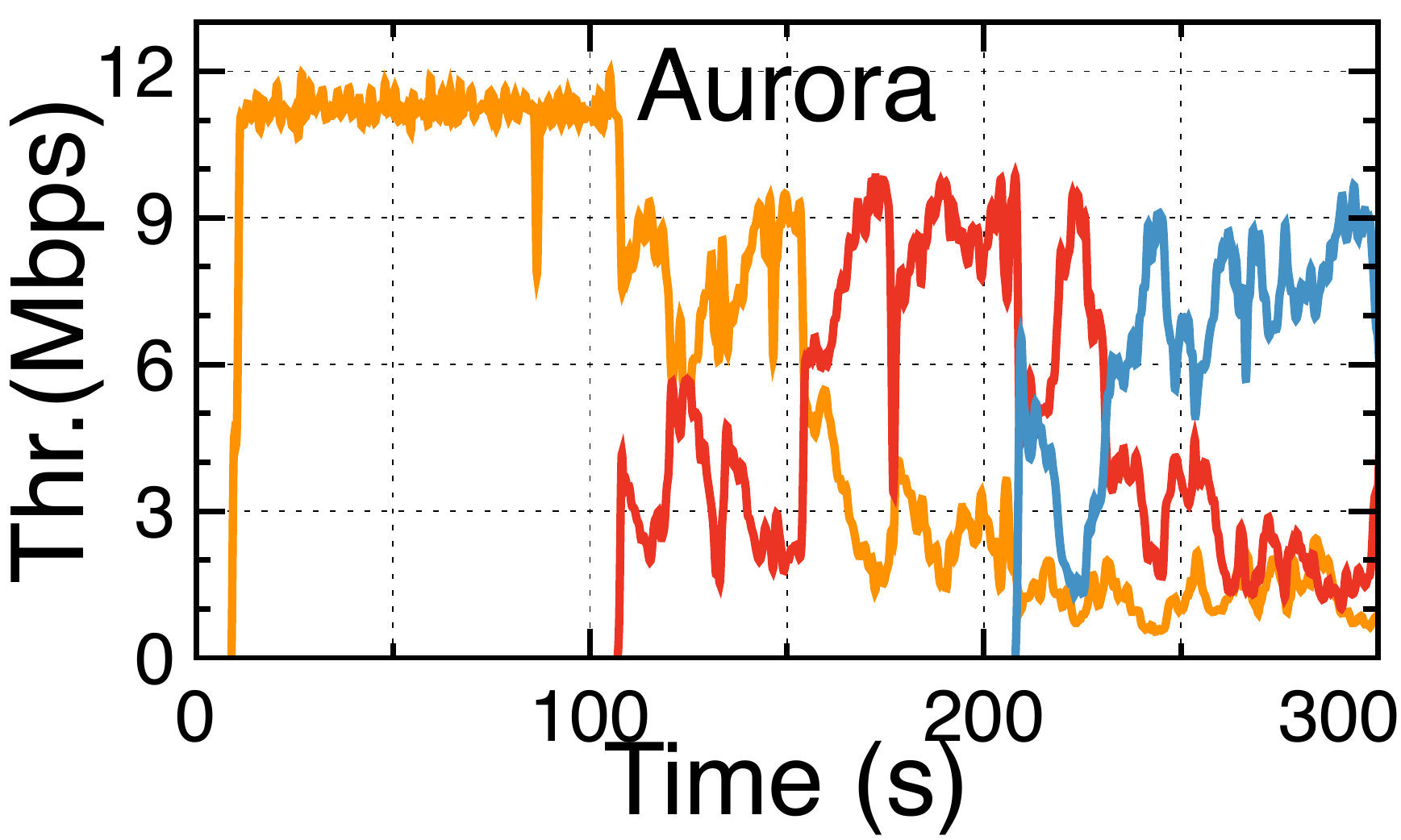}}

 \vspace{-2mm}

  \subfigure{\includegraphics[width=0.235\textwidth] {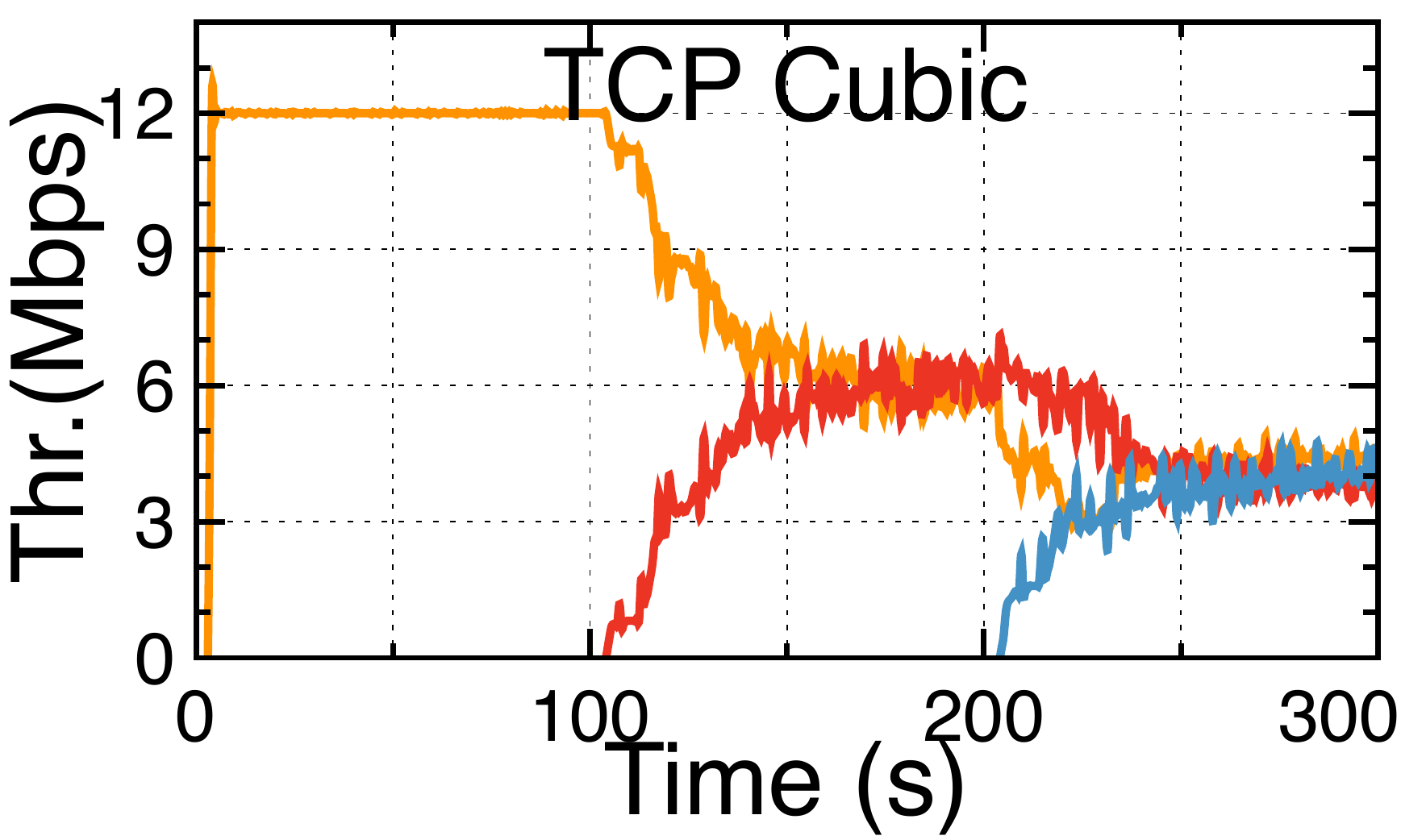}}
  \subfigure{\includegraphics[width=0.235\textwidth] {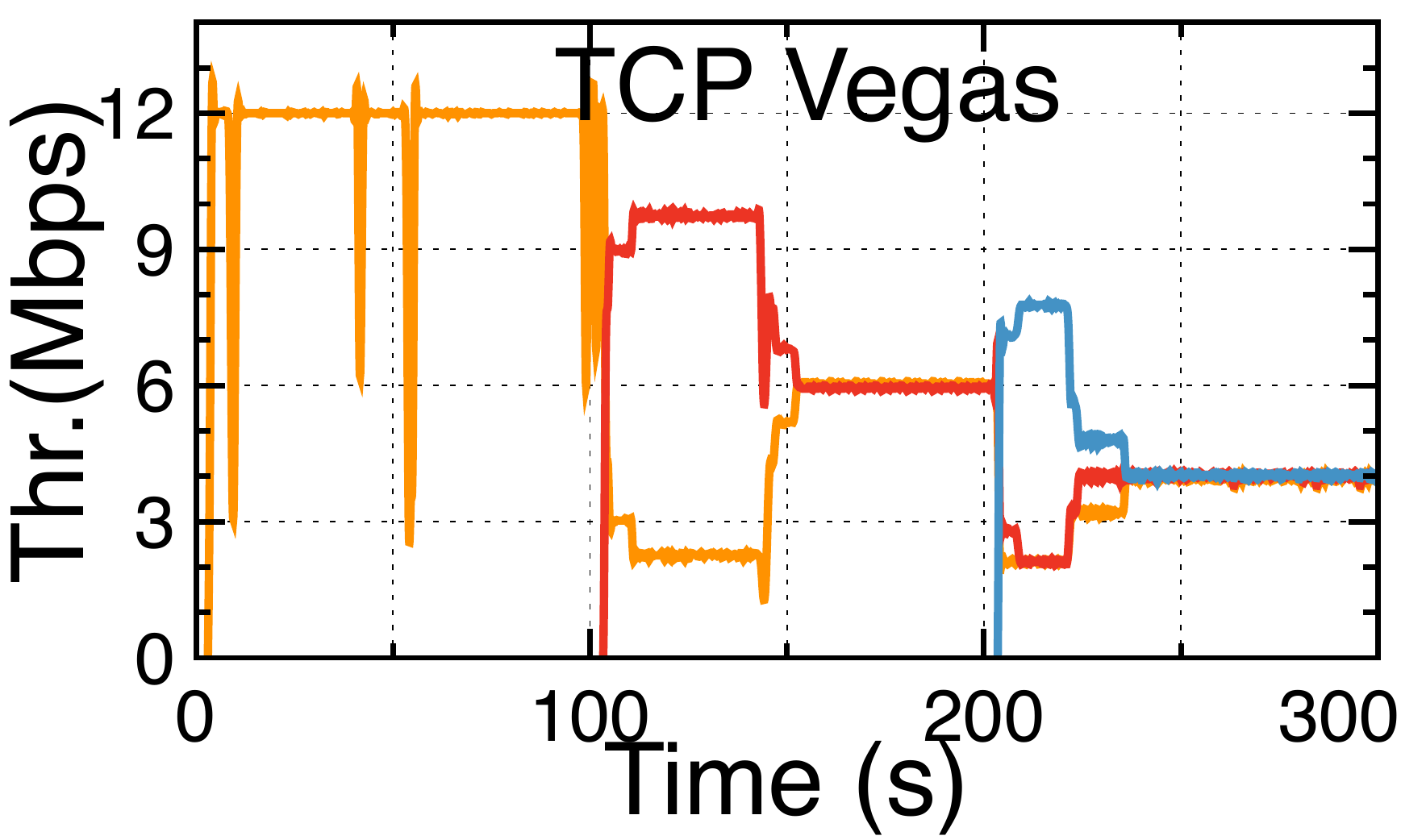}}
  \subfigure{\includegraphics[width=0.235\textwidth] {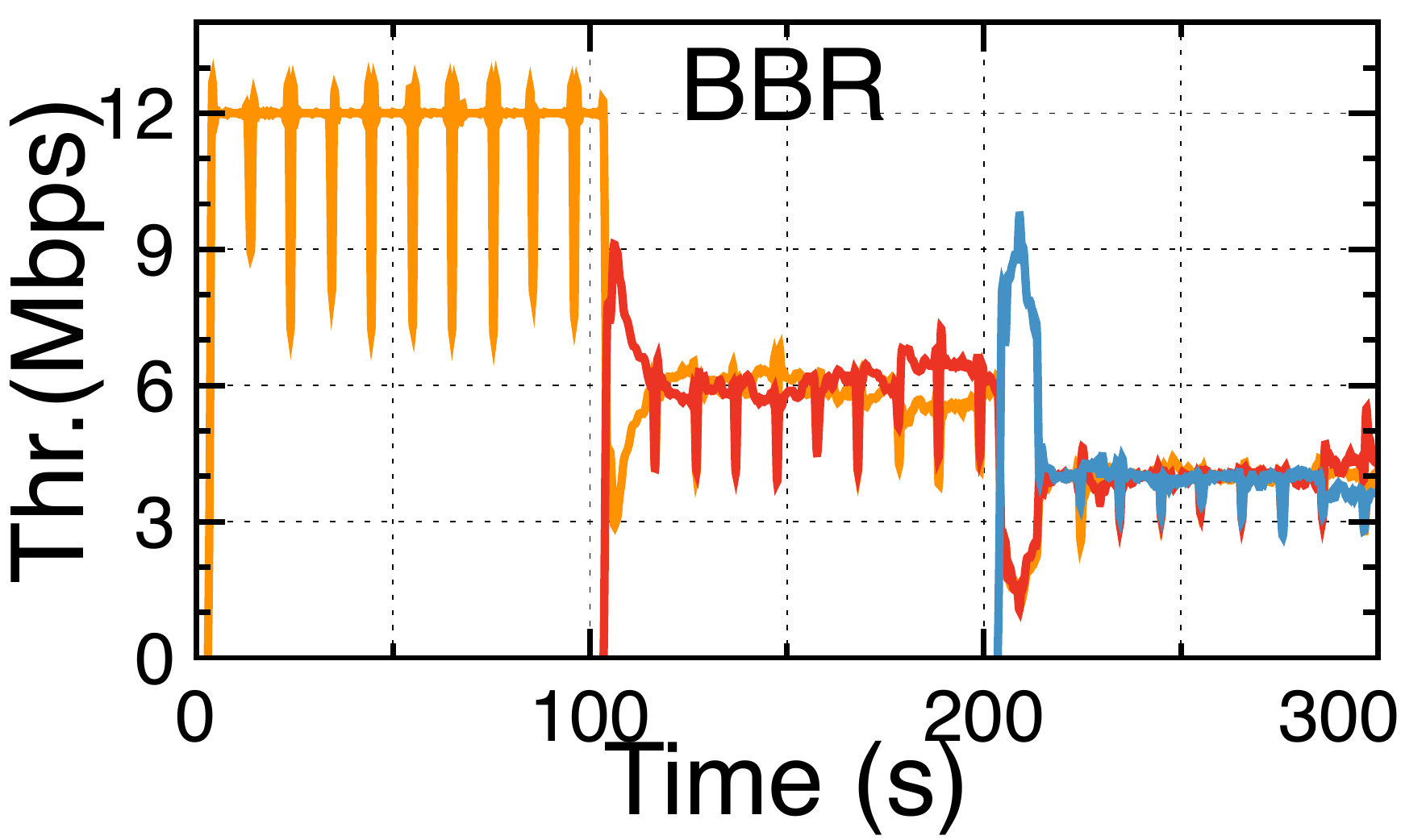}}
  \subfigure{\includegraphics[width=0.235\textwidth] {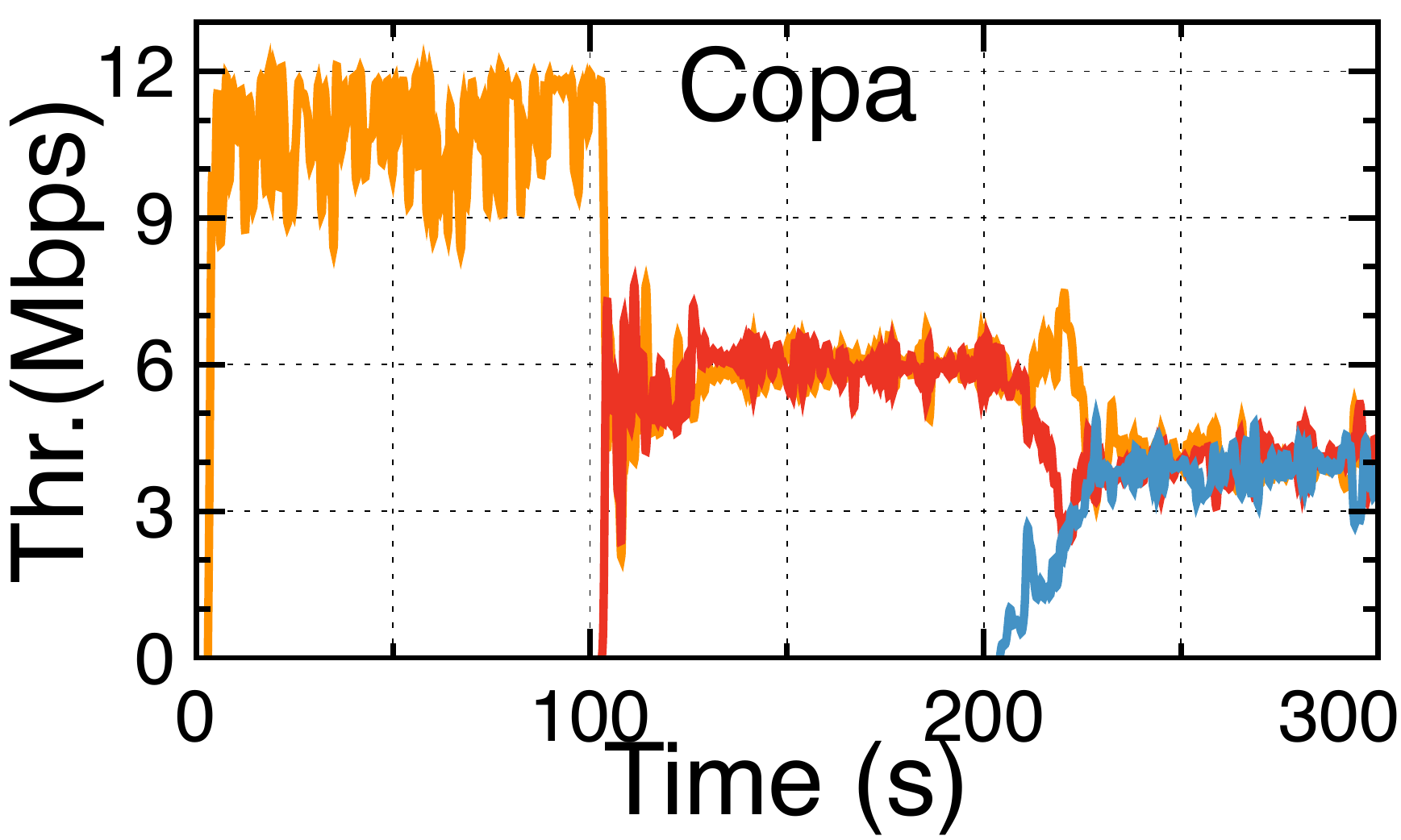}}
%\vspace{-0.25in}
\caption{Throughput dynamics of different flows competing one link for various CC}\label{fig:thrdynamics}
\end{minipage}
\begin{minipage}{0.33\textwidth}
\centering
\includegraphics[width=\textwidth]{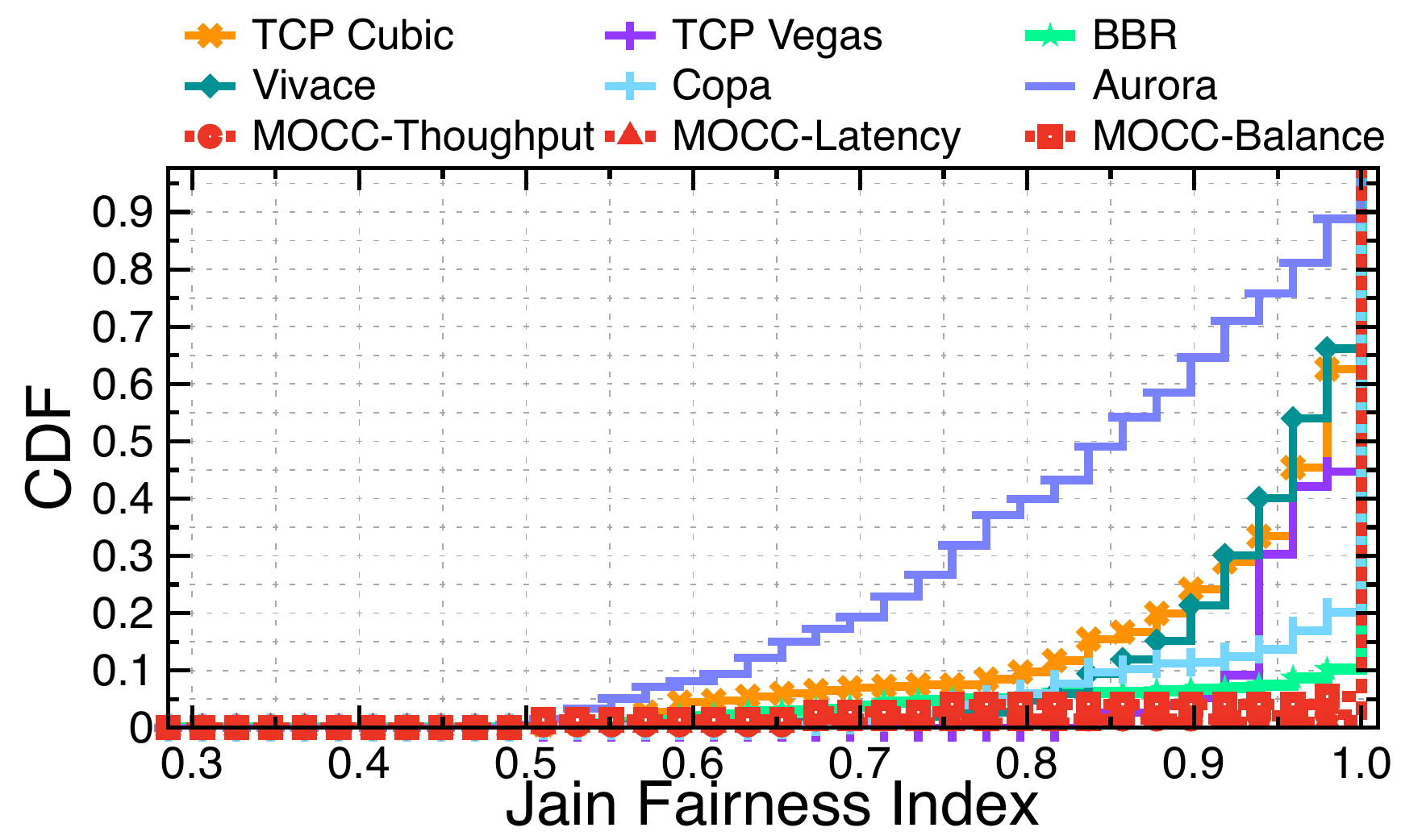}
%\vspace{-0.25in}
\caption{CDF of Jain Fairness Index under dynamics of flows.} \label{fig:jainfair2}
\end{minipage}
%\vspace{-0.1in}
\end{figure*}

\begin{figure*}[t]
\begin{minipage}{0.33\textwidth}
\centering
\includegraphics[width=0.48\textwidth]{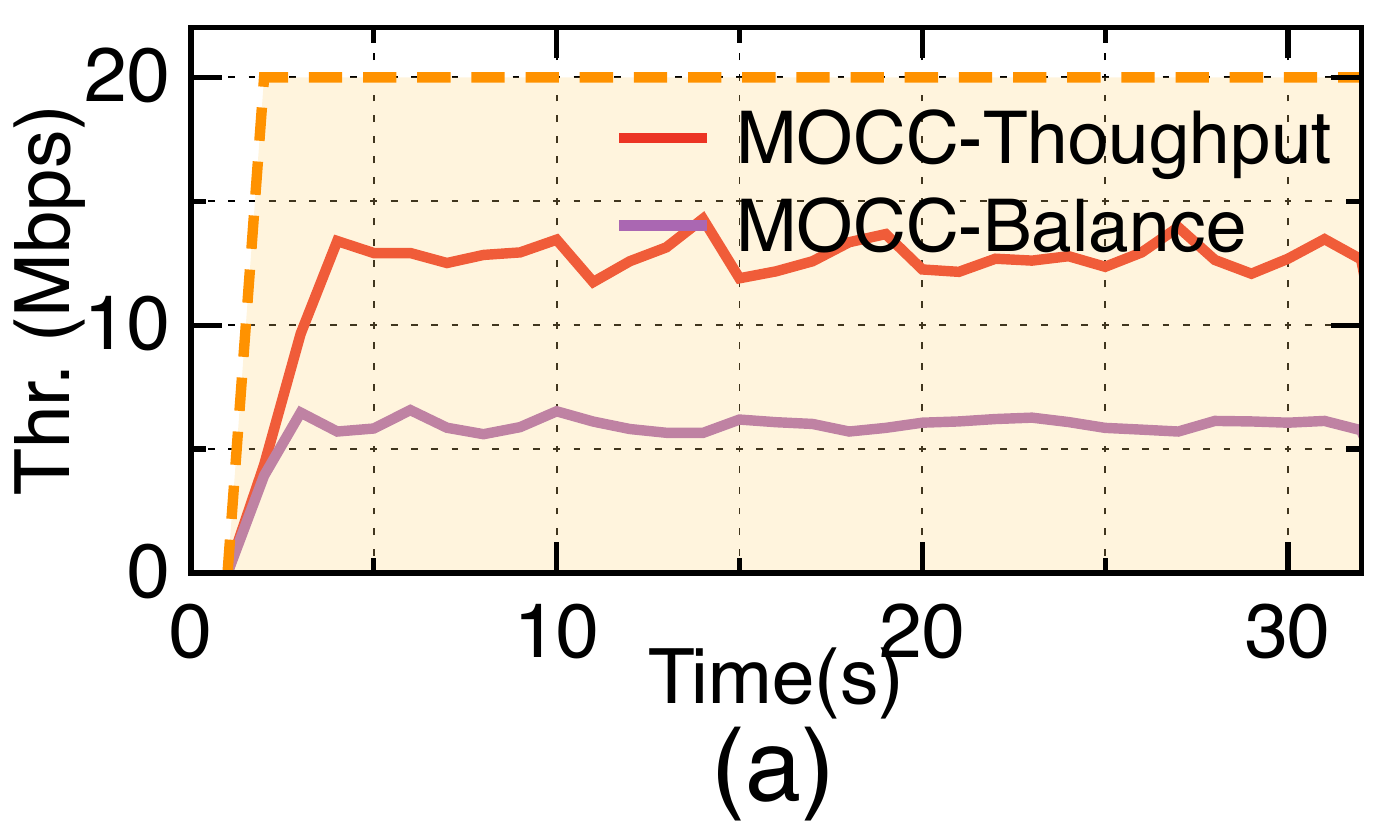}
\includegraphics[width=0.48\textwidth]{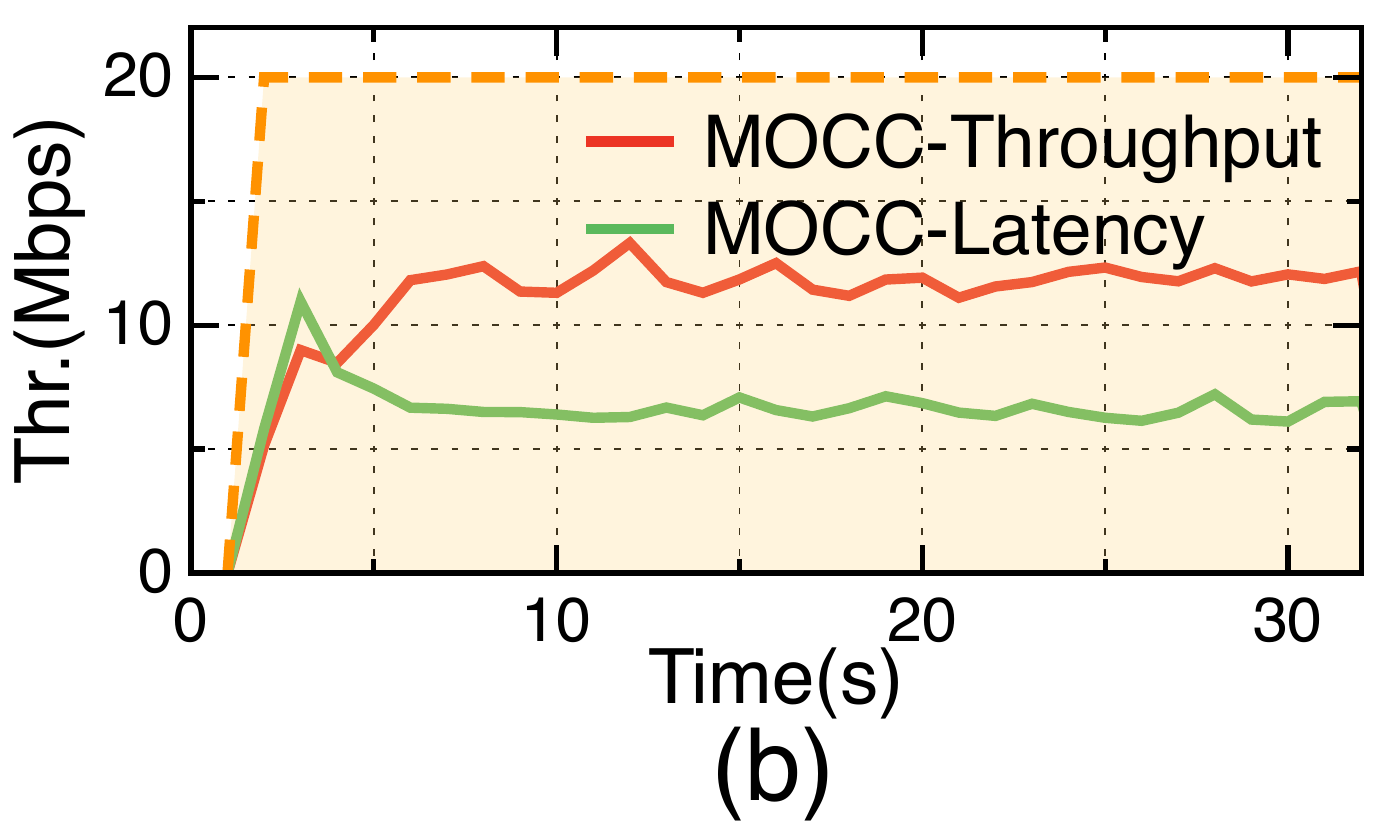}

\includegraphics[width=0.48\textwidth]{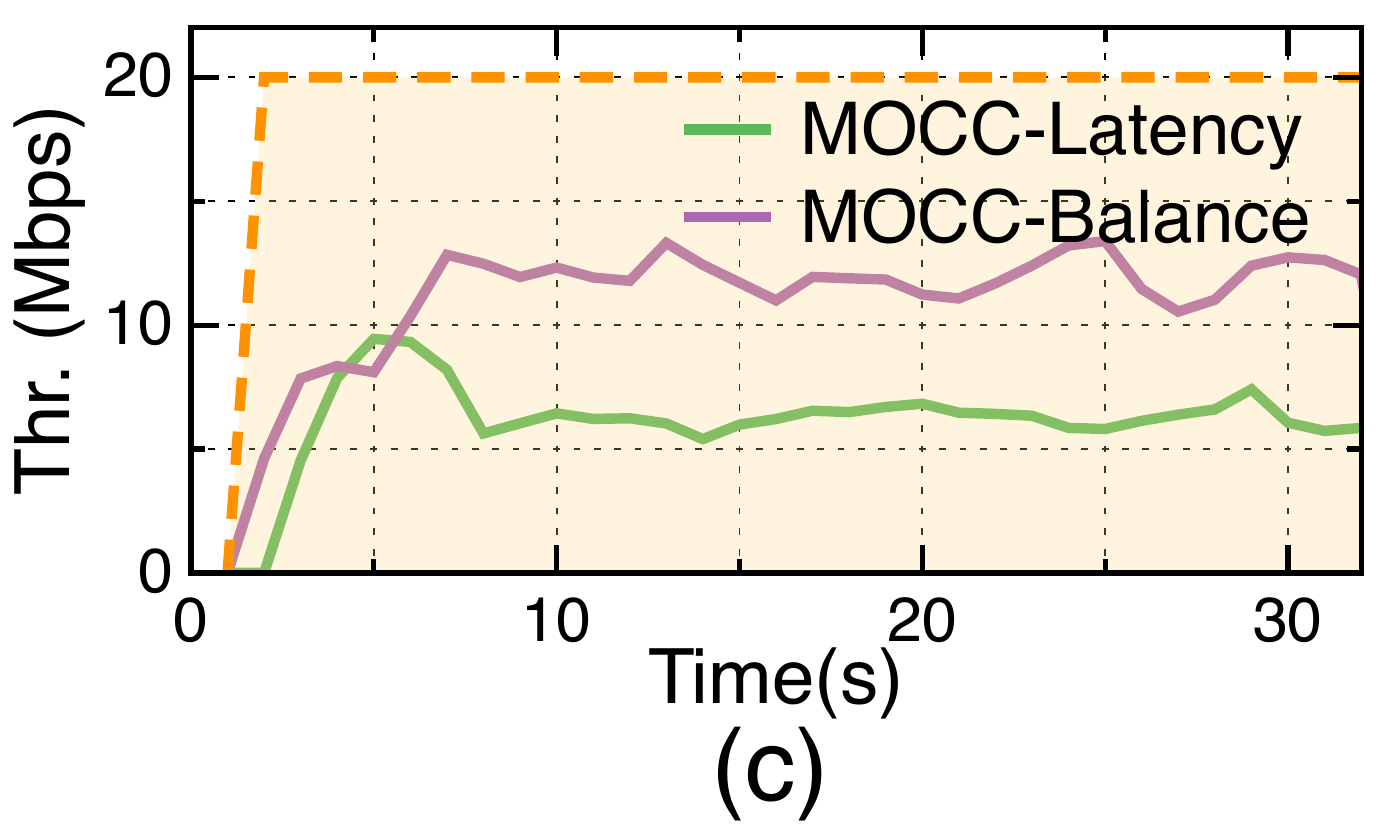}
\includegraphics[width=0.48\textwidth]{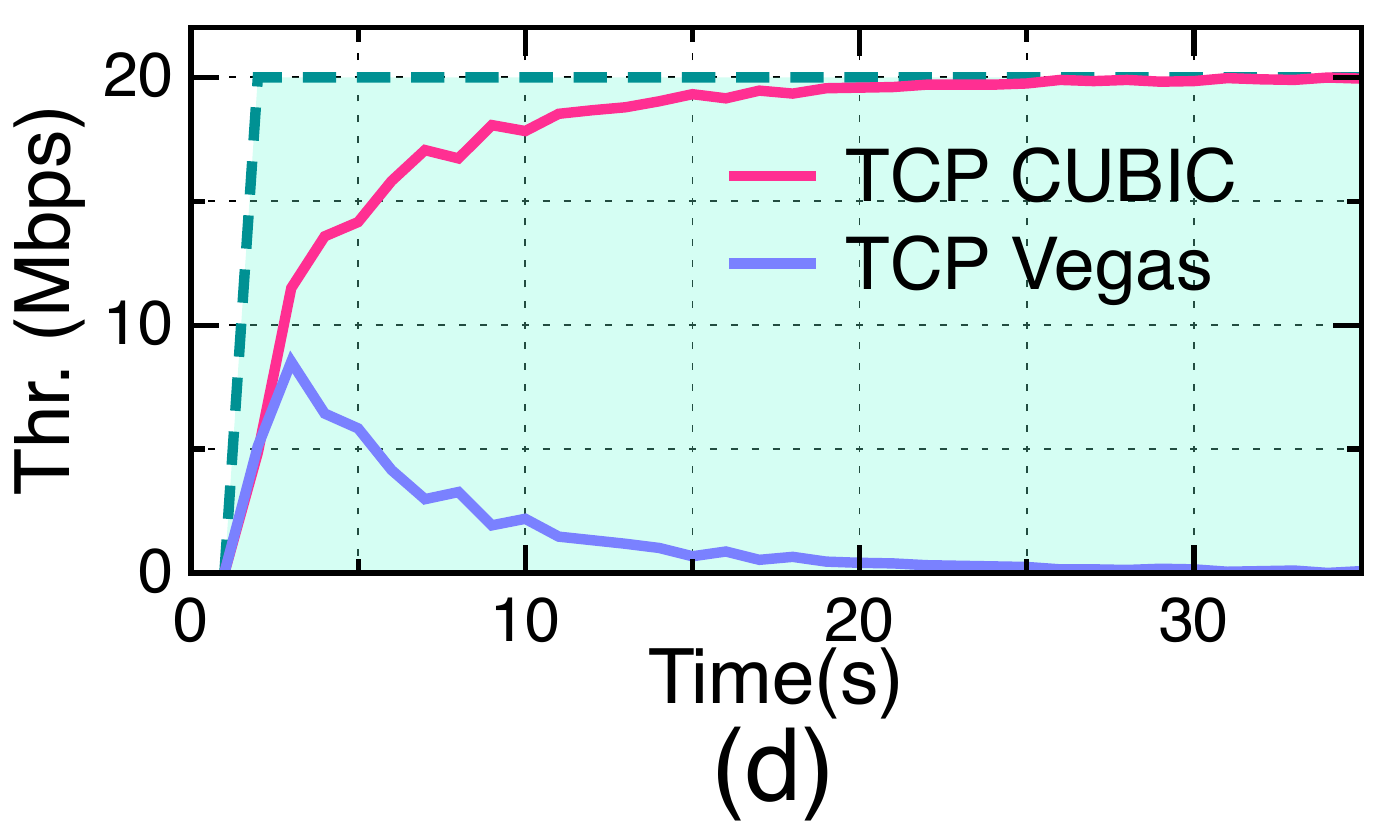}
\vspace{-0.2in}
\caption{Throughput of MOCC flows with different weights}
\label{fig:twoweight}
\end{minipage}
  \hfill
\begin{minipage}{0.33\textwidth}
\centering
\includegraphics[width=\textwidth]{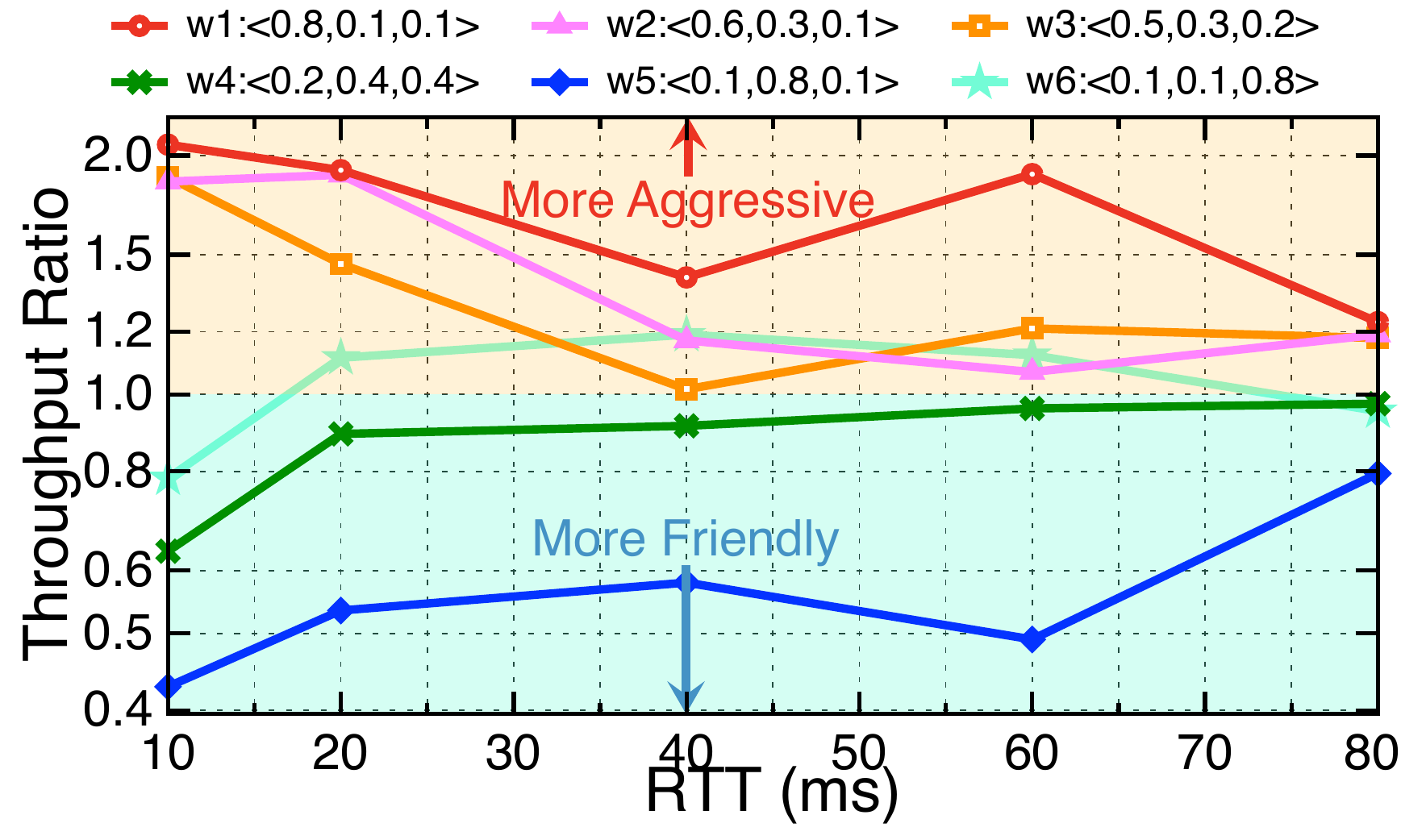}
\vspace{-0.35in}
\caption{Friendliness ratio of \mocc under different weights}
\label{fig:moccfriend}
\end{minipage}
  \hfill
\begin{minipage}{0.33\textwidth}
\includegraphics[width=\textwidth]{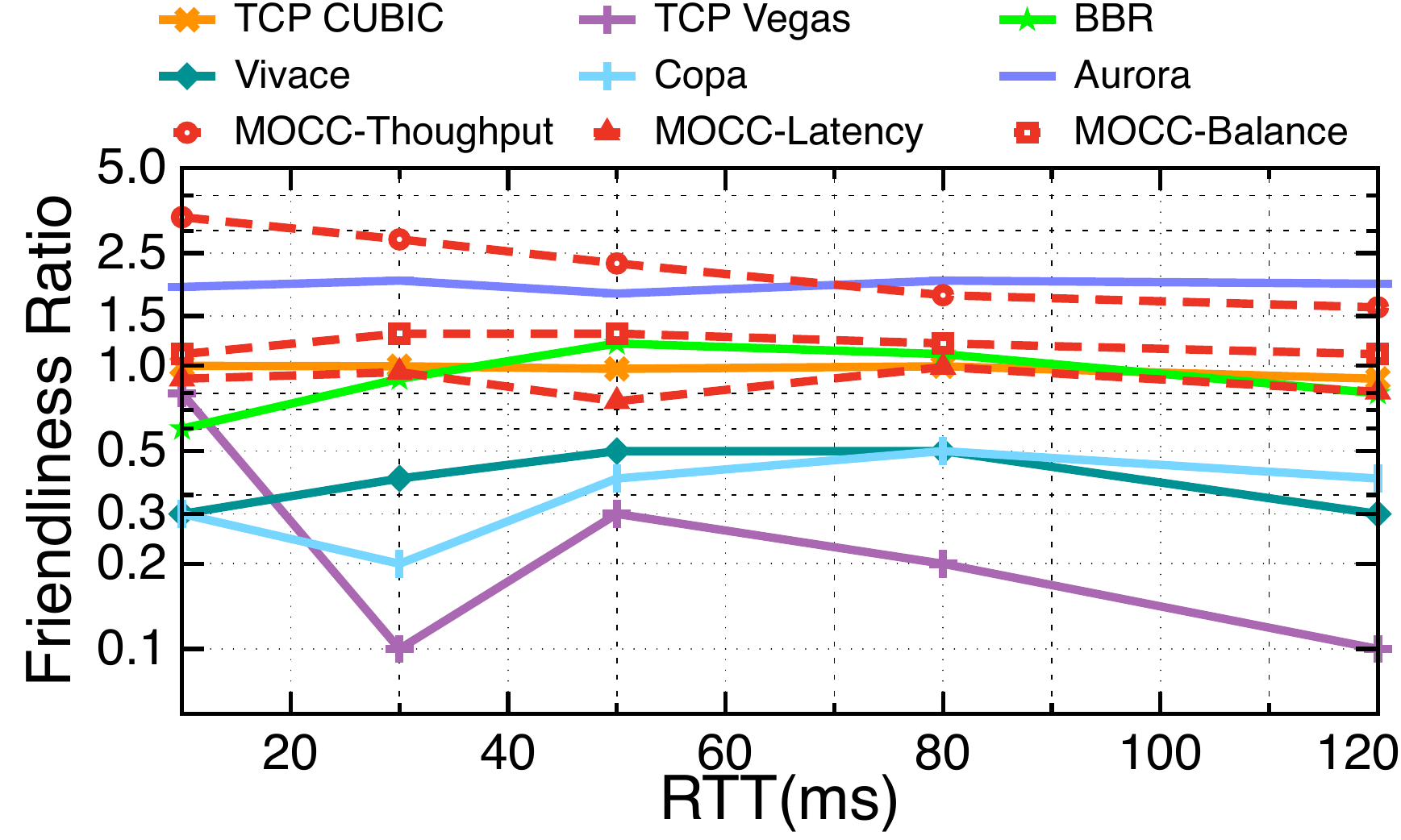}
\vspace{-0.35in}
\caption{Friendliness ratio of schemes across different RTTs}
\label{fig:friendrtt}
\end{minipage}
%\vspace{-0.2in}
\end{figure*}

We now showcase the performance of \mocc with 3 real Internet applications: video streaming, real-time communications (RTC), and bulk data transfer. These applications have different requirements, and {\em we use a single \mocc model to support all of them}. We use CCP to deploy our \mocc on Linux kernel 4.15.0-74-generic. We compare it with TCP CUBIC, Vegas and BBR which are built-in algorithms in Linux Kernel and widely used in Internet.

%\yq{compare pattern:yiqing CUBIC,vegas ,bbr}

%\parab{Video streaming:} For video streaming applications, we deployed Pensieve~\cite{pensieve}, which response the video requests via Apache server. \tianhan{This is not right! Pensieve is an ABR algorithm. More details about deployment are needed.} For video streaming tasks, \mocc applies the weight vector $\vec{w}=\langle 0.8,0.1,0.1 \rangle$ in order to achieve the required bitrate/throughput for fetching video chucks. For comparison, we used Copa~\cite{} and TCP CUBIC~\cite{} which can deliver good performance on throughput.

\parab{Video streaming:} In this experiment, we deployed a video streaming server and use ABR algorithm provided by Pensieve~\cite{pensieve}.  We applied $\vec{w}=<$$0.8,0.1,0.1$$>$ for \mocc because video streaming applications need high throughput and are not sensitive to latency due to the playback buffer. We used our browser on client to play the video from the server via both WiFi and wired networks. 

The experimental results are shown in Figure~\ref{fig:vs}. 
%Figure~\ref{fig:vs}(up) shows the average throughput of all CC algorithms.
We can see that \mocc continuously outperforms other CC algorithms in terms of throughput Figure~\ref{fig:vs} (top). Specifically, the average throughput of \mocc is $91.3\%$ (4.4 to 2.3 Mbps) higher than Vegas, $33.3\%$ (4.4 to 3.3 Mbps) higher than CUBIC, $29.4\%$ (4.4 to 3.4 Mbps) higher than BBR. Figure~\ref{fig:vs} (bottom) shows the number of chunks with different quality levels (higher is better, level 5 is the best) obtained during video streaming. The level of chunk obtained is decided by MPC algorithm and a better networking condition leads to a higher level of chunks. In our experiment, \mocc can obtain the largest number of level 5 chunks compared to others (14 in \sys vs 9 in BBR, 2 in CUBIC, and 0 in Vegas). The result further shows that \mocc can satisfy the requirement of video streaming application, outperforming the other algorithms.

%Figure \ref{fig:vs} shows the average throughput of \mocc stays higher over the whole video playback by \wait{} than TCP CUBIC and Copa as shown in Figure~\ref{fig:vs1}. Additionally, we record the number of chunks of different quality levels received through the playback in Figure~\ref{fig:vs2}, where quality level 5 denotes chunks with highest bitrate and best video quality. As shown in the result, \mocc has obtained the most number of chunks with highest video quality compared with CUBIC and Copa.

%\parab{Real-time communications (RTC):} For RTC, we adopted Salsify~\cite{salsify} which is a new design for real-time Internet video that jointly controls a video codec and a network transport protocol. Since it originally used UDP, to incorporate congestion control algorithms, we modify Salsify to apply TCP to send video frames. \tianhan{More details about deployment!}. Since RTC requires low delay transmission, \mocc applies the weight vector $\vec{w}=\langle 0.4,0.5,0.1\rangle$ to optimize for the latency while still preserving certain throughput. For comparison, we compared with TCP CUBIC and BBR\tianhan{Why?}.

\parab{Real-time communications (RTC):} We deployed Salsify (the latest real-time WebRTC)~\cite{salsify} for RTC application. We modified Salsify to work with TCP. We applied $\vec{w}=<$$0.4,0.5,0.1$$>$ for \mocc because besides throughput, RTC applications also care about latency to avoid lags. We used our browser on client side to set up a conference call with the Salsify server via both WiFi and wired networks.

Figure~\ref{fig:rtc} shows the average inter-packet delay. We observe that \mocc achieves the lowest inter-packet delay and is $21.1\%$ (3.0 to 3.8 ms) better than BBR, $63.1\%$ (3.0 to 7.9 ms) than CUBIC and $26.8\%$ (3.0 to 4.1 ms) than Vegas. The result suggests \mocc can deliver the best performance to RTC applications by keeping low inter-packet delay.

\parab{Bulk data transfer:} For bulk data transfer, we connected a server and a client by a switch and sent large file via Python. As the file transfer is throughput-hungry, we greedily applied $\vec{w}=<$$1,0,0$$>$ for \mocc. We transfer a 100MB file for 50 times. We also add a random loss rate of $0.5\%$ to the links to emulate background traffic interference.

Figure~\ref{fig:bdt} shows the results. Compared to others, \mocc achieves the lowest average file transfer completion time and is  $1.56\%$ (8.83 to 8.97 ms) than CUBIC, $1.78\%$ (8.83 to 8.99 ms) lower than BBR and $7.63\%$ (8.83 to 9.56 ms) than Vegas. Besides, \mocc maintains the most stable performance, and the standard deviation of these 50 measurements is $0.096$, while BBR is $0.154$, CUBIC $0.123$ and Vegas $0.421$, respectively. This result shows that \mocc can provide consistent high bandwidth to throughput-intensive applications.

\vspace{-0.1in}
\subsection{Fairness and Friendliness}\label{subsec:FF}
To evaluate the fairness and friendliness of \mocc, we compare \mocc with other
CC schemes using Pantheon~\cite{pantheon}. Fairness considers the scenarios where all
flows use the \emph{same} CC scheme, and friendliness considers those with \emph{different} CC schemes (including \mocc with different weights).

\parab{Fairness:} We use a canonical setting for evaluating fairness:
several flows use the same CC scheme to share a bottleneck link in a dumbbell
topology. The link is configured with 12Mbps bandwidth, 20ms RTT and 1 $\times$
BDP buffer, and three flows initiates sequentially with a 100s interval.
Figure~\ref{fig:thrdynamics} shows the throughput of different flows for each
scheme. As expected, \mocc (with the same weight) allocates bandwidth fairly between competing flows. Furthermore, it also achieves fast convergence, because it adjusts the sending rate with a
multiplicative factor as defined in Equation~\ref{eq:send}.

We also use the Jain's fairness index~\cite{jain} to quantitatively compare the fairness of different schemes for the same setup. A close-to-1 value indicates better fairness. We compute the Jain's fairness index for each second for each scheme, and we also include three variants of \mocc configured with different weights. Figure~\ref{fig:jainfair2} shows the CDF curve. From the figure, we confirm that: 1) \mocc achieves better fairness compared to other CC schemes in general, and 2) its fairness is irrespective of its weight configuration.

\parab{Friendliness:} We first evaluate the friendliness of \mocc with different weights. The setup has two flows sharing a bottleneck link of 20Mbps bandwidth, 20ms RTT and 1$\times$BDP buffer. We use three \mocc variants, which are \mocc-Throughput, \mocc-Balance, and \mocc-Latency. Figure~\ref{fig:twoweight}(a)(b)(c) show pairwise competitions of the three variants. These \mocc variants are technically \emph{different} CC schemes, and a variant with a larger $weight_{thr}$ would be more aggressive to get more bandwidth. For comparison, Figure~\ref{fig:twoweight}(d) shows the result for a TCP Cubic flow vs. a TCP Vegas flow.

\mocc is friendly in the sense that no \mocc flow will grab all bandwidth when multiple \mocc flows with different weights co-exist. This is because all \mocc flows share one objective framework, which is guaranteed to converge to a stable rate configuration~\cite{onlineconvex}. We performed another simulation to further demonstrate this point with more \mocc variants in Figure~\ref{fig:moccfriend}. We fix the bandwidth to 20Mbps and change the RTT from 10ms--90ms. The results show that the throughput ratio varies between 0.43--2.04, which confirms the friendliness of \mocc under different weights.

Finally, we evaluate the friendliness of \mocc with other TCP schemes.
We use a common setup that has two flows competing one link. Following the convention of friendliness evaluation in prior work~\cite{orca,vivace}, we fix TCP Cubic as the target CC scheme of one flow, and vary the CC scheme of the other flow to compare between them. We use the friendliness
ratio as the metric, which is defined by  $\frac{Delivery-rate-of-CC-scheme}{Delivery-rate-of-Cubic-flow}$.
Figure~\ref{fig:friendrtt} reports the friendliness
ratios of different schemes. The results indicate that
\mocc-Throughput is more aggressive in obtaining bandwidth, and \mocc-Balance and
\mocc-Latency are more friendly to TCP Cubic. In general, \mocc is comparable to other CC schemes in friendliness.

\vspace{-0.1in}
\subsection{\mocc Deep Dive}\label{subsec:DD}
Finally, we deep-dive into \mocc from some other aspects, including hyperparameter setting, CPU overhead, learning algorithm selection, and training speedup.

\begin{figure}[t]
  \centering
  \includegraphics[width=0.4\textwidth]{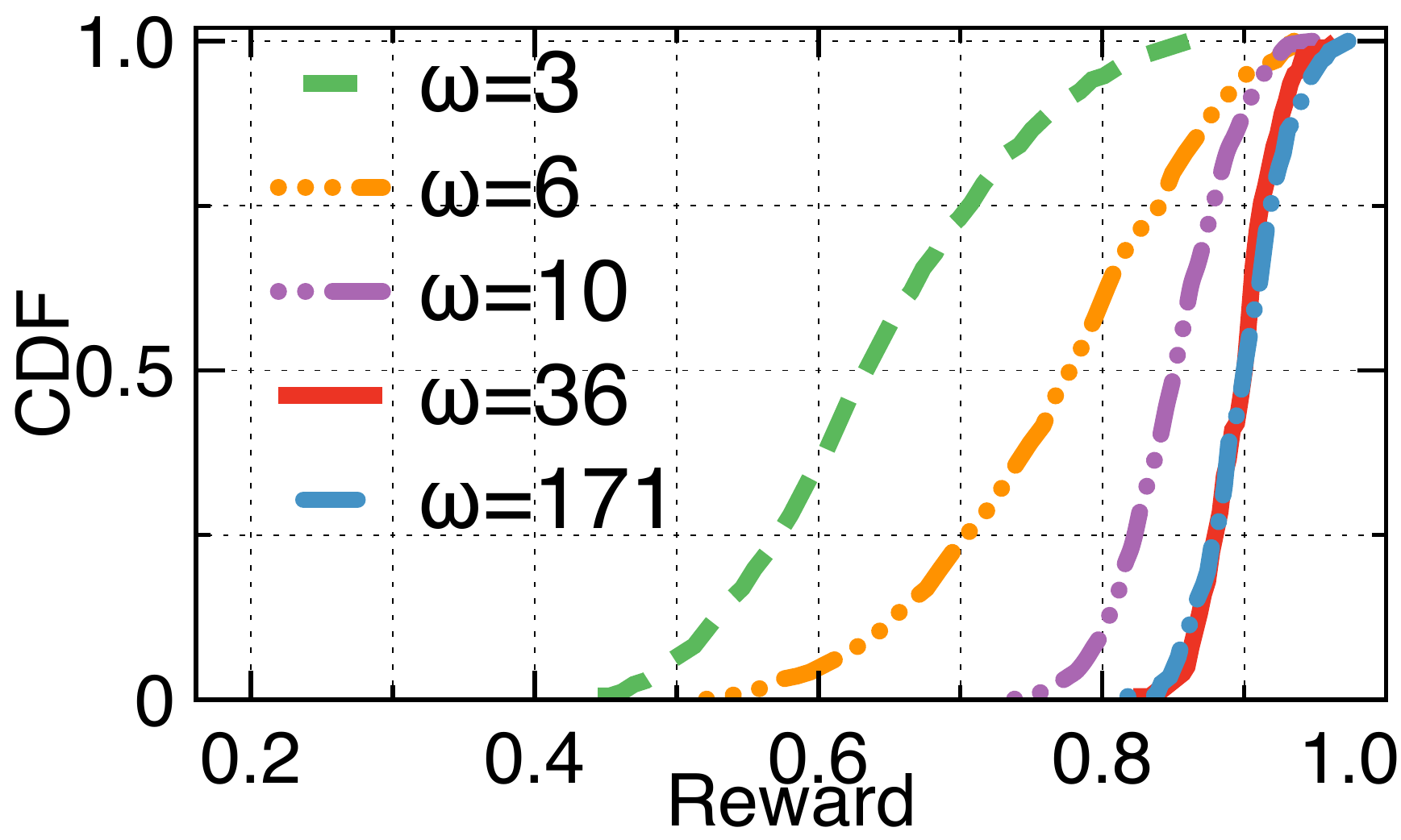}
  \vspace{-0.2in}
  \caption{Hyperparameter setting ($\omega$)}
  %\vspace{-0.2in}
  \label{fig:finegrain}
\end{figure}

\parab{Hyperparameter setting:} We explore several key hyperparameters that may affect the effectiveness of \mocc, i.e., the learning-related parameters in Table \ref{tab:para}. For history length ($\eta$) and discount factor ($\gamma$), we performed an exhaustive search and obtain similar results with~\cite{aurora}. For learning rate $\epsilon$, we followed the default value suggested in stable-baseline's PPO~\cite{ppo}, and we also tried several different values and found that $\epsilon=0.001$ indeed leads to fast convergence. For the remaining, we discuss the number of pre-grained objective weight vectors ($\omega$) that is unique to \sys.

The parameter $\omega$ causes the tradeoff between the quality of base model and the time cost of training. A larger $\omega$  brings better model quality but also increases the training time. To understand the tradeoff, we pre-train \mocc with different $\omega$ to study its performance as well as training time. Figure~\ref{fig:finegrain} (top) shows the CDF of rewards of \mocc with different number of pre-trained objectives. In general, we can see that the model quality improves as $\omega$ increases\footnote{We vary the step size of the objective weight vectors in terms of 1/4, 1/5, 1/6, 1/10, 1/20 leading to $\omega=3, 6, 12, 36, 171$.}, all the way until $\omega=36$. We find that $\omega=36$ has comparable quality as $\omega=171$, both are within 0.82 to 0.96, outperforming $\omega=3, 6, 12$ by $3\times$, $1.5\times$, $1.2\times$ on average. Meanwhile, the training time of $\omega=36$ is 5.2 hours, much shorter than $\omega=171$ (28.2 hours) and reasonably longer time than $\omega=6$ (2.6 hours). As a result, in this paper we set $\omega=36$.

\begin{figure}[t]
  \centering
  \includegraphics[width=0.4\textwidth]{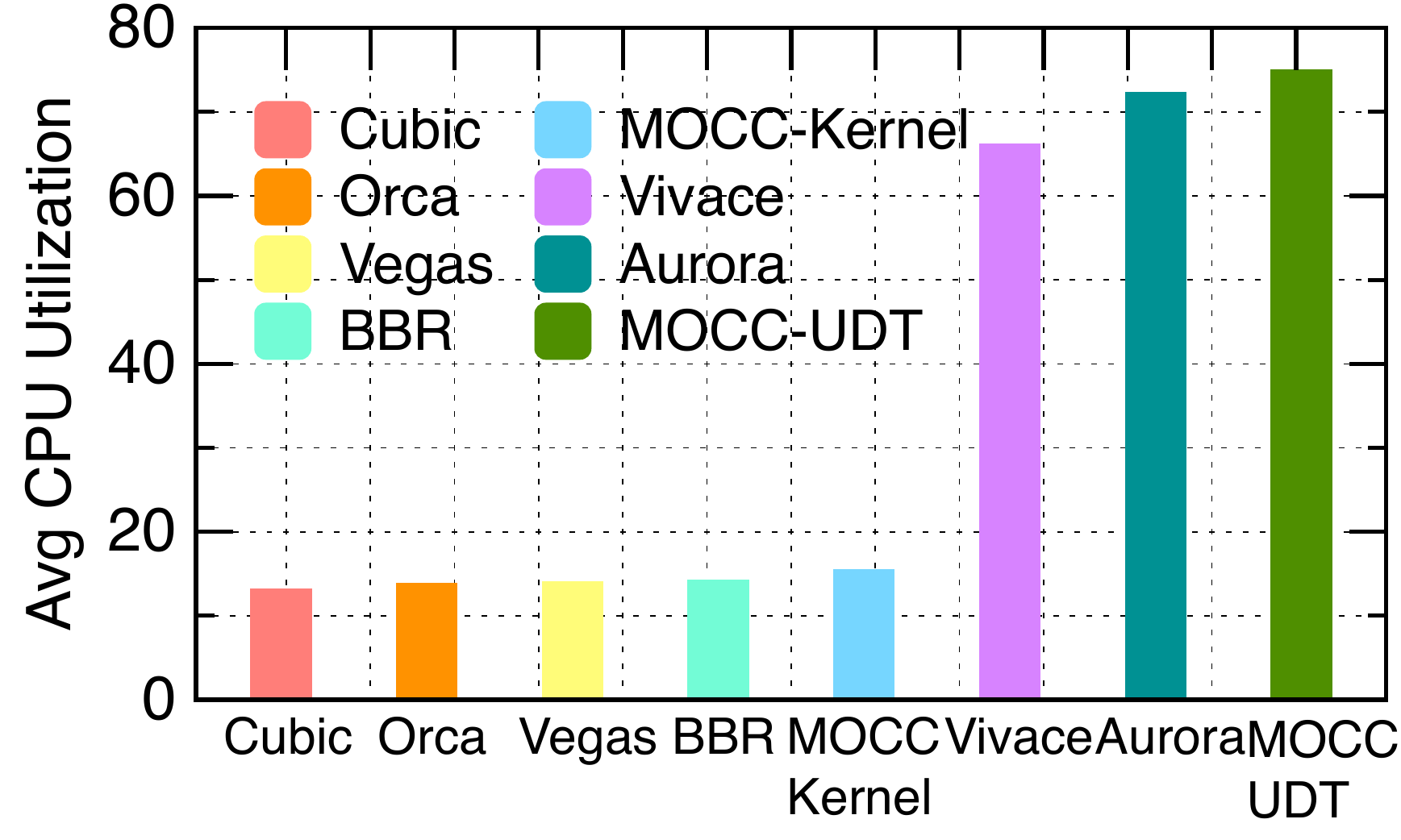}
  %\vspace{-0.2in}
  \caption{CPU Overhead of different CC schemes}
 % \vspace{-0.2in}
  \label{fig:overhead}
\end{figure}

\parab{Overhead:} We evaluate the overhead of \mocc by sending traffic on a
40Mbps link with 20ms RTT and 1 $\times$ BDP buffer. We use \textit{taskset} to
allocate processes to one CPU and report CPU utilization by \textit{htop}. We
exclude the first and the last few seconds for fair comparison. Results in
Figure~\ref{fig:overhead} show that User-space \mocc has high overhead similar
to Aurora, because \mocc agent repeats model inference in each time interval
similar to Aurora. Kernel-space \mocc achieves much lower overhead as
Orca~\cite{orca}, because with CCP, the algorithm logic is isolated from the
datapath. This decoupling provides CC feedback less frequently and significantly
reduces the CPU utilization.

%%%%%%%%%%%
\iffalse
\begin{figure}[t]
\vspace{-0.2in}
\centering
  \begin{minipage}{0.23\textwidth}
    \includegraphics[width=\textwidth]{image/magic.pdf}
    \vspace{-0.3in}
    \caption{Hyperparameter setting ($\omega$)}
    \label{fig:finegrain}
  \end{minipage}
  \hfill
  \begin{minipage}{0.23\textwidth}
    \includegraphics[width=\textwidth]{image/cpu-util.pdf}
    \vspace{-0.3in}
    \caption{CPU Overhead of different CC schemes}
    \label{fig:overhead}
  \end{minipage}
\vspace{-0.3in}
\end{figure}
\fi
%%%%%%%%%%%

\begin{figure}[t]
  \centering
  \includegraphics[width=0.4\textwidth]{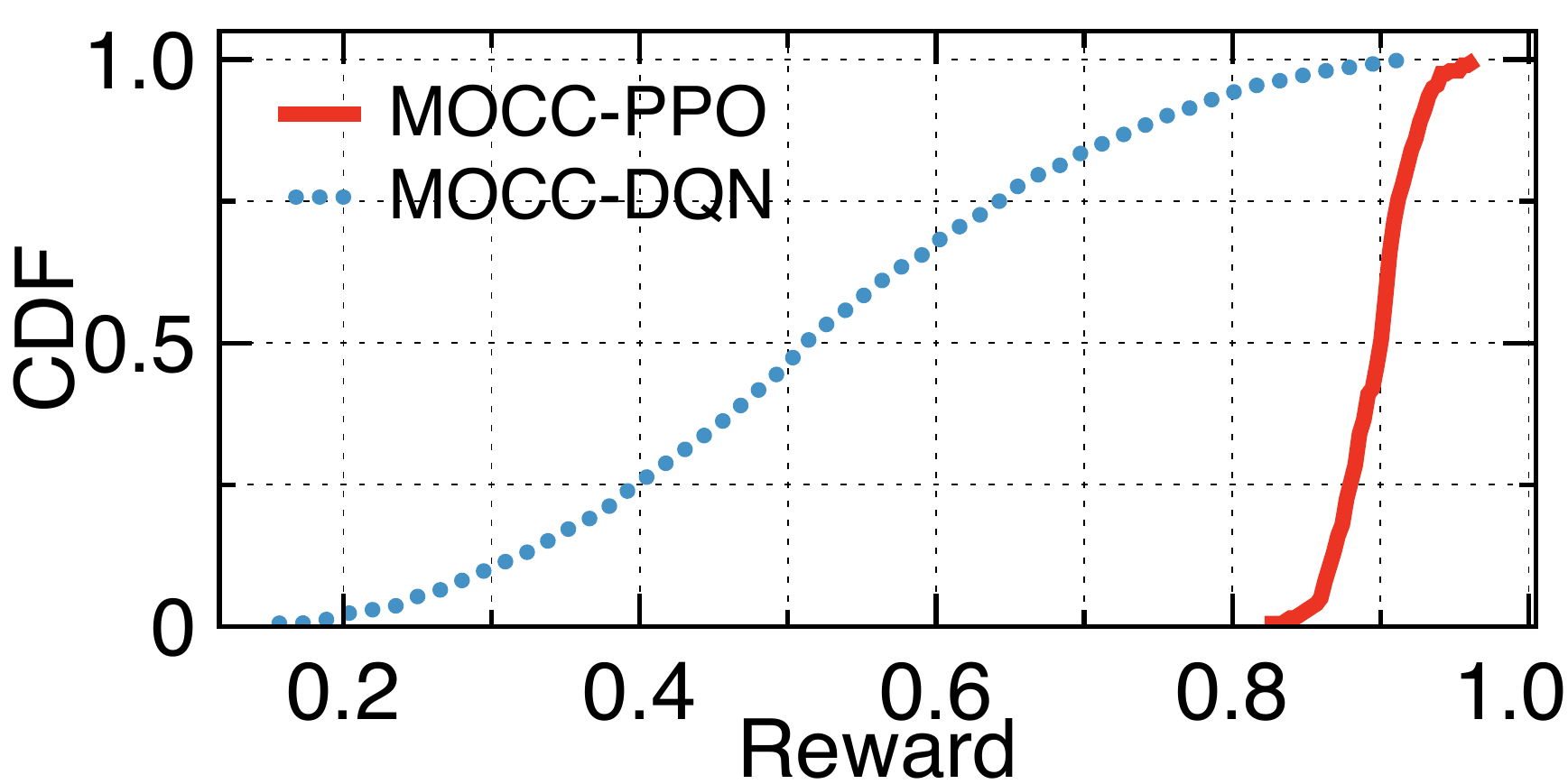}
 % \vspace{-0.2in}
  \caption{Comparison between learning algorithms}
%  \vspace{-0.2in}
  \label{fig:magic2}
\end{figure}

\parab{Learning algorithm selection:} In this paper, we chose the PPO algorithm as our RL algorithm. An alternative approach is Q-learning~\cite{rlintro}.
%Readers can refer to Appendix~\ref{ap:qlearning} for more detailed information.
In this experiment, we compare both algorithms to revisit the design decision of using PPO. For this purpose, we implemented a Q-learning version of \mocc, \mocc-DQN.
%\begin{figure}[t]
%\centering
%  \begin{minipage}{0.22\textwidth}
%    \includegraphics[width=\textwidth]{image/magic2.pdf}
%    \vspace{-0.3in}
%    \caption{Comparison between algorithms}
%    \vspace{-0.1in}
%    \label{fig:magic2}
%  \end{minipage}
%  \hfill
%  \begin{minipage}{0.24\textwidth}
%    \includegraphics[width=\textwidth]{image/transfer.pdf}
%    \vspace{-0.3in}
%    \caption{Training speedup techniques}
%    \vspace{-0.1in}
%    \label{fig:speedup}
%  \end{minipage}
%\vspace{-0.15in}
%\end{figure}
%Here we still use the method in generalized analysis in $\S$~\ref{subsec:AA}.
Figure~\ref{fig:magic2} compares \mocc-PPO with \mocc-DQN. We observe that \mocc-PPO significantly outperforms \mocc-DQN by achieving $3\times$ more rewards on average. The reason is that for CC problems, the sending rate is a continuous value. However, Q-learning scales poorly with the continuous action space, causing sub-optimal performance. On the contrary, PPO is able to output continuous action values. So we select PPO to enable a more fine-grained rate control policy.

\begin{figure}[t]
  \centering
  \includegraphics[width=0.4\textwidth]{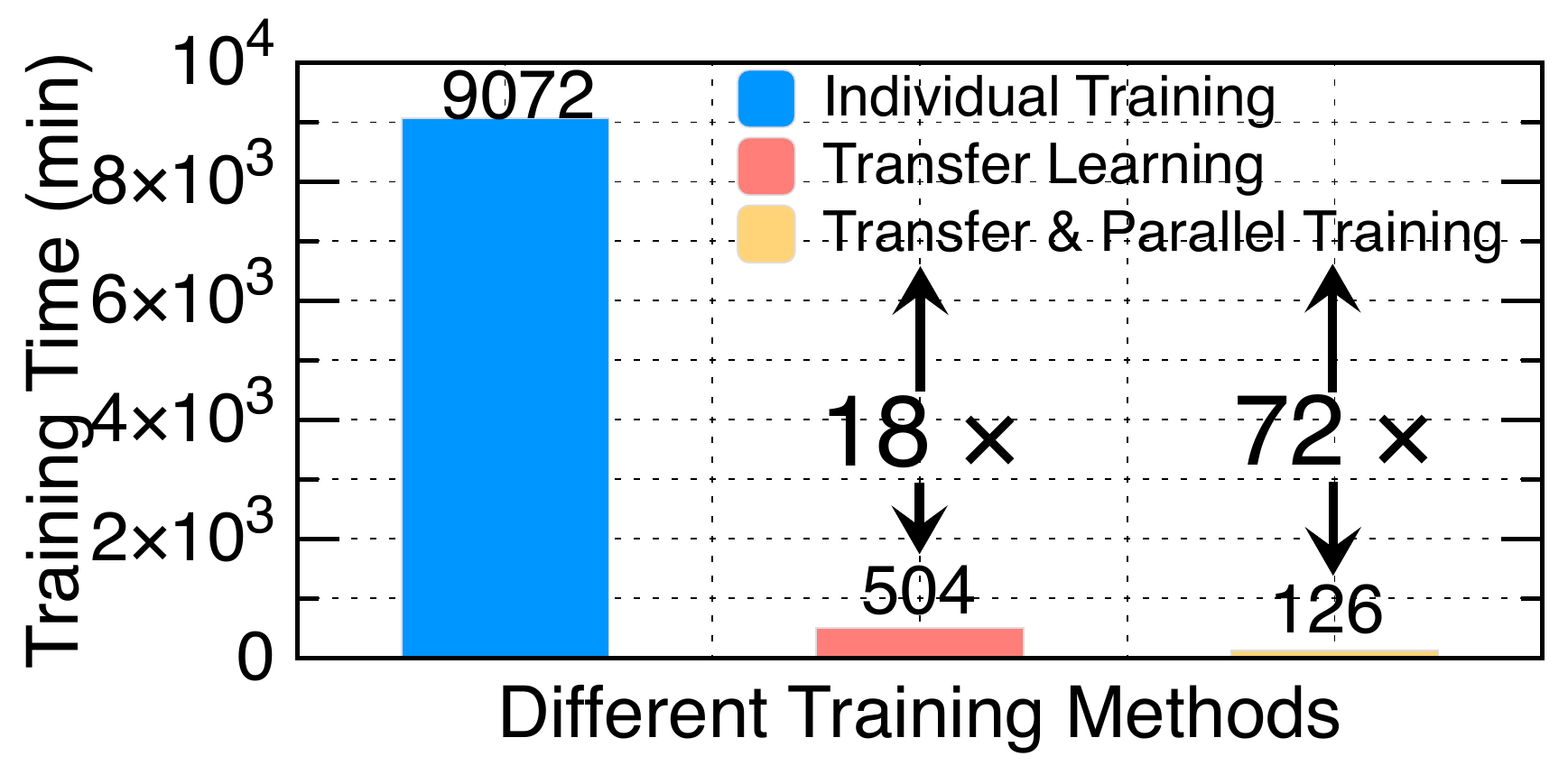}
  %\vspace{-0.2in}
  \caption{Training speedup techniques}
 % \vspace{-0.2in}
  \label{fig:speedup}
\end{figure}

\parab{Training speedup:} We evaluate the effectiveness of our training speedup techniques, including both neighborhood-based transfer learning strategy and parallel training. We train \mocc in three ways. First, we treat each single-objective as a standard RL subproblem and train them separately. Second, we use two-phase training with neighbourhood-transfer method ($\S$\ref{subsec:offline}), without parallel training. Third, based on the second one, we add parallel training. The results are shown in Figure~\ref{fig:speedup}. We observe that through transferring across neighbor objectives, we reduce the training time by $18\times$ (6 days 7.2 hours to 8.4 hours), which validates that transfer learning can significantly accelerate the training. In addition, we find parallel training can further speedup the training by $4$$\times$ (8.4 to 2.1 hours).

\vspace{-2mm}
%\vspace{-0.05in}
\section{Discussion \& Future Directions}\label{sec:discuss}
%\vspace{-0.05in}

\parab{Expressing application requirements:} In \sys, an application expresses
its requirement as a weight vector over several network-level metrics (e.g.,
throughput, latency and packet loss rate), and \sys trains a model to optimize
for the vector. Yet, applications care about application-level objectives, which may not be directly mapped to a weight vector of network-level metrics. At a high level, the weights should be set based on the application-level objectives, e.g., real-time applications should give a higher weight to latency, and bandwidth-intensive applications should give a higher weight to throughput. But how to optimally set the weights to best express an application's requirement still requires human expertise and domain knowledge. We envision a learning-based approach, which learns the mapping from an application-level objective to a weight vector, can be applied to automate this process and reduce human
efforts.
%\vspace{0.15in}

\parab{Model sharing and Federated learning:} For an unseen application,
\sys leverages transfer learning to quickly adapt its model to the new
application. Another device may have already run this application and trained a
model to optimize the performance. If different devices can share their models,
it would further reduce the adaptation time for \sys. However, sharing models
may raise privacy concerns as a trained model may unexpectedly leak a user's
traffic pattern and network condition, which could be further used to reveal the
user's other sensitive information. This setup is similar to federated learning
where a model is trained across multiple decentralized devices. Extending \sys
with privacy-preserving federated learning is an interesting future direction.

%\vspace{0.15in}

\parab{Towards a general multi-objective framework for networking:} While we
focus on congestion control in this paper, we believe the framework behind \sys is more generic and
can be applied to a wide range of networking problems~\cite{wang2018neural,neurocuts,pensieve,chen2018auto}. This framework is particularly relevant to the recent proposals
that leverage reinforcement learning to solve networking problems and
demonstrate superior performance over traditional heuristics. For example, it
can be applied to NeuroCuts~\cite{neurocuts} to learn to build packet classification
trees with multiple objectives on classification time and memory footprint, and
be applied to Pensieve~\cite{pensieve} to learn adaptive bitrate algorithms with
different Quality of Experience (QoE) metrics.
%\vspace{0.15in}

%One obvious advantage of the \mocc model is its modularity. For example, based on this framework, any multi problem can be solved efficiently by integrating any of the rexently proposed novel deep reinforcement learning-based methods. Once the trained model is available , the PF can be directly obtained by a simple feed-forward calculation of the model, in other words, the model has a moderate perform on all weights. Importantly, trained model can adapt to any change of the problem, as long as the problem settings are generated from the same distribution with the training set.\eg. the weight (per-objective preference) are both sampled from [0,1] uniformly.

%Our objective is to provide a modular framework for modelbased RL, leveraging a decomposition of the learning problem to provide reusable components that can be bootstrapped to enable fast re-training following changes in dynamics and rewards.

%\mocc has a high level of modularity and can be easily generalized to other problems:
%\mocc's RL framework is general and it can be applied to a variety of systems and objectives. For most real-world problems ,the preference between each objectives change according to different applications requirements, they are better described through multiple objectives rather than single objectives. In this case, our \mocc framework can achieve better user-defined performance.
\vspace{-2mm}

\section{Conclusion}\label{sec:conclusion}
%\vspace{-0.05in}
%\vspace{-0.2in}
This paper established a multi-objective congestion control (MOCC) framework that enables {\em one single CC algorithm} to effectively support multiple application requirements. To enable multi-objective, \sys constructs its policy network with a preference sub-network that correlates application requirements with optimal rate control policies. Furthermore, it exploits transfer learning to adapt \sys to any new applications quickly in an online manner. Extensive simulations and real Internet experiments have shown that \sys achieves all its design goals. %{\em This work does not raise any ethical issues.} 

%\begin{small}
%\vspace{-0.1in}
%\bibliographystyle{plain}
\bibliographystyle{ACM-Reference-Format}
\bibliography{reference}


\begin{thebibliography}{61}

%%% the Web address.

\ifx \showCODEN    \undefined \def \showCODEN     #1{\unskip}     \fi
\ifx \showDOI      \undefined \def \showDOI       #1{#1}\fi
\ifx \showISBNx    \undefined \def \showISBNx     #1{\unskip}     \fi
\ifx \showISBNxiii \undefined \def \showISBNxiii  #1{\unskip}     \fi
\ifx \showISSN     \undefined \def \showISSN      #1{\unskip}     \fi
\ifx \showLCCN     \undefined \def \showLCCN      #1{\unskip}     \fi
\ifx \shownote     \undefined \def \shownote      #1{#1}          \fi
\ifx \showarticletitle \undefined \def \showarticletitle #1{#1}   \fi
\ifx \showURL      \undefined \def \showURL       {\relax}        \fi
% invisible to TeX
\providecommand\bibfield[2]{#2}
\providecommand\bibinfo[2]{#2}
\providecommand\natexlab[1]{#1}
\providecommand\showeprint[2][]{arXiv:#2}

\bibitem[\protect\citeauthoryear{Abbasloo, Yen, and Chao}{Abbasloo
  et~al\mbox{.}}{2020}]%
        {orca}
\bibfield{author}{\bibinfo{person}{Soheil Abbasloo}, \bibinfo{person}{Chen-Yu
  Yen}, {and} \bibinfo{person}{H~Jonathan Chao}.}
  \bibinfo{year}{2020}\natexlab{}.
\newblock \showarticletitle{Classic meets modern: a pragmatic learning-based
  congestion control for the internet}. In
  \bibinfo{booktitle}{\emph{Proceedings of the Annual conference of the ACM
  Special Interest Group on Data Communication on the applications,
  technologies, architectures, and protocols for computer communication}}.
  \bibinfo{pages}{632--647}.
\newblock


\bibitem[\protect\citeauthoryear{Abels, Roijers, Lenaerts, Now{\'e}, and
  Steckelmacher}{Abels et~al\mbox{.}}{2019}]%
        {abels2019dynamic}
\bibfield{author}{\bibinfo{person}{Axel Abels}, \bibinfo{person}{Diederik
  Roijers}, \bibinfo{person}{Tom Lenaerts}, \bibinfo{person}{Ann Now{\'e}},
  {and} \bibinfo{person}{Denis Steckelmacher}.}
  \bibinfo{year}{2019}\natexlab{}.
\newblock \showarticletitle{Dynamic Weights in Multi-Objective Deep
  Reinforcement Learning}. In \bibinfo{booktitle}{\emph{International
  Conference on Machine Learning}}. \bibinfo{pages}{11--20}.
\newblock


\bibitem[\protect\citeauthoryear{Abels, Roijers, Lenaerts, Now{\'e}, and
  Steckelmacher}{Abels et~al\mbox{.}}{2018}]%
        {dynamic}
\bibfield{author}{\bibinfo{person}{Axel Abels}, \bibinfo{person}{Diederik~M
  Roijers}, \bibinfo{person}{Tom Lenaerts}, \bibinfo{person}{Ann Now{\'e}},
  {and} \bibinfo{person}{Denis Steckelmacher}.}
  \bibinfo{year}{2018}\natexlab{}.
\newblock \showarticletitle{Dynamic Weights in Multi-Objective Deep
  Reinforcement Learning}.
\newblock \bibinfo{journal}{\emph{arXiv preprint arXiv:1809.07803}}
  (\bibinfo{year}{2018}).
\newblock


\bibitem[\protect\citeauthoryear{Arun and Balakrishnan}{Arun and
  Balakrishnan}{2018}]%
        {copa}
\bibfield{author}{\bibinfo{person}{Venkat Arun} {and} \bibinfo{person}{Hari
  Balakrishnan}.} \bibinfo{year}{2018}\natexlab{}.
\newblock \showarticletitle{Copa: Practical delay-based congestion control for
  the internet}. In \bibinfo{booktitle}{\emph{15th $\{$USENIX$\}$ Symposium on
  Networked Systems Design and Implementation ($\{$NSDI$\}$ 18)}}.
  \bibinfo{pages}{329--342}.
\newblock


\bibitem[\protect\citeauthoryear{Brakmo, O'Malley, and Peterson}{Brakmo
  et~al\mbox{.}}{1994}]%
        {vegas}
\bibfield{author}{\bibinfo{person}{Lawrence~S Brakmo}, \bibinfo{person}{Sean~W
  O'Malley}, {and} \bibinfo{person}{Larry~L Peterson}.}
  \bibinfo{year}{1994}\natexlab{}.
\newblock \bibinfo{booktitle}{\emph{TCP Vegas: New techniques for congestion
  detection and avoidance}}.
\newblock Number~4. \bibinfo{publisher}{ACM}.
\newblock


\bibitem[\protect\citeauthoryear{Brockman, Cheung, Pettersson, Schneider,
  Schulman, Tang, and Zaremba}{Brockman et~al\mbox{.}}{2016}]%
        {openai}
\bibfield{author}{\bibinfo{person}{Greg Brockman}, \bibinfo{person}{Vicki
  Cheung}, \bibinfo{person}{Ludwig Pettersson}, \bibinfo{person}{Jonas
  Schneider}, \bibinfo{person}{John Schulman}, \bibinfo{person}{Jie Tang},
  {and} \bibinfo{person}{Wojciech Zaremba}.} \bibinfo{year}{2016}\natexlab{}.
\newblock \showarticletitle{Openai gym}.
\newblock \bibinfo{journal}{\emph{arXiv preprint arXiv:1606.01540}}
  (\bibinfo{year}{2016}).
\newblock


\bibitem[\protect\citeauthoryear{Cardwell, Cheng, Gunn, Yeganeh, and
  Jacobson}{Cardwell et~al\mbox{.}}{2016}]%
        {bbr}
\bibfield{author}{\bibinfo{person}{Neal Cardwell}, \bibinfo{person}{Yuchung
  Cheng}, \bibinfo{person}{C~Stephen Gunn}, \bibinfo{person}{Soheil~Hassas
  Yeganeh}, {and} \bibinfo{person}{Van Jacobson}.}
  \bibinfo{year}{2016}\natexlab{}.
\newblock \showarticletitle{BBR: Congestion-based congestion control}.
\newblock \bibinfo{journal}{\emph{Queue}} \bibinfo{volume}{14},
  \bibinfo{number}{5} (\bibinfo{year}{2016}), \bibinfo{pages}{20--53}.
\newblock


\bibitem[\protect\citeauthoryear{Chen, Hu, Chen, Wu, and Tsang}{Chen
  et~al\mbox{.}}{2013}]%
        {chen2013towards}
\bibfield{author}{\bibinfo{person}{Li Chen}, \bibinfo{person}{Shuihai Hu},
  \bibinfo{person}{Kai Chen}, \bibinfo{person}{Haitao Wu}, {and}
  \bibinfo{person}{Danny~HK Tsang}.} \bibinfo{year}{2013}\natexlab{}.
\newblock \showarticletitle{Towards minimal-delay deadline-driven data center
  TCP}. In \bibinfo{booktitle}{\emph{Proceedings of the Twelfth ACM Workshop on
  Hot Topics in Networks}}. \bibinfo{pages}{1--7}.
\newblock


\bibitem[\protect\citeauthoryear{Chen, Lingys, Chen, and Liu}{Chen
  et~al\mbox{.}}{2018b}]%
        {chen2018auto}
\bibfield{author}{\bibinfo{person}{Li Chen}, \bibinfo{person}{Justinas Lingys},
  \bibinfo{person}{Kai Chen}, {and} \bibinfo{person}{Feng Liu}.}
  \bibinfo{year}{2018}\natexlab{b}.
\newblock \showarticletitle{AuTO: Scaling deep reinforcement learning for
  datacenter-scale automatic traffic optimization}. In
  \bibinfo{booktitle}{\emph{Proceedings of the 2018 Conference of the ACM
  Special Interest Group on Data Communication}}. \bibinfo{pages}{191--205}.
\newblock


\bibitem[\protect\citeauthoryear{Chen, Ghadirzadeh, Bj{\"o}rkman, and
  Jensfelt}{Chen et~al\mbox{.}}{2018a}]%
        {meta}
\bibfield{author}{\bibinfo{person}{Xi Chen}, \bibinfo{person}{Ali Ghadirzadeh},
  \bibinfo{person}{M{\aa}rten Bj{\"o}rkman}, {and} \bibinfo{person}{Patric
  Jensfelt}.} \bibinfo{year}{2018}\natexlab{a}.
\newblock \showarticletitle{Meta-Learning for Multi-objective Reinforcement
  Learning}.
\newblock \bibinfo{journal}{\emph{arXiv preprint arXiv:1811.03376}}
  (\bibinfo{year}{2018}).
\newblock


\bibitem[\protect\citeauthoryear{Chen, Farley, and Ye}{Chen
  et~al\mbox{.}}{2004}]%
        {qos}
\bibfield{author}{\bibinfo{person}{Yan Chen}, \bibinfo{person}{Toni Farley},
  {and} \bibinfo{person}{Nong Ye}.} \bibinfo{year}{2004}\natexlab{}.
\newblock \showarticletitle{QoS requirements of network applications on the
  Internet}.
\newblock \bibinfo{journal}{\emph{Information Knowledge Systems Management}}
  \bibinfo{volume}{4}, \bibinfo{number}{1} (\bibinfo{year}{2004}),
  \bibinfo{pages}{55--76}.
\newblock


\bibitem[\protect\citeauthoryear{Davies}{Davies}{2016}]%
        {qos1}
\bibfield{author}{\bibinfo{person}{Ron Davies}.}
  \bibinfo{year}{2016}\natexlab{}.
\newblock \bibinfo{booktitle}{\emph{5G Network Technology: Putting Europe at
  the Leading Edge}}.
\newblock \bibinfo{publisher}{EPRS, European Parliamentary Research Service,
  Members' Research Service}.
\newblock


\bibitem[\protect\citeauthoryear{Dong, Li, Zarchy, Godfrey, and Schapira}{Dong
  et~al\mbox{.}}{2015}]%
        {allegro}
\bibfield{author}{\bibinfo{person}{Mo Dong}, \bibinfo{person}{Qingxi Li},
  \bibinfo{person}{Doron Zarchy}, \bibinfo{person}{P~Brighten Godfrey}, {and}
  \bibinfo{person}{Michael Schapira}.} \bibinfo{year}{2015}\natexlab{}.
\newblock \showarticletitle{$\{$PCC$\}$: Re-architecting Congestion Control for
  Consistent High Performance}. In \bibinfo{booktitle}{\emph{12th
  $\{$USENIX$\}$ Symposium on Networked Systems Design and Implementation
  ($\{$NSDI$\}$ 15)}}. \bibinfo{pages}{395--408}.
\newblock


\bibitem[\protect\citeauthoryear{Dong, Meng, Zarchy, Arslan, Gilad, Godfrey,
  and Schapira}{Dong et~al\mbox{.}}{2018}]%
        {vivace}
\bibfield{author}{\bibinfo{person}{Mo Dong}, \bibinfo{person}{Tong Meng},
  \bibinfo{person}{Doron Zarchy}, \bibinfo{person}{Engin Arslan},
  \bibinfo{person}{Yossi Gilad}, \bibinfo{person}{Brighten Godfrey}, {and}
  \bibinfo{person}{Michael Schapira}.} \bibinfo{year}{2018}\natexlab{}.
\newblock \showarticletitle{$\{$PCC$\}$ Vivace: Online-Learning Congestion
  Control}. In \bibinfo{booktitle}{\emph{15th $\{$USENIX$\}$ Symposium on
  Networked Systems Design and Implementation ($\{$NSDI$\}$ 18)}}.
  \bibinfo{pages}{343--356}.
\newblock


\bibitem[\protect\citeauthoryear{Durward, Levine, Nemeth, Prettegiani, and
  Tweedie}{Durward et~al\mbox{.}}{1997}]%
        {vr}
\bibfield{author}{\bibinfo{person}{James Durward}, \bibinfo{person}{Jonathan
  Levine}, \bibinfo{person}{Michael Nemeth}, \bibinfo{person}{Jerry
  Prettegiani}, {and} \bibinfo{person}{Ian~T Tweedie}.}
  \bibinfo{year}{1997}\natexlab{}.
\newblock \bibinfo{title}{Virtual reality network with selective distribution
  and updating of data to reduce bandwidth requirements}.
\newblock
\newblock
\newblock
\shownote{US Patent 5,659,691.}


\bibitem[\protect\citeauthoryear{Floyd, Henderson, Gurtov, et~al\mbox{.}}{Floyd
  et~al\mbox{.}}{1999}]%
        {newreno}
\bibfield{author}{\bibinfo{person}{Sally Floyd}, \bibinfo{person}{Tom
  Henderson}, \bibinfo{person}{Andrei Gurtov}, {et~al\mbox{.}}}
  \bibinfo{year}{1999}\natexlab{}.
\newblock \showarticletitle{The NewReno modification to TCP’s fast recovery
  algorithm}.
\newblock  (\bibinfo{year}{1999}).
\newblock


\bibitem[\protect\citeauthoryear{Fluckiger}{Fluckiger}{1995}]%
        {fluckiger1995understanding}
\bibfield{author}{\bibinfo{person}{Fran{\c{c}}ois Fluckiger}.}
  \bibinfo{year}{1995}\natexlab{}.
\newblock \bibinfo{booktitle}{\emph{Understanding networked multimedia:
  applications and technology}}.
\newblock \bibinfo{publisher}{Prentice Hall International (UK) Ltd.}
\newblock


\bibitem[\protect\citeauthoryear{Fouladi, Emmons, Orbay, Wu, Wahby, and
  Winstein}{Fouladi et~al\mbox{.}}{2018}]%
        {salsify}
\bibfield{author}{\bibinfo{person}{Sadjad Fouladi}, \bibinfo{person}{John
  Emmons}, \bibinfo{person}{Emre Orbay}, \bibinfo{person}{Catherine Wu},
  \bibinfo{person}{Riad~S Wahby}, {and} \bibinfo{person}{Keith Winstein}.}
  \bibinfo{year}{2018}\natexlab{}.
\newblock \showarticletitle{Salsify: low-latency network video through tighter
  integration between a video codec and a transport protocol}. In
  \bibinfo{booktitle}{\emph{15th $\{$USENIX$\}$ Symposium on Networked Systems
  Design and Implementation ($\{$NSDI$\}$ 18)}}. \bibinfo{pages}{267--282}.
\newblock


\bibitem[\protect\citeauthoryear{Furht}{Furht}{2011}]%
        {ar}
\bibfield{author}{\bibinfo{person}{Borko Furht}.}
  \bibinfo{year}{2011}\natexlab{}.
\newblock \bibinfo{booktitle}{\emph{Handbook of augmented reality}}.
\newblock \bibinfo{publisher}{Springer Science \& Business Media}.
\newblock


\bibitem[\protect\citeauthoryear{Gao, Wei, and Zhou}{Gao et~al\mbox{.}}{2020}]%
        {qoe1}
\bibfield{author}{\bibinfo{person}{Yun Gao}, \bibinfo{person}{Xin Wei}, {and}
  \bibinfo{person}{Liang Zhou}.} \bibinfo{year}{2020}\natexlab{}.
\newblock \showarticletitle{Personalized QoE improvement for networking video
  service}.
\newblock \bibinfo{journal}{\emph{IEEE Journal on Selected Areas in
  Communications}} \bibinfo{volume}{38}, \bibinfo{number}{10}
  (\bibinfo{year}{2020}), \bibinfo{pages}{2311--2323}.
\newblock


\bibitem[\protect\citeauthoryear{Gu and Grossman}{Gu and Grossman}{2007}]%
        {udt}
\bibfield{author}{\bibinfo{person}{Yunhong Gu} {and} \bibinfo{person}{Robert~L
  Grossman}.} \bibinfo{year}{2007}\natexlab{}.
\newblock \showarticletitle{UDT: UDP-based data transfer for high-speed wide
  area networks}.
\newblock \bibinfo{journal}{\emph{Computer Networks}} \bibinfo{volume}{51},
  \bibinfo{number}{7} (\bibinfo{year}{2007}), \bibinfo{pages}{1777--1799}.
\newblock


\bibitem[\protect\citeauthoryear{Ha, Rhee, and Xu}{Ha et~al\mbox{.}}{2008}]%
        {cubic}
\bibfield{author}{\bibinfo{person}{Sangtae Ha}, \bibinfo{person}{Injong Rhee},
  {and} \bibinfo{person}{Lisong Xu}.} \bibinfo{year}{2008}\natexlab{}.
\newblock \showarticletitle{CUBIC: a new TCP-friendly high-speed TCP variant}.
\newblock \bibinfo{journal}{\emph{ACM SIGOPS operating systems review}}
  \bibinfo{number}{5} (\bibinfo{year}{2008}), \bibinfo{pages}{64--74}.
\newblock


\bibitem[\protect\citeauthoryear{Hausknecht and Stone}{Hausknecht and
  Stone}{2015}]%
        {dqn}
\bibfield{author}{\bibinfo{person}{Matthew Hausknecht} {and}
  \bibinfo{person}{Peter Stone}.} \bibinfo{year}{2015}\natexlab{}.
\newblock \showarticletitle{Deep recurrent q-learning for partially observable
  mdps}. In \bibinfo{booktitle}{\emph{2015 AAAI Fall Symposium Series}}.
\newblock


\bibitem[\protect\citeauthoryear{Hazan}{Hazan}{2019}]%
        {onlineconvex}
\bibfield{author}{\bibinfo{person}{Elad Hazan}.}
  \bibinfo{year}{2019}\natexlab{}.
\newblock \showarticletitle{Introduction to online convex optimization}.
\newblock \bibinfo{journal}{\emph{arXiv preprint arXiv:1909.05207}}
  (\bibinfo{year}{2019}).
\newblock


\bibitem[\protect\citeauthoryear{Huo, Wang, Xu, Li, Ding, and Wang}{Huo
  et~al\mbox{.}}{2019}]%
        {qoe2}
\bibfield{author}{\bibinfo{person}{Liangyu Huo}, \bibinfo{person}{Zulin Wang},
  \bibinfo{person}{Mai Xu}, \bibinfo{person}{Yong Li}, \bibinfo{person}{Zhiguo
  Ding}, {and} \bibinfo{person}{Hao Wang}.} \bibinfo{year}{2019}\natexlab{}.
\newblock \showarticletitle{A Meta-Learning Framework for Learning Multi-User
  Preferences in QoE Optimization of DASH}.
\newblock \bibinfo{journal}{\emph{IEEE Transactions on Circuits and Systems for
  Video Technology}} \bibinfo{volume}{30}, \bibinfo{number}{9}
  (\bibinfo{year}{2019}), \bibinfo{pages}{3210--3225}.
\newblock


\bibitem[\protect\citeauthoryear{Jain, Durresi, and Babic}{Jain
  et~al\mbox{.}}{1999}]%
        {jain}
\bibfield{author}{\bibinfo{person}{Raj Jain}, \bibinfo{person}{Arjan Durresi},
  {and} \bibinfo{person}{Gojko Babic}.} \bibinfo{year}{1999}\natexlab{}.
\newblock \showarticletitle{Throughput fairness index: An explanation}. In
  \bibinfo{booktitle}{\emph{ATM Forum contribution}},
  Vol.~\bibinfo{volume}{99}.
\newblock


\bibitem[\protect\citeauthoryear{Jay, Rotman, Godfrey, Schapira, and Tamar}{Jay
  et~al\mbox{.}}{2019}]%
        {aurora}
\bibfield{author}{\bibinfo{person}{Nathan Jay}, \bibinfo{person}{Noga Rotman},
  \bibinfo{person}{Brighten Godfrey}, \bibinfo{person}{Michael Schapira}, {and}
  \bibinfo{person}{Aviv Tamar}.} \bibinfo{year}{2019}\natexlab{}.
\newblock \showarticletitle{A Deep Reinforcement Learning Perspective on
  Internet Congestion Control}. In \bibinfo{booktitle}{\emph{International
  Conference on Machine Learning ICML}}. \bibinfo{pages}{3050--3059}.
\newblock


\bibitem[\protect\citeauthoryear{Jin, Wei, and Low}{Jin et~al\mbox{.}}{2004}]%
        {fast}
\bibfield{author}{\bibinfo{person}{Cheng Jin}, \bibinfo{person}{David~X Wei},
  {and} \bibinfo{person}{Steven~H Low}.} \bibinfo{year}{2004}\natexlab{}.
\newblock \showarticletitle{FAST TCP: motivation, architecture, algorithms,
  performance}. In \bibinfo{booktitle}{\emph{IEEE INFOCOM 2004}},
  Vol.~\bibinfo{volume}{4}. IEEE, \bibinfo{pages}{2490--2501}.
\newblock


\bibitem[\protect\citeauthoryear{Karlsson}{Karlsson}{1996}]%
        {mmnetwork}
\bibfield{author}{\bibinfo{person}{Gunnar Karlsson}.}
  \bibinfo{year}{1996}\natexlab{}.
\newblock \showarticletitle{Quality requirements for multimedia network
  services}. In \bibinfo{booktitle}{\emph{Proceedings of Radiovetenskap ach
  kommunikation}}. \bibinfo{pages}{96--100}.
\newblock


\bibitem[\protect\citeauthoryear{Kenyon and Nightingale}{Kenyon and
  Nightingale}{1992}]%
        {kenyon1992audiovisual}
\bibfield{author}{\bibinfo{person}{Nicholas~D Kenyon} {and} \bibinfo{person}{C
  Nightingale}.} \bibinfo{year}{1992}\natexlab{}.
\newblock \bibinfo{booktitle}{\emph{Audiovisual telecommunications}}.
\newblock \bibinfo{publisher}{Chapman \& Hall, Ltd.}
\newblock


\bibitem[\protect\citeauthoryear{Kingma and Ba}{Kingma and Ba}{2014}]%
        {adam}
\bibfield{author}{\bibinfo{person}{Diederik~P Kingma} {and}
  \bibinfo{person}{Jimmy Ba}.} \bibinfo{year}{2014}\natexlab{}.
\newblock \showarticletitle{Adam: A method for stochastic optimization}.
\newblock \bibinfo{journal}{\emph{arXiv preprint arXiv:1412.6980}}
  (\bibinfo{year}{2014}).
\newblock


\bibitem[\protect\citeauthoryear{Langley, Riddoch, Wilk, Vicente, Krasic,
  Zhang, Yang, Kouranov, Swett, Iyengar, Bailey, Dorfman, Roskind, Kulik,
  Westin, Tenneti, Shade, Hamilton, Vasiliev, Chang, and Shi}{Langley
  et~al\mbox{.}}{2017}]%
        {quic}
\bibfield{author}{\bibinfo{person}{Adam Langley}, \bibinfo{person}{Alistair
  Riddoch}, \bibinfo{person}{Alyssa Wilk}, \bibinfo{person}{Antonio Vicente},
  \bibinfo{person}{Charles Krasic}, \bibinfo{person}{Dan Zhang},
  \bibinfo{person}{Fan Yang}, \bibinfo{person}{Fedor Kouranov},
  \bibinfo{person}{Ian Swett}, \bibinfo{person}{Janardhan Iyengar},
  \bibinfo{person}{Jeff Bailey}, \bibinfo{person}{Jeremy Dorfman},
  \bibinfo{person}{Jim Roskind}, \bibinfo{person}{Joanna Kulik},
  \bibinfo{person}{Patrik Westin}, \bibinfo{person}{Raman Tenneti},
  \bibinfo{person}{Robbie Shade}, \bibinfo{person}{Ryan Hamilton},
  \bibinfo{person}{Victor Vasiliev}, \bibinfo{person}{Wan-Teh Chang}, {and}
  \bibinfo{person}{Zhongyi Shi}.} \bibinfo{year}{2017}\natexlab{}.
\newblock \showarticletitle{The {QUIC} Transport Protocol: Design and
  {Internet}-Scale Deployment}. In \bibinfo{booktitle}{\emph{ACM SIGCOMM}}.
\newblock


\bibitem[\protect\citeauthoryear{Li, Zhang, and Wang}{Li et~al\mbox{.}}{2019}]%
        {neighbour}
\bibfield{author}{\bibinfo{person}{Kaiwen Li}, \bibinfo{person}{Tao Zhang},
  {and} \bibinfo{person}{Rui Wang}.} \bibinfo{year}{2019}\natexlab{}.
\newblock \showarticletitle{Deep Reinforcement Learning for Multi-objective
  Optimization}.
\newblock \bibinfo{journal}{\emph{arXiv preprint arXiv:1906.02386}}
  (\bibinfo{year}{2019}).
\newblock


\bibitem[\protect\citeauthoryear{Liang, Liaw, Nishihara, Moritz, Fox, Gonzalez,
  Goldberg, and Stoica}{Liang et~al\mbox{.}}{2017}]%
        {rllib}
\bibfield{author}{\bibinfo{person}{Eric Liang}, \bibinfo{person}{Richard Liaw},
  \bibinfo{person}{Robert Nishihara}, \bibinfo{person}{Philipp Moritz},
  \bibinfo{person}{Roy Fox}, \bibinfo{person}{Joseph Gonzalez},
  \bibinfo{person}{Ken Goldberg}, {and} \bibinfo{person}{Ion Stoica}.}
  \bibinfo{year}{2017}\natexlab{}.
\newblock \showarticletitle{Ray rllib: A composable and scalable reinforcement
  learning library}.
\newblock \bibinfo{journal}{\emph{arXiv preprint arXiv:1712.09381}}
  (\bibinfo{year}{2017}).
\newblock


\bibitem[\protect\citeauthoryear{Liang, Zhu, Jin, and Stoica}{Liang
  et~al\mbox{.}}{2019}]%
        {neurocuts}
\bibfield{author}{\bibinfo{person}{Eric Liang}, \bibinfo{person}{Hang Zhu},
  \bibinfo{person}{Xin Jin}, {and} \bibinfo{person}{Ion Stoica}.}
  \bibinfo{year}{2019}\natexlab{}.
\newblock \showarticletitle{Neural packet classification}.
\newblock In \bibinfo{booktitle}{\emph{Proceedings of the ACM Special Interest
  Group on Data Communication}}. \bibinfo{pages}{256--269}.
\newblock


\bibitem[\protect\citeauthoryear{Liu, Xu, and Hu}{Liu et~al\mbox{.}}{2014a}]%
        {liu2014multiobjective}
\bibfield{author}{\bibinfo{person}{Chunming Liu}, \bibinfo{person}{Xin Xu},
  {and} \bibinfo{person}{Dewen Hu}.} \bibinfo{year}{2014}\natexlab{a}.
\newblock \showarticletitle{Multiobjective reinforcement learning: A
  comprehensive overview}.
\newblock \bibinfo{journal}{\emph{IEEE Transactions on Systems, Man, and
  Cybernetics: Systems}} \bibinfo{volume}{45}, \bibinfo{number}{3}
  (\bibinfo{year}{2014}), \bibinfo{pages}{385--398}.
\newblock


\bibitem[\protect\citeauthoryear{Liu, Xu, and Hu}{Liu et~al\mbox{.}}{2014b}]%
        {morl}
\bibfield{author}{\bibinfo{person}{Chunming Liu}, \bibinfo{person}{Xin Xu},
  {and} \bibinfo{person}{Dewen Hu}.} \bibinfo{year}{2014}\natexlab{b}.
\newblock \showarticletitle{Multiobjective reinforcement learning: A
  comprehensive overview}.
\newblock \bibinfo{journal}{\emph{IEEE Transactions on Systems, Man, and
  Cybernetics: Systems}} \bibinfo{volume}{45}, \bibinfo{number}{3}
  (\bibinfo{year}{2014}), \bibinfo{pages}{385--398}.
\newblock


\bibitem[\protect\citeauthoryear{Mangiante, Klas, Navon, GuanHua, Ran, and
  Silva}{Mangiante et~al\mbox{.}}{2017}]%
        {2017vr}
\bibfield{author}{\bibinfo{person}{Simone Mangiante}, \bibinfo{person}{Guenter
  Klas}, \bibinfo{person}{Amit Navon}, \bibinfo{person}{Zhuang GuanHua},
  \bibinfo{person}{Ju Ran}, {and} \bibinfo{person}{Marco~Dias Silva}.}
  \bibinfo{year}{2017}\natexlab{}.
\newblock \showarticletitle{Vr is on the edge: How to deliver 360 videos in
  mobile networks}. In \bibinfo{booktitle}{\emph{Proceedings of the Workshop on
  Virtual Reality and Augmented Reality Network}}. \bibinfo{pages}{30--35}.
\newblock


\bibitem[\protect\citeauthoryear{Mnih, Badia, Mirza, Graves, Lillicrap, Harley,
  Silver, and Kavukcuoglu}{Mnih et~al\mbox{.}}{2016}]%
        {a3c}
\bibfield{author}{\bibinfo{person}{Volodymyr Mnih},
  \bibinfo{person}{Adria~Puigdomenech Badia}, \bibinfo{person}{Mehdi Mirza},
  \bibinfo{person}{Alex Graves}, \bibinfo{person}{Timothy Lillicrap},
  \bibinfo{person}{Tim Harley}, \bibinfo{person}{David Silver}, {and}
  \bibinfo{person}{Koray Kavukcuoglu}.} \bibinfo{year}{2016}\natexlab{}.
\newblock \showarticletitle{Asynchronous methods for deep reinforcement
  learning}. In \bibinfo{booktitle}{\emph{International conference on machine
  learning}}. \bibinfo{pages}{1928--1937}.
\newblock


\bibitem[\protect\citeauthoryear{Moritz, Nishihara, Wang, Tumanov, Liaw, Liang,
  Elibol, Yang, Paul, Jordan, et~al\mbox{.}}{Moritz et~al\mbox{.}}{2018}]%
        {ray}
\bibfield{author}{\bibinfo{person}{Philipp Moritz}, \bibinfo{person}{Robert
  Nishihara}, \bibinfo{person}{Stephanie Wang}, \bibinfo{person}{Alexey
  Tumanov}, \bibinfo{person}{Richard Liaw}, \bibinfo{person}{Eric Liang},
  \bibinfo{person}{Melih Elibol}, \bibinfo{person}{Zongheng Yang},
  \bibinfo{person}{William Paul}, \bibinfo{person}{Michael~I Jordan},
  {et~al\mbox{.}}} \bibinfo{year}{2018}\natexlab{}.
\newblock \showarticletitle{Ray: A distributed framework for emerging
  $\{$AI$\}$ applications}. In \bibinfo{booktitle}{\emph{13th $\{$USENIX$\}$
  Symposium on Operating Systems Design and Implementation ($\{$OSDI$\}$ 18)}}.
  \bibinfo{pages}{561--577}.
\newblock


\bibitem[\protect\citeauthoryear{Mossalam, Assael, Roijers, and
  Whiteson}{Mossalam et~al\mbox{.}}{2016}]%
        {mossalam2016multi}
\bibfield{author}{\bibinfo{person}{Hossam Mossalam}, \bibinfo{person}{Yannis~M
  Assael}, \bibinfo{person}{Diederik~M Roijers}, {and} \bibinfo{person}{Shimon
  Whiteson}.} \bibinfo{year}{2016}\natexlab{}.
\newblock \showarticletitle{Multi-objective deep reinforcement learning}.
\newblock \bibinfo{journal}{\emph{arXiv preprint arXiv:1610.02707}}
  (\bibinfo{year}{2016}).
\newblock


\bibitem[\protect\citeauthoryear{Narayan, Cangialosi, Raghavan, Goyal,
  Narayana, Mittal, Alizadeh, and Balakrishnan}{Narayan et~al\mbox{.}}{2018}]%
        {ccp}
\bibfield{author}{\bibinfo{person}{Akshay Narayan}, \bibinfo{person}{Frank
  Cangialosi}, \bibinfo{person}{Deepti Raghavan}, \bibinfo{person}{Prateesh
  Goyal}, \bibinfo{person}{Srinivas Narayana}, \bibinfo{person}{Radhika
  Mittal}, \bibinfo{person}{Mohammad Alizadeh}, {and} \bibinfo{person}{Hari
  Balakrishnan}.} \bibinfo{year}{2018}\natexlab{}.
\newblock \showarticletitle{Restructuring endpoint congestion control}. In
  \bibinfo{booktitle}{\emph{Proceedings of the 2018 Conference of the ACM
  Special Interest Group on Data Communication}}. ACM, \bibinfo{pages}{30--43}.
\newblock


\bibitem[\protect\citeauthoryear{Natarajan and Tadepalli}{Natarajan and
  Tadepalli}{2005}]%
        {natarajan2005dynamic}
\bibfield{author}{\bibinfo{person}{Sriraam Natarajan} {and}
  \bibinfo{person}{Prasad Tadepalli}.} \bibinfo{year}{2005}\natexlab{}.
\newblock \showarticletitle{Dynamic preferences in multi-criteria reinforcement
  learning}. In \bibinfo{booktitle}{\emph{Proceedings of the 22nd international
  conference on Machine learning}}. \bibinfo{pages}{601--608}.
\newblock


\bibitem[\protect\citeauthoryear{Nussbaumer, Patel, Schaffa, and
  Sterbenz}{Nussbaumer et~al\mbox{.}}{1995}]%
        {vsnetwork}
\bibfield{author}{\bibinfo{person}{J-P Nussbaumer}, \bibinfo{person}{Baiju~V.
  Patel}, \bibinfo{person}{Frank Schaffa}, {and} \bibinfo{person}{James P.~G.
  Sterbenz}.} \bibinfo{year}{1995}\natexlab{}.
\newblock \showarticletitle{Networking requirements for interactive video on
  demand}.
\newblock \bibinfo{journal}{\emph{IEEE Journal on Selected Areas in
  Communications}} \bibinfo{volume}{13}, \bibinfo{number}{5}
  (\bibinfo{year}{1995}), \bibinfo{pages}{779--787}.
\newblock


\bibitem[\protect\citeauthoryear{Peesapati, Schwanda, Schultz, Lepage, Jeong,
  and Cosley}{Peesapati et~al\mbox{.}}{2010}]%
        {pensieve}
\bibfield{author}{\bibinfo{person}{S~Tejaswi Peesapati},
  \bibinfo{person}{Victoria Schwanda}, \bibinfo{person}{Johnathon Schultz},
  \bibinfo{person}{Matt Lepage}, \bibinfo{person}{So-yae Jeong}, {and}
  \bibinfo{person}{Dan Cosley}.} \bibinfo{year}{2010}\natexlab{}.
\newblock \showarticletitle{Pensieve: supporting everyday reminiscence}. In
  \bibinfo{booktitle}{\emph{Proceedings of the SIGCHI Conference on Human
  Factors in Computing Systems}}. ACM, \bibinfo{pages}{2027--2036}.
\newblock


\bibitem[\protect\citeauthoryear{Schaul, Horgan, Gregor, and Silver}{Schaul
  et~al\mbox{.}}{2015}]%
        {schaul2015universal}
\bibfield{author}{\bibinfo{person}{Tom Schaul}, \bibinfo{person}{Daniel
  Horgan}, \bibinfo{person}{Karol Gregor}, {and} \bibinfo{person}{David
  Silver}.} \bibinfo{year}{2015}\natexlab{}.
\newblock \showarticletitle{Universal value function approximators}. In
  \bibinfo{booktitle}{\emph{International conference on machine learning}}.
  \bibinfo{pages}{1312--1320}.
\newblock


\bibitem[\protect\citeauthoryear{Schulman, Levine, Abbeel, Jordan, and
  Moritz}{Schulman et~al\mbox{.}}{2015}]%
        {schulman2015trust}
\bibfield{author}{\bibinfo{person}{John Schulman}, \bibinfo{person}{Sergey
  Levine}, \bibinfo{person}{Pieter Abbeel}, \bibinfo{person}{Michael Jordan},
  {and} \bibinfo{person}{Philipp Moritz}.} \bibinfo{year}{2015}\natexlab{}.
\newblock \showarticletitle{Trust region policy optimization}. In
  \bibinfo{booktitle}{\emph{International conference on machine learning}}.
  \bibinfo{pages}{1889--1897}.
\newblock


\bibitem[\protect\citeauthoryear{Schulman, Wolski, Dhariwal, Radford, and
  Klimov}{Schulman et~al\mbox{.}}{2017}]%
        {ppo}
\bibfield{author}{\bibinfo{person}{John Schulman}, \bibinfo{person}{Filip
  Wolski}, \bibinfo{person}{Prafulla Dhariwal}, \bibinfo{person}{Alec Radford},
  {and} \bibinfo{person}{Oleg Klimov}.} \bibinfo{year}{2017}\natexlab{}.
\newblock \showarticletitle{Proximal policy optimization algorithms}.
\newblock \bibinfo{journal}{\emph{arXiv preprint arXiv:1707.06347}}
  (\bibinfo{year}{2017}).
\newblock


\bibitem[\protect\citeauthoryear{Silveira, Margi, Gonzalez, Favero,
  Vilcachagua, Bressan, and Ruggiero}{Silveira et~al\mbox{.}}{1999}]%
        {silveira1999multimedia}
\bibfield{author}{\bibinfo{person}{Regina~Melo Silveira},
  \bibinfo{person}{C{\'\i}ntia~Borges Margi}, \bibinfo{person}{LG Gonzalez},
  \bibinfo{person}{E Favero}, \bibinfo{person}{OD Vilcachagua},
  \bibinfo{person}{Gra{\c{c}}a Bressan}, {and} \bibinfo{person}{Wilson~Vicente
  Ruggiero}.} \bibinfo{year}{1999}\natexlab{}.
\newblock \showarticletitle{A Multimedia on Demand System for Distance
  Education}. In \bibinfo{booktitle}{\emph{International Conference on
  Technology and Distance Education, Fort Lauderdale-Florida}}.
\newblock


\bibitem[\protect\citeauthoryear{Sivaraman, Winstein, Thaker, and
  Balakrishnan}{Sivaraman et~al\mbox{.}}{2014}]%
        {remy}
\bibfield{author}{\bibinfo{person}{Anirudh Sivaraman}, \bibinfo{person}{Keith
  Winstein}, \bibinfo{person}{Pratiksha Thaker}, {and} \bibinfo{person}{Hari
  Balakrishnan}.} \bibinfo{year}{2014}\natexlab{}.
\newblock \showarticletitle{An experimental study of the learnability of
  congestion control}. In \bibinfo{booktitle}{\emph{ACM SIGCOMM Computer
  Communication Review}}.
\newblock


\bibitem[\protect\citeauthoryear{Sutton, Barto, et~al\mbox{.}}{Sutton
  et~al\mbox{.}}{1998a}]%
        {sutton1998introduction}
\bibfield{author}{\bibinfo{person}{Richard~S Sutton}, \bibinfo{person}{Andrew~G
  Barto}, {et~al\mbox{.}}} \bibinfo{year}{1998}\natexlab{a}.
\newblock \bibinfo{booktitle}{\emph{Introduction to reinforcement learning}}.
  Vol.~\bibinfo{volume}{135}.
\newblock \bibinfo{publisher}{MIT press Cambridge}.
\newblock


\bibitem[\protect\citeauthoryear{Sutton, Barto, et~al\mbox{.}}{Sutton
  et~al\mbox{.}}{1998b}]%
        {rlintro}
\bibfield{author}{\bibinfo{person}{Richard~S Sutton}, \bibinfo{person}{Andrew~G
  Barto}, {et~al\mbox{.}}} \bibinfo{year}{1998}\natexlab{b}.
\newblock \bibinfo{booktitle}{\emph{Introduction to reinforcement learning}}.
  Vol.~\bibinfo{volume}{2}.
\newblock \bibinfo{publisher}{MIT press Cambridge}.
\newblock


\bibitem[\protect\citeauthoryear{Szuprowicz}{Szuprowicz}{1995}]%
        {szuprowicz1995multimedia}
\bibfield{author}{\bibinfo{person}{Bohdan~O Szuprowicz}.}
  \bibinfo{year}{1995}\natexlab{}.
\newblock \bibinfo{booktitle}{\emph{Multimedia networking}}.
\newblock \bibinfo{publisher}{McGraw-Hill, Inc.}
\newblock


\bibitem[\protect\citeauthoryear{Tan, Song, Zhang, and Sridharan}{Tan
  et~al\mbox{.}}{2006}]%
        {compound}
\bibfield{author}{\bibinfo{person}{Kun Tan}, \bibinfo{person}{Jingmin Song},
  \bibinfo{person}{Qian Zhang}, {and} \bibinfo{person}{Murari Sridharan}.}
  \bibinfo{year}{2006}\natexlab{}.
\newblock \showarticletitle{A compound TCP approach for high-speed and long
  distance networks}. In \bibinfo{booktitle}{\emph{Proceedings IEEE INFOCOM
  2006. 25TH IEEE International Conference on Computer Communications}}. IEEE,
  \bibinfo{pages}{1--12}.
\newblock


\bibitem[\protect\citeauthoryear{Taylor and Stone}{Taylor and Stone}{2009}]%
        {transferrl}
\bibfield{author}{\bibinfo{person}{Matthew~E Taylor} {and}
  \bibinfo{person}{Peter Stone}.} \bibinfo{year}{2009}\natexlab{}.
\newblock \showarticletitle{Transfer learning for reinforcement learning
  domains: A survey}.
\newblock \bibinfo{journal}{\emph{Journal of Machine Learning Research}}
  \bibinfo{volume}{10}, \bibinfo{number}{Jul} (\bibinfo{year}{2009}),
  \bibinfo{pages}{1633--1685}.
\newblock


\bibitem[\protect\citeauthoryear{Wang, Cui, Xiao, Wang, Yang, Chen, and
  Zhu}{Wang et~al\mbox{.}}{2018}]%
        {wang2018neural}
\bibfield{author}{\bibinfo{person}{Mowei Wang}, \bibinfo{person}{Yong Cui},
  \bibinfo{person}{Shihan Xiao}, \bibinfo{person}{Xin Wang},
  \bibinfo{person}{Dan Yang}, \bibinfo{person}{Kai Chen}, {and}
  \bibinfo{person}{Jun Zhu}.} \bibinfo{year}{2018}\natexlab{}.
\newblock \showarticletitle{Neural network meets DCN: Traffic-driven topology
  adaptation with deep learning}.
\newblock \bibinfo{journal}{\emph{Proceedings of the ACM on Measurement and
  Analysis of Computing Systems}} \bibinfo{volume}{2}, \bibinfo{number}{2}
  (\bibinfo{year}{2018}), \bibinfo{pages}{1--25}.
\newblock


\bibitem[\protect\citeauthoryear{White}{White}{2001}]%
        {mdp}
\bibfield{author}{\bibinfo{person}{CC White}.} \bibinfo{year}{2001}\natexlab{}.
\newblock \bibinfo{booktitle}{\emph{Markov decision processes}}.
\newblock \bibinfo{publisher}{Springer}.
\newblock


\bibitem[\protect\citeauthoryear{Winstein, Sivaraman, and
  Balakrishnan}{Winstein et~al\mbox{.}}{2013}]%
        {selfinflict}
\bibfield{author}{\bibinfo{person}{Keith Winstein}, \bibinfo{person}{Anirudh
  Sivaraman}, {and} \bibinfo{person}{Hari Balakrishnan}.}
  \bibinfo{year}{2013}\natexlab{}.
\newblock \showarticletitle{Stochastic forecasts achieve high throughput and
  low delay over cellular networks}. In \bibinfo{booktitle}{\emph{Presented as
  part of the 10th $\{$USENIX$\}$ Symposium on Networked Systems Design and
  Implementation ($\{$NSDI$\}$ 13)}}. \bibinfo{pages}{459--471}.
\newblock


\bibitem[\protect\citeauthoryear{Yan, Ma, Hill, Raghavan, Wahby, Levis, and
  Winstein}{Yan et~al\mbox{.}}{2018}]%
        {pantheon}
\bibfield{author}{\bibinfo{person}{Francis~Y Yan}, \bibinfo{person}{Jestin Ma},
  \bibinfo{person}{Greg~D Hill}, \bibinfo{person}{Deepti Raghavan},
  \bibinfo{person}{Riad~S Wahby}, \bibinfo{person}{Philip Levis}, {and}
  \bibinfo{person}{Keith Winstein}.} \bibinfo{year}{2018}\natexlab{}.
\newblock \showarticletitle{Pantheon: the training ground for Internet
  congestion-control research}. In \bibinfo{booktitle}{\emph{2018
  $\{$USENIX$\}$ Annual Technical Conference ($\{$USENIX$\}$$\{$ATC$\}$ 18)}}.
\newblock


\bibitem[\protect\citeauthoryear{Yang, Sun, and Narasimhan}{Yang
  et~al\mbox{.}}{2019}]%
        {yang2019generalized}
\bibfield{author}{\bibinfo{person}{Runzhe Yang}, \bibinfo{person}{Xingyuan
  Sun}, {and} \bibinfo{person}{Karthik Narasimhan}.}
  \bibinfo{year}{2019}\natexlab{}.
\newblock \showarticletitle{A Generalized Algorithm for Multi-Objective
  Reinforcement Learning and Policy Adaptation}. In
  \bibinfo{booktitle}{\emph{Advances in Neural Information Processing
  Systems}}. \bibinfo{pages}{14610--14621}.
\newblock


\bibitem[\protect\citeauthoryear{Zeng, Bai, Chen, Chen, Han, Zhu, and Cui}{Zeng
  et~al\mbox{.}}{2019}]%
        {zeng2019congestion}
\bibfield{author}{\bibinfo{person}{Gaoxiong Zeng}, \bibinfo{person}{Wei Bai},
  \bibinfo{person}{Ge Chen}, \bibinfo{person}{Kai Chen},
  \bibinfo{person}{Dongsu Han}, \bibinfo{person}{Yibo Zhu}, {and}
  \bibinfo{person}{Lei Cui}.} \bibinfo{year}{2019}\natexlab{}.
\newblock \showarticletitle{Congestion control for cross-datacenter networks}.
  In \bibinfo{booktitle}{\emph{2019 IEEE 27th International Conference on
  Network Protocols (ICNP)}}. IEEE, \bibinfo{pages}{1--12}.
\newblock


\end{thebibliography}
%\end{small}

%\newpage
\appendix
%\input{appendix.tex}
%\clearpage

\section*{Appendix}

\section{Background on Multi-Objective Reinforcement Learning}\label{ap:morl}
We briefly review multi-objective reinforcement learning (MORL) techniques~\cite{morl,schaul2015universal,mossalam2016multi,yang2019generalized,abels2019dynamic} we used in this paper. In particular, we introduce \textbf{two design choices} and \textbf{two enhancements} in order to adopt MORL for the problem of \sys. To the best of our knowledge, \mocc is the first work to solve the multi-objective CC problem by adopting the MORL framework. %For more details about MORL, please see the thorough survey about MORL~\cite{morl} and recent works~\cite{schaul2015universal,mossalam2016multi,yang2019generalized,abels2019dynamic}.
%More details can be found in~\cite{morl,yang2019generalized,abels2019dynamic}.

As a fast-growing research area, MORL is a generalized RL framework used for solving multi-objective Markov decision process (MOMDP). MOMDP extends Markov decision process (MDP) by incorporating multiple optimization objectives. A MOMDP can be formalized by the tuple $<S,A,P,\vec{r},\Omega,f_{\Omega}>$. $S$ and $A$ are the state space and action space. $P(s'\in S|s\in S,a\in A)$ define the state transition probability. $\vec{r}(s,a)$ consists a vector of reward functions w.r.t. each objective repectively. $\Omega$ defines the preference space and $f_{\Omega}(\vec{r})$ is the preference function producing the integrated reward value with given preference and collected objective-specific rewards. 
With $\Omega$ fixed to a single preference, a MOMDP degrades to a standard MDP and can be solved with single-objective RL algorithm. %In this paper, we consider the preference space:
%\begin{equation*} 
%\Omega:\{<w_{thr},w_{lat},w_{loss}>|w_{thr} + w_{lat} + w_{loss}=1,w_{i}\in (0,1)\}
%\end{equation*}
Under the linear preference function $f_{\Omega}(\vec{r})=\vec{w}^T \vec{r}$ in \mocc, the optimal policy set for our MOMDP is called a convex coverage set (CCS), which contains all optimal polices for any given preference $\vec{w}$. The learning goal is to recover the entire CCS, the optimal policy set for all possible application requirements, and apply corresponding one for any given application requirement $w$.

In this paper, to use MORL for \mocc, we make the following two design choices:

\parab{We use multiple-policy approach to optimize for each application preference rather than average preference of all applications.} Existing MORL algorithms can be divided into two groups: single-policy approaches and multiple-policy approaches~\cite{morl}. While single-policy approaches learn a single policy to optimize the average performance among different objectives, multiple-policy approaches learn and maintain a set of optimal policies. A general method adopted by multiple-policy approaches is to collect policies by running standard RL algorithm over different preferences~\cite{natarajan2005dynamic}. To simultaneously support multiple existing applications and quickly adapt to new arrival applications while not compromise the performance of old applications, \mocc adopts multiple-policy approach to learn and maintain multiple policies for every application requirement.

\parab{We use policy-based rather than value-based algorithms to optimize for CC where the decision space is continuous instead of discrete.} The training approach for RL can be divided into two groups: value-based approaches and policy-based approaches. Value-based approaches estimate the value of each state, and take action with highest value estimation. Policy-based approaches directly learn the optimal policy for the task. Generally, policy-based algorithms outperform value-based ones for continuous control problems because policy models can directly output continuous action~\cite{ppo}. Recent MORL algorithms are all value-based approaches \cite{schaul2015universal,abels2019dynamic,yang2019generalized}. However, to better fit the continuous property of the sending rate in CC, \mocc adopts policy-based algorithm PPO and transfers MORL structures from value-based to policy-based.

Besides, based on the above design choices, we also make following two enhancements towards MORL:

\parab{Enhance PPO with requirement replay learning algorithm in order not to compromise performance of old application.} DQN-based Conditional Network (CN) proposed in \cite{abels2019dynamic} adopts new training policy to adapt to new policy as well as maintain previously learned policies. Based on it, we design the requirement replay learning mechanism for PPO during online learning (\S\ref{subsec:online}) to recall old applications.

\parab{Enhance MORL training with transfer learning to accelerate training speed.} Furthermore, with the prospect shown in \cite{schaul2015universal} of using transfer learning to solve new objective faster from similar learned objectives, we design the two-phase offline training scheme with the neighbourhood-based algorithm (\S\ref{subsec:offline}) to unleash the full power of transfer learning and significantly speedup our training (\S\ref{subsec:DD}).

\section{Neighborhood-based algorithm}\label{ap:neighbor}
In the fast traversing phase of offline training (\S\ref{subsec:offline}), to speedup the training speed, we design a neighborhood-based algorithm to leverage the solutions of its neighboring RLs.

Our algorithm is based on Dijkstra's shortest path algorithm. By constructing an undirected graph $G$ from candidate objectives, we reorder objectives according to their distances from the bootstrapped ones. We construct $G$ with vertices representing candidate weight vectors (all weight vectors satisfying $w_{thr} + w_{lat} + w_{loss}$=$1$, $w_{i}\in (0,1)$, at a given step size), and edges representing the neighborhood relationships. We define two weight vectors to be neighbor if they have difference in at most two dimensions and the difference is within unit step size. For example, at the step size of 0.1, $<$$0.2, 0.4, 0.4$$>$ and $<$$0.2, 0.5, 0.3$$>$ are neighbors, $<$$0.2, 0.4, 0.4$$>$ and $<$$0.1, 0.5, 0.4$$>$ are neighbors, but $<$$0.2, 0.4, 0.4$$>$ and $<$$0.1, 0.3, 0.6$$>$ are not neighbors. We add edges between neighbors and set all edge weights to be 1.
%The defined distance between vertices fits our goal to transfer knowledge between neighbor objectives. %It is different from the Euclidean distance between weight vectors, so we need a graph-based algorithm to sort objectives.

%$<1,0,0>$, $<0,1,0>$, $<0,0,1>$ only have two neighbors, and other vertices have six different neighbors. 

Algorithm \ref{alg:nei} presents the pseudocode for our neighborhood-based algorithm on $G$. We iterate on each bootstrapped objective/vertices and apply Dijkstra's algorithm: For the current bootstrapped vertices $o$, the algorithm extracts the nearest unvisited vertices, puts it into the list $L$, and updates its unvisited neighbors' distances from $o$. Finally, $L$ contains a sorted list of objectives that can be used as the training order for our \mocc model. 

To accelerate the fast traversing phase, we chose the bootstrapped objectives $<$$0.6,0.3,0.1$$>$, $<$$0.1,0.6,0.3$$>$, $<$$0.3,0.1,0.6$$>$ to cover different application requirements as much as possible. Figure~\ref{fig:nei} illustrates the traversing path.

\begin{algorithm}[t]
  \LinesNumbered 
  \SetInd{0.25em}{1em}
  \caption{Neighborhood-based Objective Sorting Algorithm}\label{alg:nei}
  \SetKwData{Left}{left}\SetKwData{This}{this}\SetKwData{Up}{up}
  \SetKwFunction{Union}{Union}\SetKwFunction{FindCompress}{FindCompress}
  \SetKwInOut{Input}{input}\SetKwInOut{Output}{output}
  \Input{The undirected objective graph $G=(V,E)$, the bootstrapped vertices $O$}
  \Output{The sorted objective list $L$}
  $L\leftarrow[]$\;
  \ForEach{$v\in V$}{
    $v.visited\leftarrow False$\;
    \For{$i\leftarrow 1$ \KwTo $|O|$}{
      \If {$v$ has edge with bootstrapped vertices $O$}{
        $v.d[i]\leftarrow 1$\;
      }\Else{
        $v.d[i]\leftarrow \infty$\;}
    }
  }
  \For{$i\leftarrow 1$ \KwTo $|O|$}{
    $visits\leftarrow \ceil{\frac{|V|}{|O|}}$\;
    \If{$O[i].visited=False$}{
      Append $O[i]$ to L\;
      $O[i].visited\leftarrow True$\;
      $visits\leftarrow visits-1$\;
    }
    \While{$visits > 0$ and $L$ is not full}{
      Find $u\in V$ with minimum $u.d[i]$ and $u.visited=False$\;
      Append $u$ to $L$\;
      $u.visited =True$\;
      $visits\leftarrow visits-1$\;
      \ForEach{w $\in$ neighbors of u}{
        \If{$w.visited=False$ and $u.d[i] + 1 < w.d[i]$}{
          $w.d[i]\leftarrow u.d[i] + 1$\;
        }
      }
    }
  }
\end{algorithm}

%\section{Online Algorithm}\label{appendix:online}
%The online training algorithms of \mocc is basically the same with that in $S$\ref{subsec:offline}. However, different from offline training procedure, we cannot control the sequence of coming applications and objectives. Without maintenance mechanism for trained policies, \mocc may overfit to the present \yq{frequent appeared} applications' requirements and forget learned policies for past ones. 

%To avoid the phenomenon, inspired by \cite{dynamic}, we modify the training objective function to "replay" history application objectives. During the online learning, \mocc records all coming applications' objectives (weight vectors) in a list. For one online training step, a trajectory is trained both on the current objective and a random previously encountered objective sampled from the list. In other words, we compute the loss for a given sample as the sum of the losses on the active weight vector $w_t$ and on $w_j$ randomly sampled from the list.

%\begin{equation}
%\begin{aligned}
%& L(\theta,W^{(t)})= \frac{1}{2} \times [ L(\theta,W^{(t)}) +L(\theta,W^{(j)}) ] \\
%& L(\theta,W)= \\
%& \hat{\mathbb{E}}_{t}\left[\min \left(r_{t,W}(\theta) \hat{A}_{t,W}, \operatorname{clip}\left(r_{t,W}(\theta), 1-\epsilon, 1+\epsilon\right) \hat{A}_{t,W}\right)\right]
%\end{aligned}
%\label{eq:online}
%\end{equation}

\begin{figure}[t]
  \centering
  \includegraphics[width=8cm]{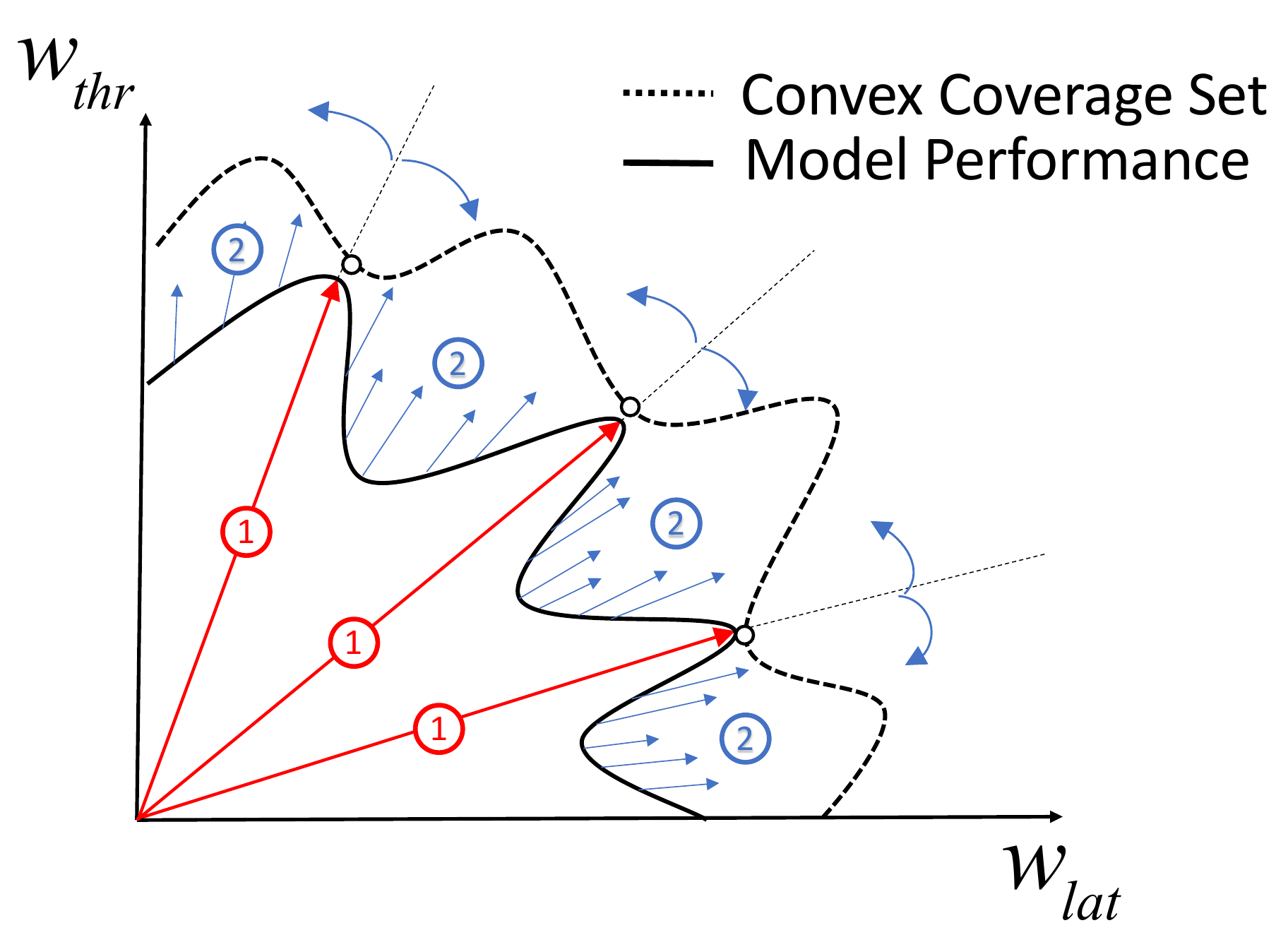}
  \vspace{-0.1in}
  \caption{The illustration of how \mocc's performance is improved during the two-phase offline training on the preference space. For the fisrt bootstrapping phase, \mocc learns the optimal policies on a few pivot points. Transferred from these learned policies, we iterate on other objectives in a  neighborhood-based way (Appendix~\ref{ap:neighbor}) to improve the overall performance in the fast traversal phase.}\label{fig:pf}
  \vspace{-0.1in}
\end{figure}

\section{Two-phase offline training illustration}\label{ap:twophase}
We use Figure~\ref{fig:pf} to illustrate why the two-phase training can effectively accelerate the training. Here we only consider the two-dimensional preference space with two performance metrics, throughput and latency, for visual simplicity. Each point in this figure shows a certain objective (a combination of throughput and latency requirements) and the distance to the origin point shows the effectiveness (\ie, how optimally the model can act) of the model.

The dash curve is convex converge set (CCS), which represents the optimal solution of the task. The solid curve represents the effectiveness of our model. The goal of training is to make the effectiveness (solid line) of our model to approach the optimality (dash line). 

After the bootstrapping phase, we have a set of solutions for certain set of objectives, shown as pivot points in the figure. These points are: 1) uniformly distributed, and 2) very close to optimality. The training is also fast because the set is small. Then, in the fast traversing phase, we will determine the rest of the points. Because we already have those pivot points, we can have a very good starting point during training and the effectiveness will not be far away from optimality. Meanwhile, the training can be effectively accelerated by using the information provided by pivot points.

\end{document}